\documentclass[a4paper,10pt]{article}
\usepackage[utf8]{inputenc}
\usepackage{amsthm}
\usepackage{amsfonts}
\usepackage{amssymb}
\usepackage{amsmath}
\usepackage{bbold}
\allowdisplaybreaks
\usepackage{mathtools}
\usepackage[english]{babel}
\usepackage{color}
\usepackage{slashed}
\usepackage{enumerate}
\usepackage{graphicx}
\usepackage{systeme}
\usepackage{dsfont}
\usepackage{relsize}
\usepackage[margin=0.90in]{geometry}
\usepackage{float}

\usepackage[labelformat=simple]{subcaption}

\usepackage{bmpsize}
\usepackage{epstopdf}
\usepackage{bbm}
\usepackage{titling}
\usepackage{authblk}

\usepackage[absolute]{textpos}
\usepackage{physics}
\usepackage[hang,flushmargin]{footmisc}

\usepackage{xfrac}
\usepackage{nicefrac}
\usepackage{esvect}
\usepackage{cases}
\usepackage{empheq}
\usepackage{stmaryrd}
\usepackage{cancel}
\usepackage[linguistics]{forest}
\usepackage{url}
\usepackage[font=small]{caption}
\usepackage{multirow}
\usepackage[percent]{overpic}
\usepackage{tikz}

\usepackage{setspace}
\usepackage{xcolor,etoolbox}
\usepackage[dvipsnames]{xcolor}

\usepackage{lipsum}
\usepackage[giveninits=true,sortcites=true,date=year,maxbibnames=99,doi=false,isbn=false,url=false,eprint=false]{biblatex}
\renewbibmacro{in:}{}
\DeclareFieldFormat{pages}{#1}
\AtEveryBibitem{%
  \clearlist{language}%
}
\usepackage{csquotes}

\title{Transmissibility, boundary-guided waves, and representative unit cell choice in finite-sized metamaterials}
\renewcommand\thefootnote{\arabic{footnote}}  % arabic numbers
\makeatother
\author{%
  Plastiras Demetriou\textsuperscript{1}%
  \quad
  and
  \quad
  Gianluca Rizzi\textsuperscript{1,*}%
}
\thanksmarkseries{arabic}
\date{\today}
\begin{document}
\maketitle
%
%%%%%%%%%%%%%%%%%%%%%%%%%%%%%%%%%%%%%%%%%%%%%%%%%%%%%%%%%%%%
% define affiliation footnote
\footnotetext[1]{Faculty of Architecture and Civil Engineering, TU Dortmund, August-Schmidt-Str. 8, 44227 Dortmund, Germany}

% Symbol footnote (for corresponding author)
\begingroup
\renewcommand{\thefootnote}{\fnsymbol{footnote}}
\footnotetext[1]{
  Corresponding author. Email: gianluca.rizzi@tu-dortmund.de
}
\endgroup
%%%%%%%%%%%%%%%%%%%%%%%%%%%%%%%%%%%%%%%%%%%%%%%%%%%%%%%%%%%%
%
%
%
%
%
%%%%%%%%%%%%%%%%%%%%%%%%%%%%%%%%%%%%%%%%%%%%%%%%%%%%%%%%%%%%
%%%%%%%%%%%%%%%%%%%%%%%%%%%%%%%%%%%%%%%%%%%%%%%%%%%%%%%%%%%%
\begin{abstract}
  The implications of selecting different unit cells are often overlooked in both direct studies of microstructured materials and their homogenized equivalents.
  Investigating the effects of unit cell selection is crucial not only for understanding boundary phenomena but also for identifying which finite-sized metamaterial performs best for a given purpose (e.g., achieving zero or minimal transmissibility in the band-gap range).
  This study examines how the choice of a representative unit cell in periodic metamaterials influences boundary effects and, consequently, the transmissibility of finite-size samples, while providing a strategy based on eigenfrequency calculations that enables a priori optimization.
\end{abstract}

%
%
%
%
%%%%%%%%%%%%%%%%%%%%%%%%%%%%%%%%%%%%%%%%%%%%%%%%%%%%%%%%%%%%
%%%%%%%%%%%%%%%%%%%%%%%%%%%%%%%%%%%%%%%%%%%%%%%%%%%%%%%%%%%%
\section{Introduction}
\label{sec:intro}
%%%%%%%%%%%%%%%%%%%%%%%%%%%%%%%%%%%%%%%%%%%%%%%%%%%%%%%%%%%%
%%%%%%%%%%%%%%%%%%%%%%%%%%%%%%%%%%%%%%%%%%%%%%%%%%%%%%%%%%%%
In recent years, mechanical metamaterials have attracted significant attention for their ability to manipulate mechanical waves in ways that exceed the capabilities of classical materials.
The potential applications of mechanical metamaterials are broad, including vibration isolation \cite{al2022advances,otlu2023three,li2017design, yuksel2020realization, nadejde2023pushing}, energy harvesting \cite{lee2022piezoelectric, de2020experimental,li2017design, zhao2022graded}, and many others \cite{vo2022blast,haid2023mechanical, buckmann2014elasto,zanotto2022metamaterial, chen2020light, pishvar2020foundations, kolken2017auxetic, cheng2022design, koh2011generalized}.

One of the most interesting topics is the study of mechanical metamaterials capable of generating band-gaps, which can be used for shielding \cite{miniaci2016large, oh2017elastic, rizzi2022metamaterial, wu2020mechanical}, vibration control \cite{baravelli2013internally, shen2025vibration, ji2021vibration}, and soundproofing applications \cite{hermann2024design, lai2001engineering, yang2024low, krushynska2014towards}, among others.
Usually, to analyze the dispersion properties of a metamaterial, a standard Bloch–Floquet analysis \cite{cool2024guide,iorio2024roton,nadejde2025mechanisms,carrillo2025symmetry,maurin2018probability} is performed on a Representative Unit Cell (RUC), the smallest portion of the metamaterial that can tile the plane by translation.
However, the information obtained from this analysis often falls short when considering metamaterials of finite size.
An important reason is that the truncation of periodicity due to the finite size dictates the shape of the boundaries, and consequently the boundary effects that may arise \cite{ren2006electronic,demetriou2024reduced,hermann2024design, ramirez2023surface,demetriou2025effective,HERMANN2026106102}.
Boundary-localized modes may appear in the band-gap range and act as unintended waveguides, thereby reducing the effectiveness of the metamaterial.
To quantitatively assess this effectiveness, the \textit{transmissibility coefficient} ($T$) is typically employed.

The main goal of this paper is to investigate how different choices of Representative Unit Cells, which give rise to different finite-sized samples, affect their effective transmissibility.
Furthermore, it is proposed a set of tools based on eigenfrequency analyses that enable the a priori characterization and optimization of RUCs.

The main novelty lies in the development of a new procedure for the a priori selection of RUCs that effectively attenuate wave propagation within band-gap frequency ranges.
The proposed approach is based on evaluating the energy distribution associated with the eigenmodes of the structure in order to identify spurious boundary-guided waves that may arise within the band-gap ranges and compromise their attenuation performance.
By detecting and minimizing the influence of these localized modes, the methodology enables the design of finite-size structures with improved wave attenuation capabilities, ultimately leading to more effective acoustic and vibration shields.

The article is structured as follows:
Section~\ref{sec:intro} introduces the topic and motivation;
Section~\ref{sec:cutNshape} describes the general procedure for generating a RUC;
Section~\ref{sec:transmission} defines transmissibility and presents the test setup;
Section~\ref{sec:4reso} focuses on a four-resonator unit cell metamaterial, including RUC choice, transmissibility, and boundary effects, as well as a discussion;
Section~\ref{sec:circ} covers a square-circular-hole unit cell metamaterial, also including RUC choice, transmissibility, and boundary effects, as well as a discussion;
Section~\ref{sec:clover} presents an example of an irregular shape sample, a four-leaf clover;
finally, Section~\ref{sec:conclusions} presents the conclusions.

%%%%%%%%%%%%%%%%%%%%%%%%%%%%%%%%%%%%%%%%%%%%%%%%%%%%%%%%%%%%
\begin{figure}[!htbp]
  \centering
  \includegraphics[width=0.7\textwidth]{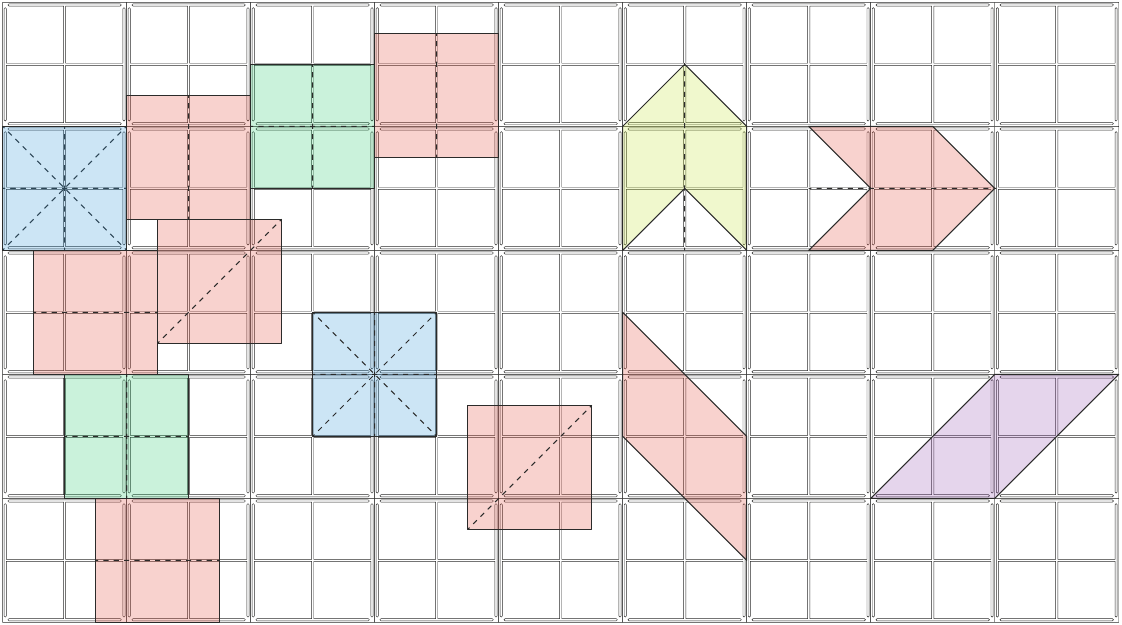}
  \\[10pt]
  % \resizebox{0.8\textwidth}{!}{
  \begin{forest}
    for tree={
    l sep=5pt,        % horizontal spacing
    s sep=1pt,        % vertical spacing
    parent anchor=south,
    child anchor=north,
    edge={-},
    edge path={
        % vertical from center of parent bottom to midpoint
        \noexpand\path[\forestoption{edge}]
        (!u.south) coordinate (tmp) -- (!v.north |- tmp) -- (!v.north)
        \forestoption{edge label};
      }
    }
    [RUC
      % [Shape
      [\tikz{\draw (0,0) rectangle (0.8,0.8);}
        [RUC$_{\rm s}^{\scalebox{0.5}{$\square$}}$
          [RUC$_{\rm s}^\alpha$\\*
              \parbox{1cm}{\centering\includegraphics[width=1cm]{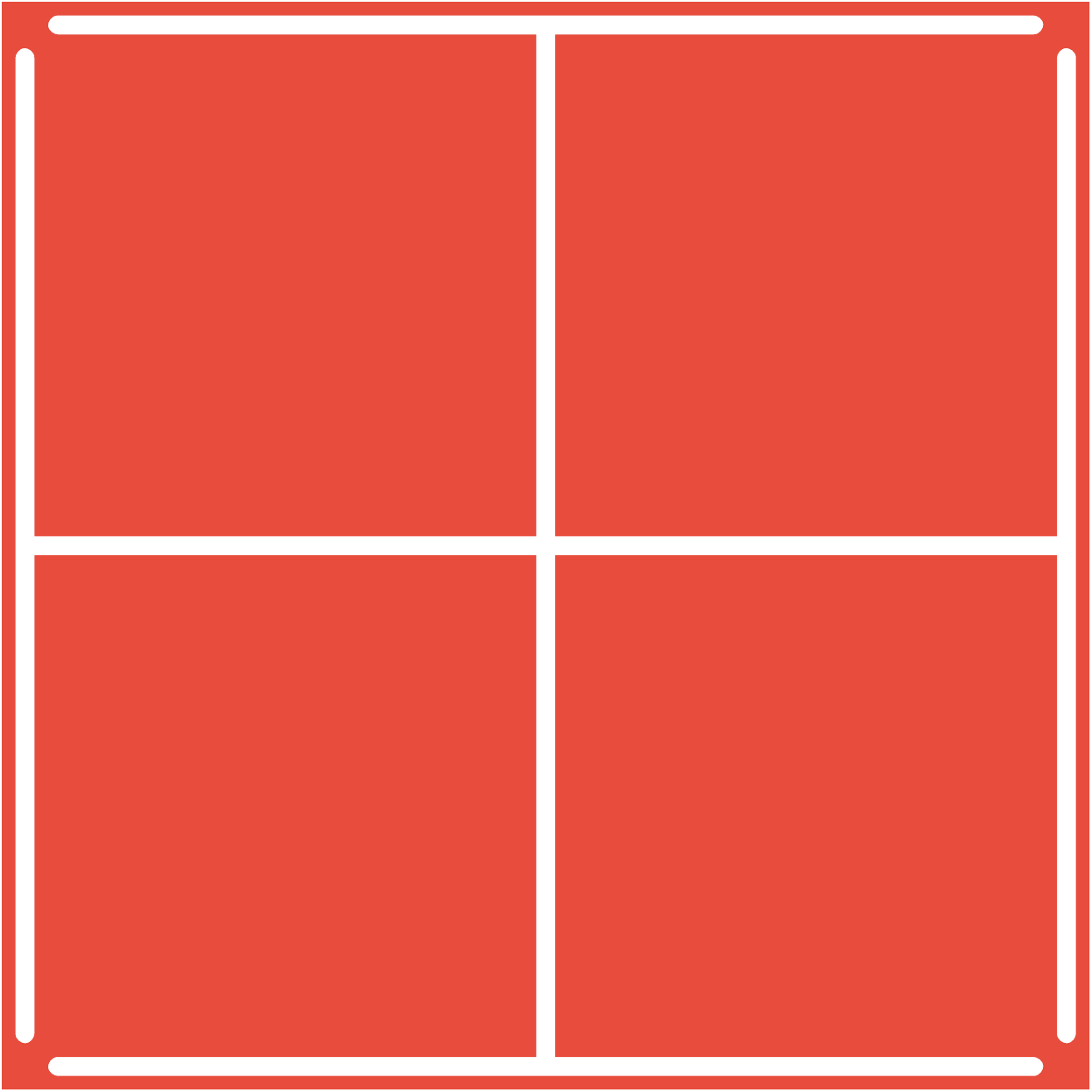}}]
            [RUC$_{\rm s}^\beta$\\*
              \parbox{1cm}{\centering\includegraphics[width=1cm]{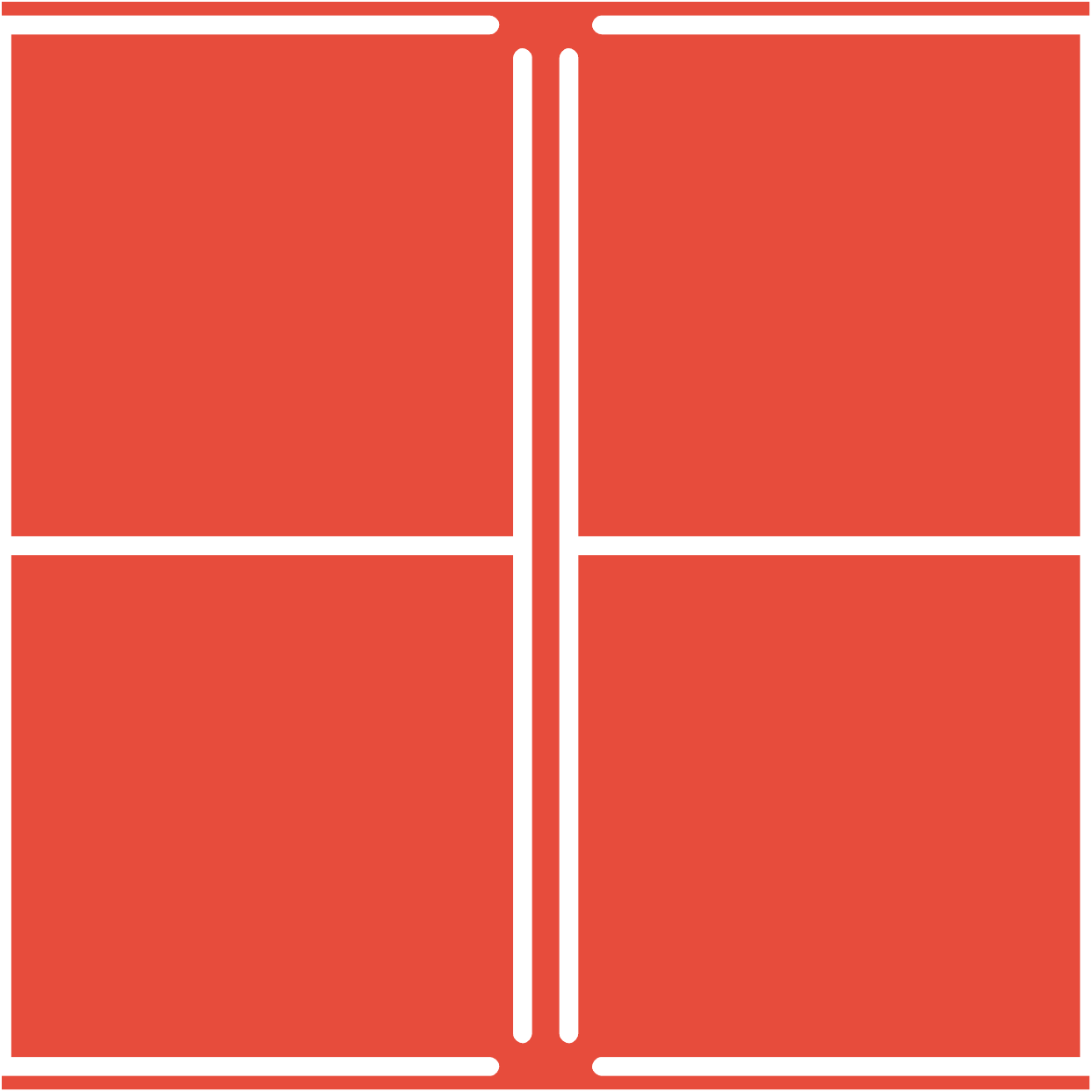}}]
            [{$\cdots$}]
        ]
      ]
      [\tikz{
          \draw (0,0.8) -- (0,1.6) -- (0.8,0.8) -- (0.8,0) -- cycle;
        }
        [RUC$_{\rm p}^{\scalebox{0.5}{$\square$}}$
          [RUC$_{\rm p}^\alpha$\\*
              \parbox{2cm}{\centering\includegraphics[width=2cm,angle=90]{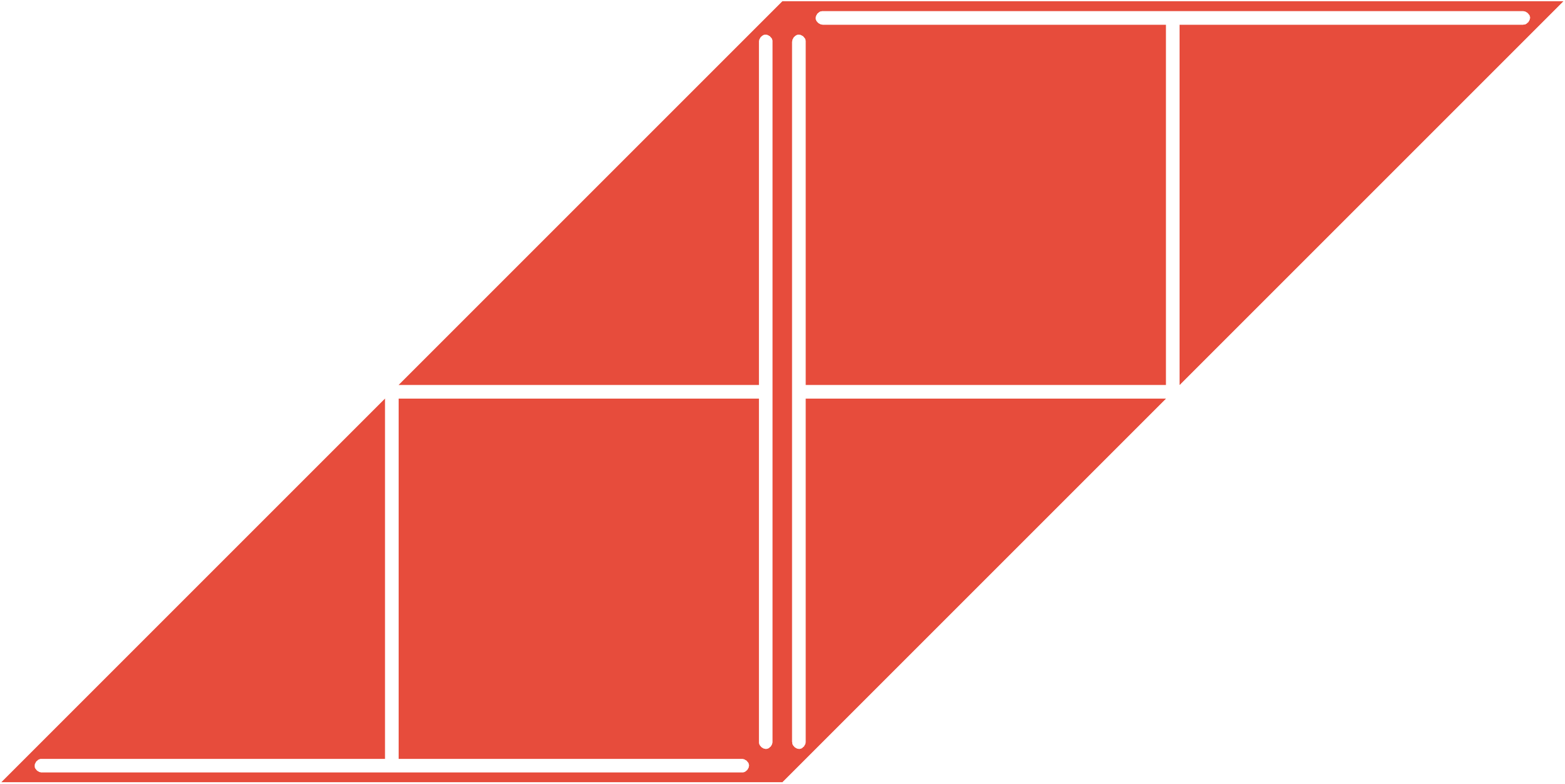}}]
            [RUC$_{\rm p}^\beta$\\*
              \parbox{2cm}{\centering\includegraphics[width=2cm,angle=90]{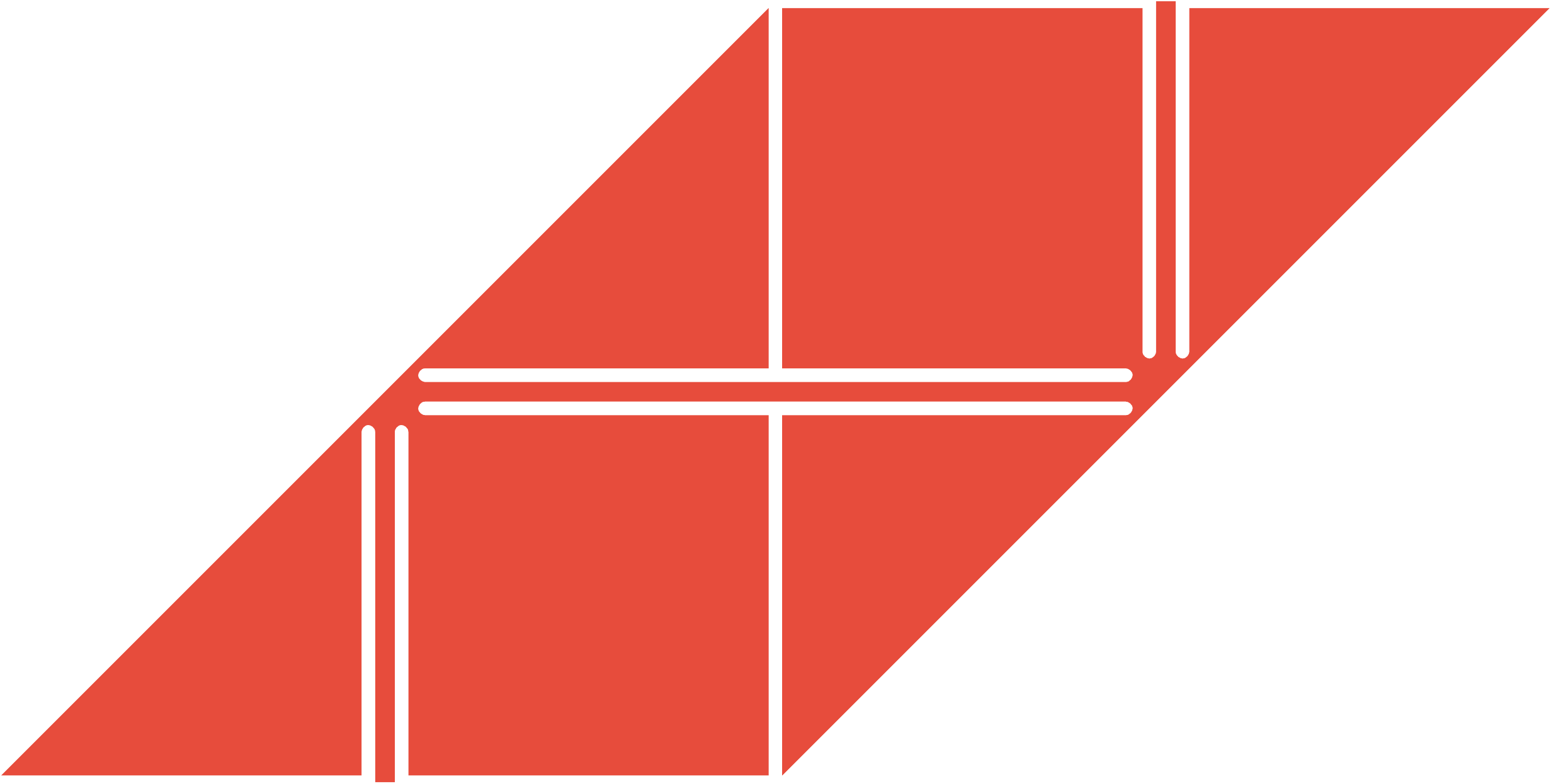}}]
            [{$\cdots$}]
        ]
      ]
      [\tikz{
          \draw (0,0) -- (0,0.8) -- (0.4,1.2) -- (0.8,0.8) -- (0.8,0) -- (0.4,0.4) -- cycle;
        }
        [RUC$_{\rm a}^{\scalebox{0.5}{$\square$}}$ % a/2 + r/Sqrt[2]
          [RUC$_{\rm a}^\alpha$\\*
              \parbox{1.5cm}{\centering\includegraphics[width=1cm]{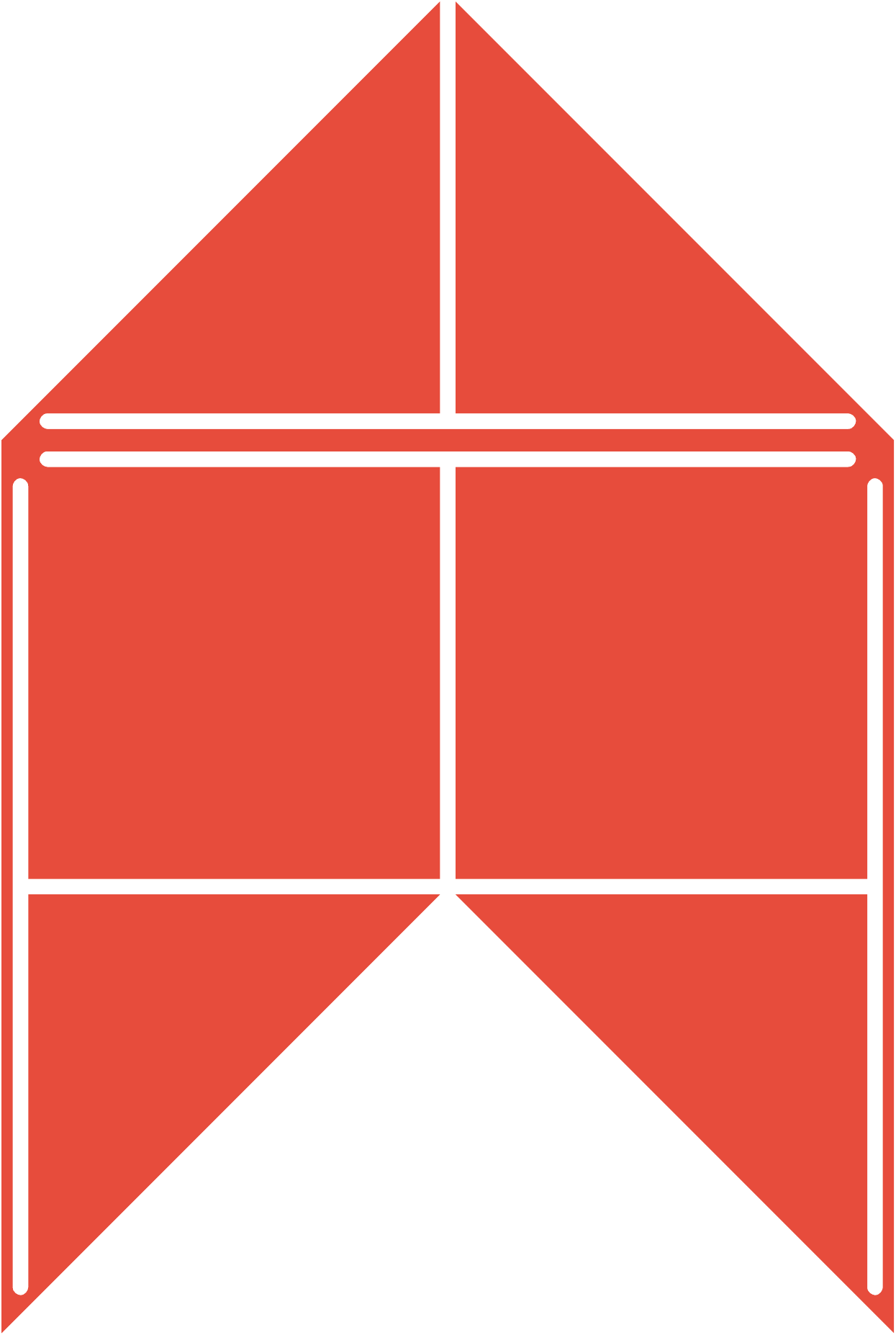}}]
            [RUC$_{\rm a}^\beta$\\*
              \parbox{1.5cm}{\centering\includegraphics[width=1cm]{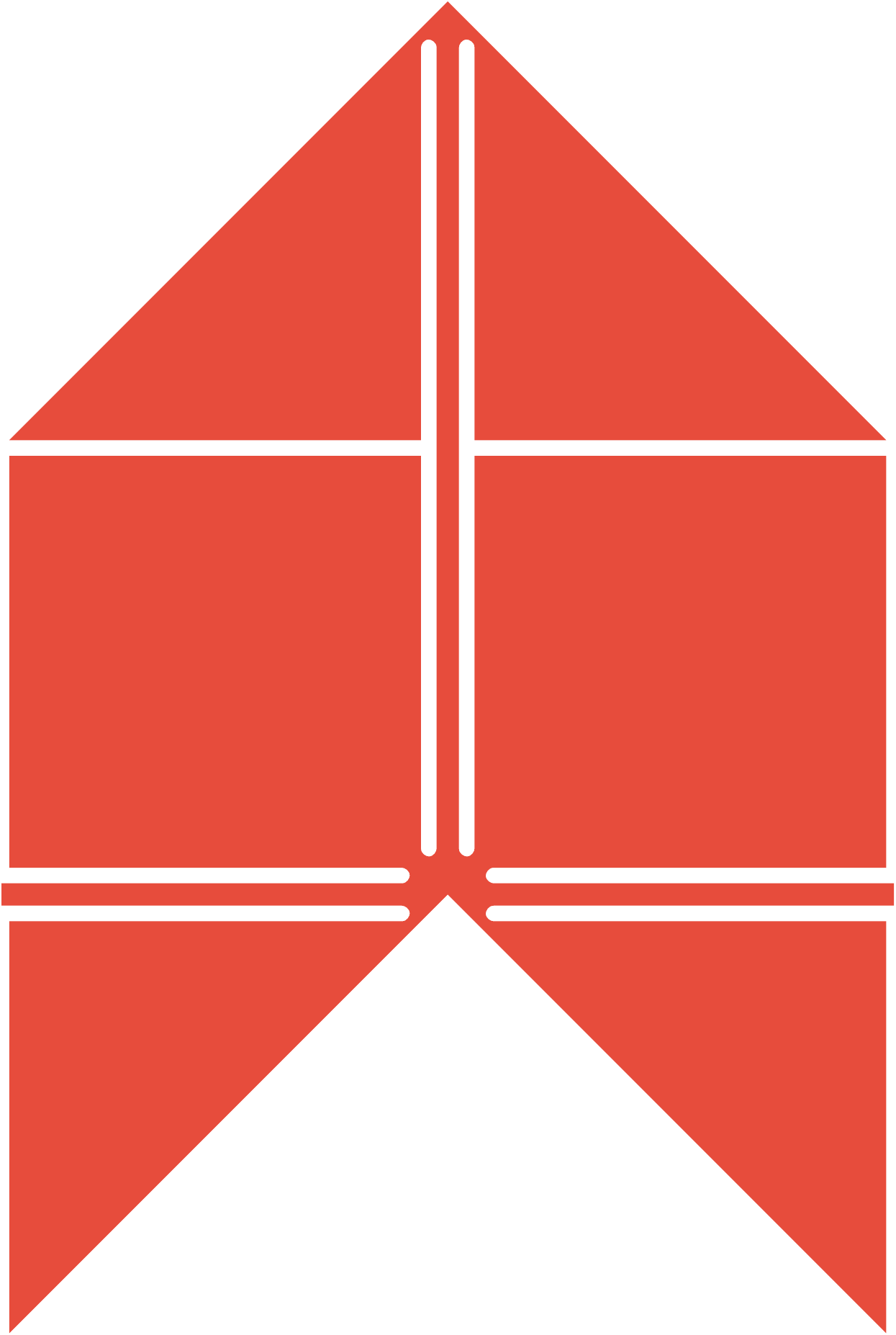}}]
            [{$\cdots$}]
        ]
      ]
      [{$\cdots$}
          [RUC$_{\scalebox{0.5}{$\square$}}^{\scalebox{0.5}{$\square$}}$
            [{$\cdots$}]
              [{$\cdots$}]
              [{$\cdots$}]
          ]
      ]
    % ]
    ]
  \end{forest}
  % }
  \caption{
    (\textit{Top})
    Different possible choices of RUC:
    the red ones are not admissible for finite-size tests due to floating parts;
    the purple one belongs to the triclinic class of symmetry;
    the yellow one belongs to the monoclinic class of symmetry;
    the green ones belong to the orthotropic class of symmetry;
    the blue ones belong to the tetragonal class of symmetry.
    (\textit{Bottom})
    Scheme for classifying the type of RUC, where the subscript Latin letter represents the shape of the RUC (square, parallelogram, arrow, etc.) and the superscript Greek letter the ``cut'' of material chosen for that shape.
  }
  \label{fig:RUC_tree}
\end{figure}
%%%%%%%%%%%%%%%%%%%%%%%%%%%%%%%%%%%%%%%%%%%%%%%%%%%%%%%%%%%%

%
%
%
%
%%%%%%%%%%%%%%%%%%%%%%%%%%%%%%%%%%%%%%%%%%%%%%%%%%%%%%%%%%%%
%%%%%%%%%%%%%%%%%%%%%%%%%%%%%%%%%%%%%%%%%%%%%%%%%%%%%%%%%%%%
\section{The general procedure for generating a RUC}
\label{sec:cutNshape}
%%%%%%%%%%%%%%%%%%%%%%%%%%%%%%%%%%%%%%%%%%%%%%%%%%%%%%%%%%%%
%%%%%%%%%%%%%%%%%%%%%%%%%%%%%%%%%%%%%%%%%%%%%%%%%%%%%%%%%%%%

The behavior of an infinite lattice is represented by the Bloch-Floquet (B-F) curves, which predict zero transmissibility within the band-gap ranges.
The size and position of the band gaps are determined by the geometrical and material properties of the RUC and are independent of the particular choice of RUC.
Indeed, provided that the correct Bloch-Floquet boundary conditions are applied, all valid RUCs represent the same infinite domain.
The Bloch-Floquet analysis is also unaffected by the apparent symmetry of the selected RUC, which may be lower than the actual symmetry of the structure. Again, provided that the correct Bloch-Floquet boundary conditions are imposed, the analysis reflects only the intrinsic symmetry of the metamaterial, for example the tetragonal symmetry class of the unit cells considered in this work.

However, when dealing with a finite sized domain, the choice of RUC, that is, the manner in which the infinite lattice is truncated, can significantly affect the transmissibility behavior, potentially in all propagation directions.
This occurs because the finite domain has boundaries that can support the propagation of boundary waves, an effect that classical Bloch-Floquet analysis cannot inherently capture.

An infinitely extended periodic metamaterial in two (or three) dimensions can be generated from a RUC by tessellation, where ``representative'' indicates that all geometrical and material information is contained within this minimal portion of the domain, and ``tessellation'' refers to covering the plane (or space, in 3D) by translation along integer multiples of two (or three) primitive vectors.
Primitive vectors are implicitly defined according to the tiling rule and therefore are not unique \cite{Ashcroft1976}.

A RUC is also not unique, as several choices can reproduce the same periodic architecture (see Fig.~\ref{fig:RUC_tree}).
However, in this work, any RUC containing disconnected or floating parts is avoided, since while such configurations may still be compatible with an ideal infinite tiling, they become problematic once the structure is truncated to a finite sample, where geometric continuity is essential.

Figure~\ref{fig:RUC_tree} shows a schematic of the possible choices of the RUC for a metamaterial whose highest possible symmetry class is tetragonal.
Different choices of RUC can arise from a two-step procedure: first selecting a shape of minimal area, and then positioning that shape on the lattice, ``cutting out'' a possible RUC.
That minimal area is also the one that must be used in a Bloch–Floquet analysis; otherwise, the dispersion diagrams will contain additional spurious modes \cite{mukherjee2015phononic}.
Henceforth, only 2D materials with a maximum tetragonal symmetry class are considered.

%
%
%
%
%%%%%%%%%%%%%%%%%%%%%%%%%%%%%%%%%%%%%%%%%%%%%%%%%%%%%%%%%%%%
%%%%%%%%%%%%%%%%%%%%%%%%%%%%%%%%%%%%%%%%%%%%%%%%%%%%%%%%%%%%
\subsection{Shapes and cuts of the RUC}
%%%%%%%%%%%%%%%%%%%%%%%%%%%%%%%%%%%%%%%%%%%%%%%%%%%%%%%%%%%%
%%%%%%%%%%%%%%%%%%%%%%%%%%%%%%%%%%%%%%%%%%%%%%%%%%%%%%%%%%%%

Within Bloch–Floquet analysis, the symmetry class of a periodic metamaterial determines the geometry and extent of the irreducible Brillouin zone \cite{maurin2018probability} as well as the associated dispersion characteristics \cite{carrillo2025symmetry,iorio2024roton,nadejde2023pushing}, yet for finite samples these band-structure descriptors alone are insufficient to accurately predict transmission behavior, so complementary finite-size approaches are required to avoid mischaracterization.

For a metamaterial with tetragonal symmetry, the RUC is typically chosen to be square (see Fig.~\ref{fig:RUC_tree}), as it has the smallest area, allowing for an unambiguous Bloch–Floquet analysis while preserving the lattice symmetry class \cite{Ashcroft1976}.

%%%%%%%%%%%%%%%%%%%%%%%%%%%%%%%%%%%%%%%%%%%%%%%%%%%%%%%%%%%%
\begin{figure}[!htbp]
  \centering
  \includegraphics[width=0.8\textwidth]{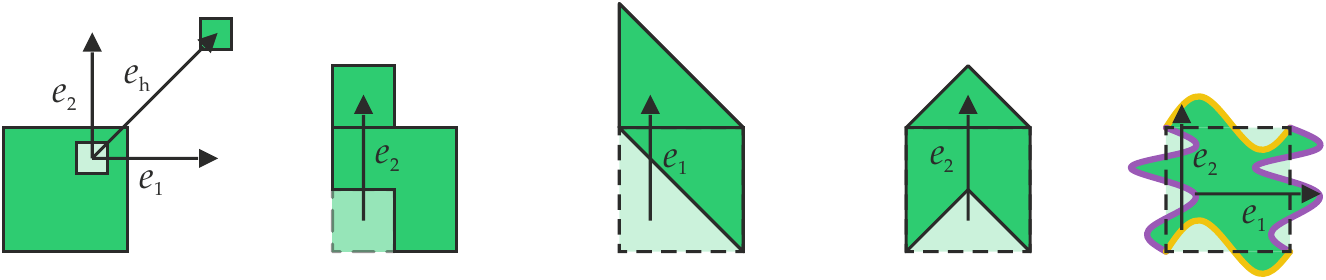}\\[10pt]
  \includegraphics[width=0.8\textwidth]{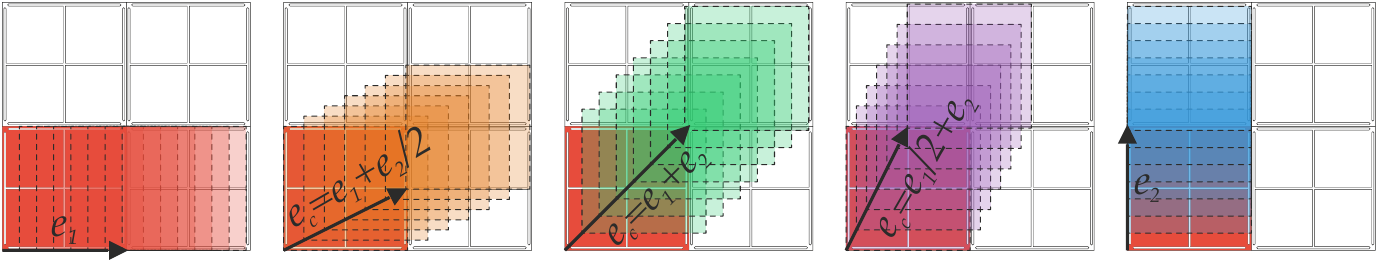}
  \caption{
    (\textit{Top}) Procedure for creating different shapes of the RUC for a tetragonal class of symmetry starting from a square, where the vectors $e_{1}=(L,0)^{\rm T}$ and $e_{2}=(0,L)^{\rm T}$ are two primitive vectors.
    (\textit{Bottom}) Procedure for creating different RUC cuts for a given square shape.
  }
  \label{fig:RUC_proc}
\end{figure}
%%%%%%%%%%%%%%%%%%%%%%%%%%%%%%%%%%%%%%%%%%%%%%%%%%%%%%%%%%%%

However, other equal-area shapes can be chosen (see Fig.~\ref{fig:RUC_proc}) by following a simple procedure \cite{Ashcroft1976}, in which (i) one starts with a valid unit cell shape (e.g., a square), (ii) removes a portion of the domain, and (iii) restores the removed geometry by translating it along a linear combination of integer multiples of the primitive vectors, as shown in Eq.(\ref{eq:diff_shape}),
\begin{equation}
  e_{\rm h} = a\,e_{1} + b\,e_{2}\, ,
  \qquad \text{where} \qquad
  (a,b) \in \mathbb{Z}\, ,
  \label{eq:diff_shape}
\end{equation}
where $e_{1} = (L,0)^{\rm T}$ and $e_{2} = (0,L)^{\rm T}$, with $L$ denoting the periodicity length, and (iv) identifies the resulting shapes as valid if they do not contain floating or disconnected parts.

An easy sub-case of this general procedure is illustrated in the last two panels on top of Fig.~\ref{fig:RUC_proc}.
In this case, starting from a valid unit cell shape, only pairs of boundaries where Bloch–Floquet periodic boundary conditions apply are modified, treating one boundary as the source and the other as the destination. Any geometry added or removed on the source boundary is correspondingly removed or added on the destination boundary.
This sub-case has the advantage of avoiding a class of shapes that are guaranteed to have disconnected parts, i.e., floating parts (see the unit cell shape of the leftmost panel of Fig.~\ref{fig:RUC_proc}).

Once Floquet-periodic conditions are applied to the correct boundaries, the location from which the RUC is ``cut out'' of the metamaterial is irrelevant for the Bloch–Floquet analysis, since it still represents an infinitely extended periodic domain.

For any fixed unit cell shape, sliding it across the lattice and cutting out the geometry produces a RUC, as illustrated in Fig.~\ref{fig:RUC_tree}.
To obtain all distinct cuts after choosing a shape, it is sufficient to translate it by linear combinations of suitable primitive vectors, as shown in Eq.(\ref{eq:diff_cuts}),
\begin{equation}
  e_{\rm c} = c\,e_{1} + d\,e_{2}\, ,
  \qquad \text{where} \qquad
  \begin{cases}
    0 \leq c < 1\, , \\
    0 \leq d < 1\, ,
  \end{cases}
  \qquad \text{and} \qquad
  (c,d) \in \mathbb{R}\, ,
  \label{eq:diff_cuts}
\end{equation}
in order to avoid repeated cuts (see Fig.~\ref{fig:RUC_proc}).

%
%
%
%
%%%%%%%%%%%%%%%%%%%%%%%%%%%%%%%%%%%%%%%%%%%%%%%%%%%%%%%%%%%%
\section{Transmissibility definition and test setup}
\label{sec:transmission}
%%%%%%%%%%%%%%%%%%%%%%%%%%%%%%%%%%%%%%%%%%%%%%%%%%%%%%%%%%%%

In this section, two linear elastic metamaterials are introduced, and their transmissibility (see Eq.~(\ref{eq:trans_defi}) for the definition) is evaluated in finite-size frequency-domain tests over a prescribed frequency range.
Particular emphasis is placed on the band-gap region to assess how the choice of the RUC influences wave propagation.
When a system is excited by a vibration applied at an input boundary ($B_{\rm in}$), the wave reaching an output boundary ($B_{\rm out}$) generally differs in both amplitude and phase, depending on the material and geometric properties and on the excitation frequency.
The \textit{transmissibility} ($T$) quantifies the fraction of the input vibration transmitted through the system to the output boundary and is defined as
\begin{equation}
  T = \frac{\lvert f_{\text{out}} \rvert}{\lvert f_{\text{in}} \rvert} > 0 \, ,
  \label{eq:trans_defi}
\end{equation}
and where $f_{\star}$ denotes a suitable measure of the input and output excitation.
Physically, the transmissibility $T$ has the following interpretations
\begin{equation}
  \begin{array}{lllllllllllll}
     &  & T > 1   &  & \text{(vibration amplification)} \, , &
     &  & T = 1   &  & \text{(unchanged transmission)} \, ,    \\
     &  & T < 1   &  & \text{(vibration attenuation)} \, ,   &
     &  & T \ll 1 &  & \text{(insulation)} \, .
  \end{array}
  \label{eq:trans_inter}
\end{equation}
In this work, particular attention is devoted to assessing whether, within the band-gap ranges, $T$ falls within the insulation regime ($T \ll 1$), as defined in Eq.(\ref{eq:trans_inter}).
%%%%%%%%%%%%%%%%%%%%%%%%%%%%%%%%%%%%%%%%%%%%%%%%%%%%%%%%%%%%
\begin{figure}[!htbp]
  \centering
  \includegraphics[width=0.8\textwidth]{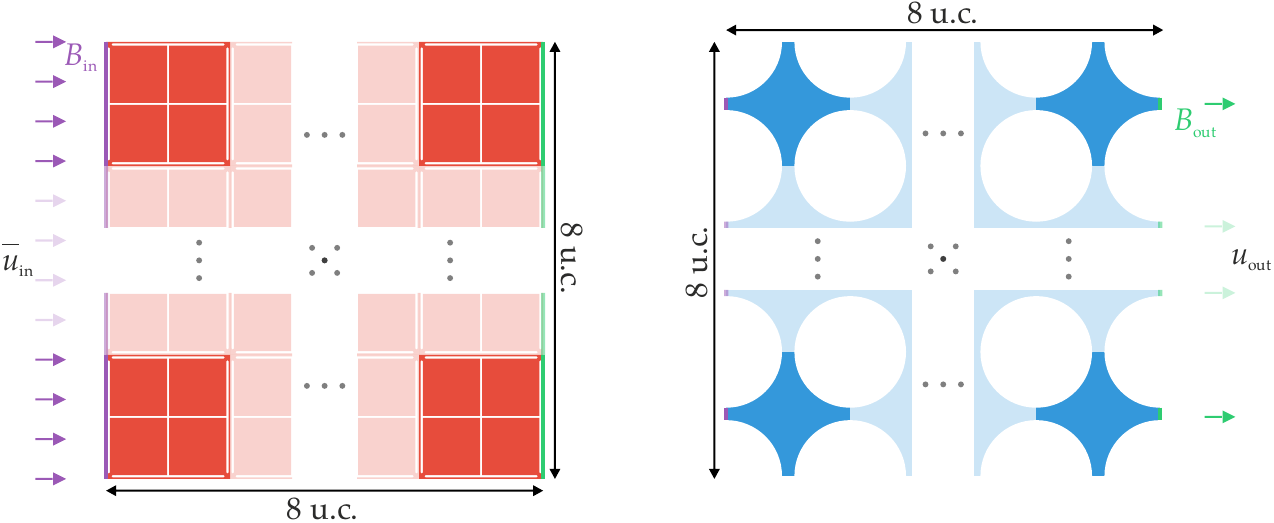}
  \caption{
    Schematics of the test described in Sec.~\ref{sec:4reso} and Sec.~\ref{sec:circ} for two RUCs: on the \textit{left}, the four-resonator metamaterial (RUC$_{\rm s}^{\alpha}$); on the \textit{right}, the square-circular-hole metamaterial (RUC$_{\rm s}^{\delta}$).
    An $8 \times 8$ array of RUCs is considered, and a displacement $\overline{u}_{\rm in} = \overline{u}_0 \, \frac{e_1}{L}$ is prescribed on the left boundary ($B_{\rm in}$), while all other boundaries are traction free.
  }
  \label{fig:BC}
\end{figure}
%%%%%%%%%%%%%%%%%%%%%%%%%%%%%%%%%%%%%%%%%%%%%%%%%%%%%%%%%%%%
To assess this, the displacement field is chosen as the excitation measure, as shown in Eq.~(\ref{eq:transmission}), and, more specifically, $T$ is evaluated as
\begin{equation}
  T=\dfrac{
    \dfrac{1}{\left| B_{\rm o} \right|}\int\limits_{B_{\rm o}} \sqrt{\left| u_1 \right|^2 + \left| u_2 \right|^2}
  }{
    \dfrac{1}{\left| B_{\rm i} \right|}\int\limits_{B_{\rm i}} \sqrt{\left| \overline{u}_1 \right|^2 + \left| u_2 \right|^2}
  } \, ,
  \label{eq:transmission}
\end{equation}
where $u_1$ and $u_2$ are the components of the displacement field ($u$), $\overline{u}_1$ is the component of the prescribed input displacement applied on the loaded boundary ($\overline{u}_{\rm in}$), and $B_{\rm i}$ and $B_{\rm o}$ denote the input and output boundaries, respectively (see Eq.(\ref{eq:load}) and the scheme in Fig.~\ref{fig:BC}). The samples are $8 \times 8$ arrays of RUCs, and it should be noted that on $B_{\rm i}$ the component $u_2$ is not constrained.
\begin{equation}
  \overline{u}_{\rm in} = \overline{u}_0 \,\frac{e_1}{L}
  \qquad\qquad
  \text{with}
  \qquad\qquad
  \overline{u}_0 = \frac{L \cdot \text{``u.c.''}}{100} = 4 \text{[mm]} \, .
  \label{eq:load}
\end{equation}
Since the analysis is performed in the frequency domain, the displacement components are complex, and taking the absolute value gives their magnitude, corresponding to the maximum amplitude of each component at a given frequency. This choice is conservative for the output excitation, as it also includes transverse displacement, which occurs even though the input excitation is purely horizontal, because the boundary is free to move transversely. Consequently, it is reasonable to adopt the same measure on the input boundary.

All results presented in the following sections were obtained through frequency-domain simulations conducted using COMSOL Multiphysics\textsuperscript{\textregistered}, and a mesh convergence study was performed to ensure the accuracy and reliability of the numerical results.

%
%
%
%
%%%%%%%%%%%%%%%%%%%%%%%%%%%%%%%%%%%%%%%%%%%%%%%%%%%%%%%%%%%%
\section{Four-resonator unit cell metamaterial}
\label{sec:4reso}
\begin{figure}[!htbp]
  \centering
  \includegraphics[width=0.8\textwidth]{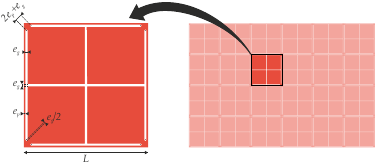}\\*
  \vspace*{12pt}
  \renewcommand{\arraystretch}{1.3}
  \centering
  \begin{tabular}{ccccccc}
    \hline
    $L$ [mm] & $e_{\rm g}$ [mm] & $e_{\rm p}$ [mm] & $\rho$  [kg/m$^3$] & $E$ [GPa] & $\nu$ [-] & $\eta$ [-]
    \\
    \hline
    50       & 0.875            & 0.625            & 4400               & 112       & 0.34      & 0.005
    \\
    \hline
  \end{tabular}
  \\[10pt]
  \begin{minipage}{0.185\textwidth}
    \centering
    \includegraphics[width=\linewidth]{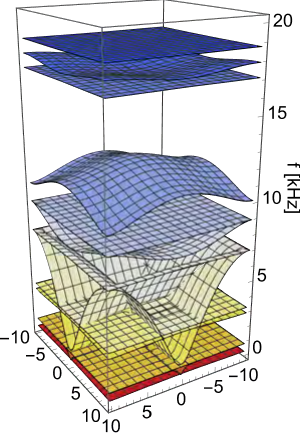}
  \end{minipage}
  \hfill
  \begin{minipage}{0.305\textwidth}
    \centering
    \includegraphics[width=\linewidth]{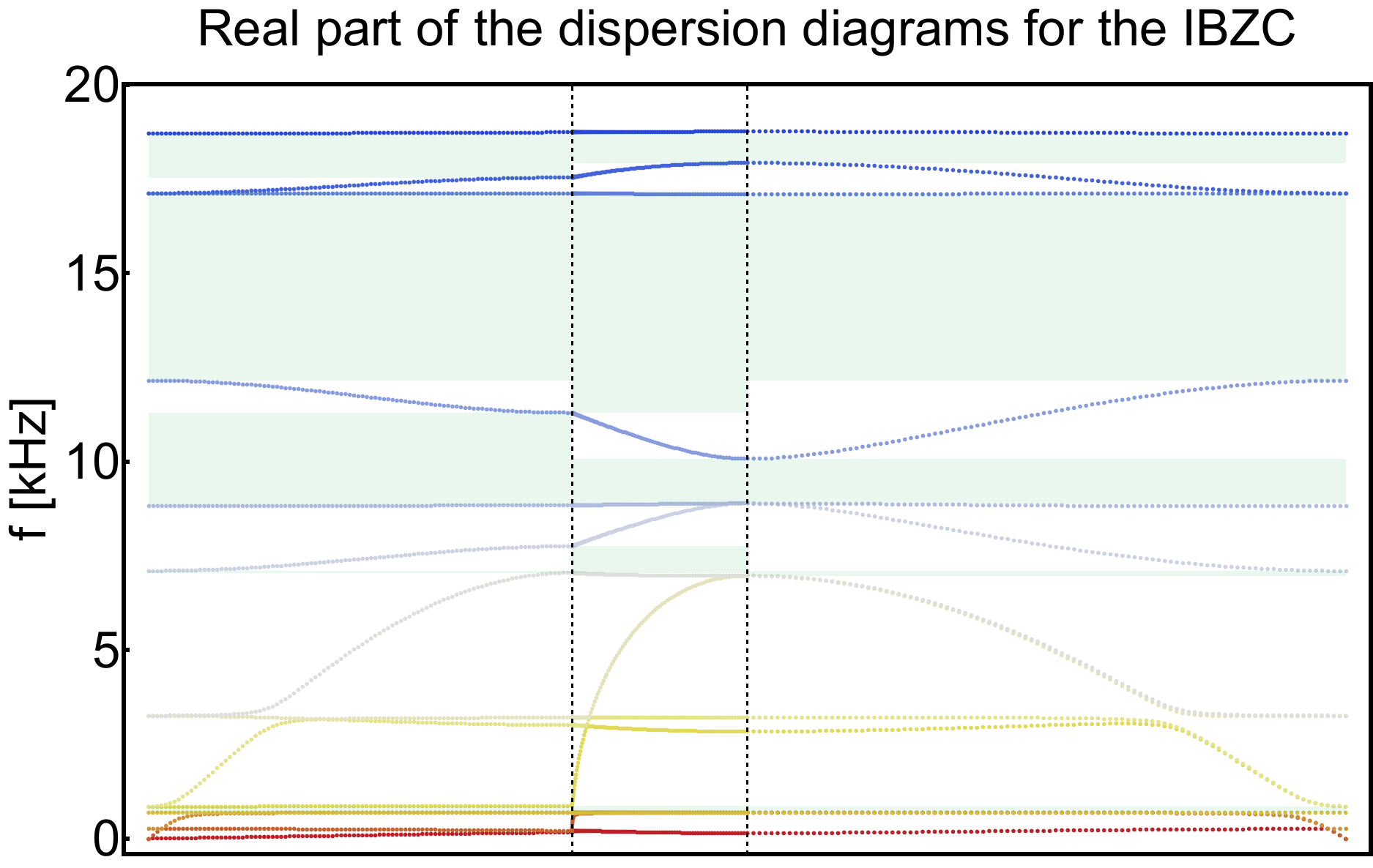}\\*
    \includegraphics[width=\linewidth]{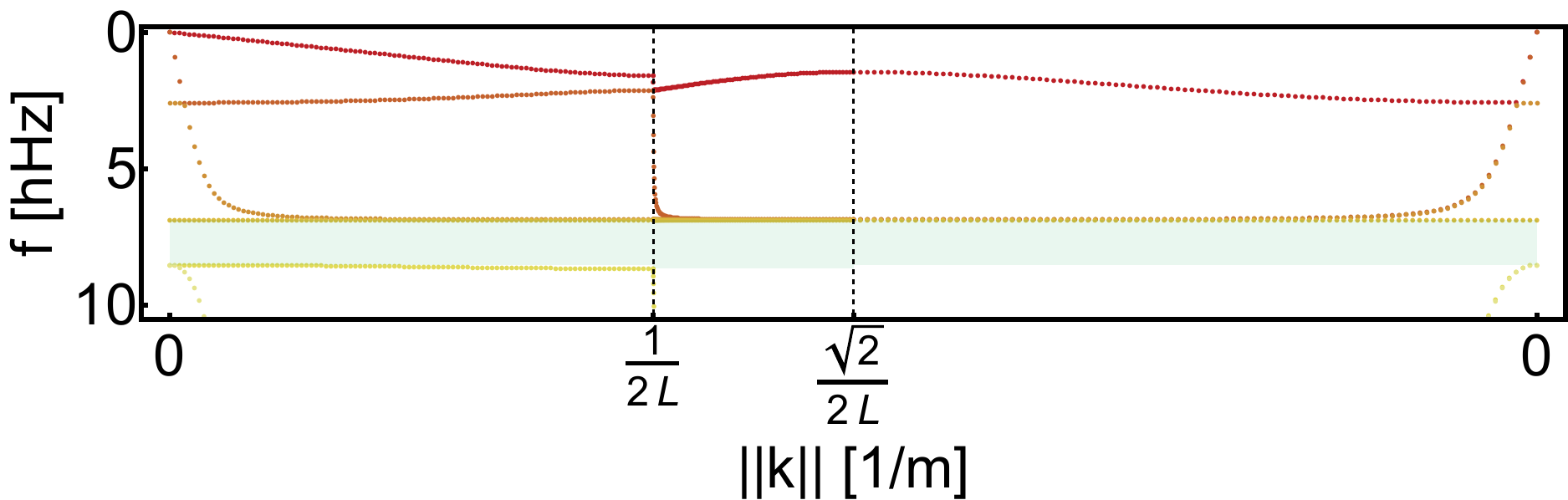}
  \end{minipage}
  \hfill
  \begin{minipage}{0.305\textwidth}
    \centering
    \includegraphics[width=\linewidth]{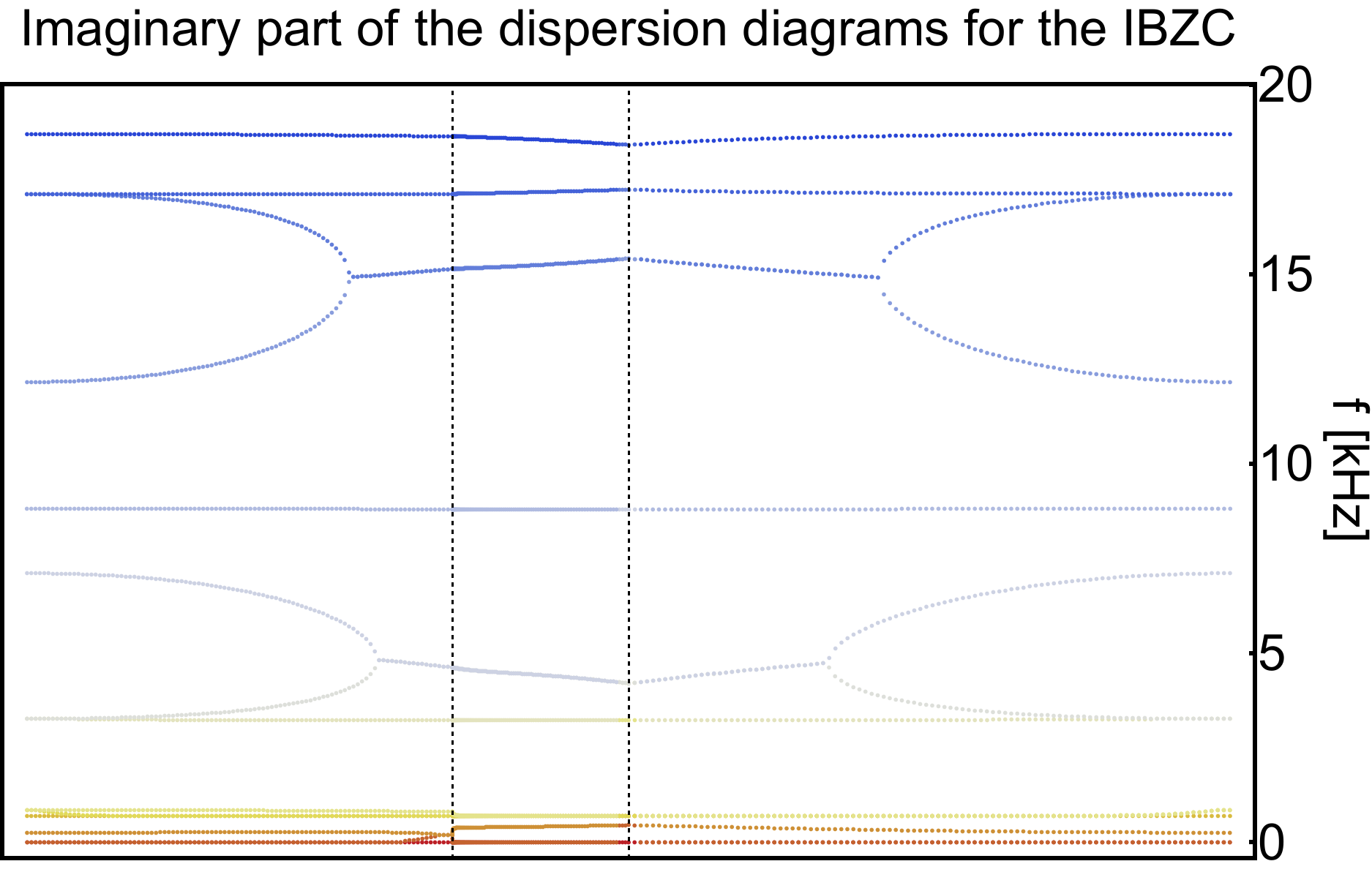}\\*
    \includegraphics[width=\linewidth]{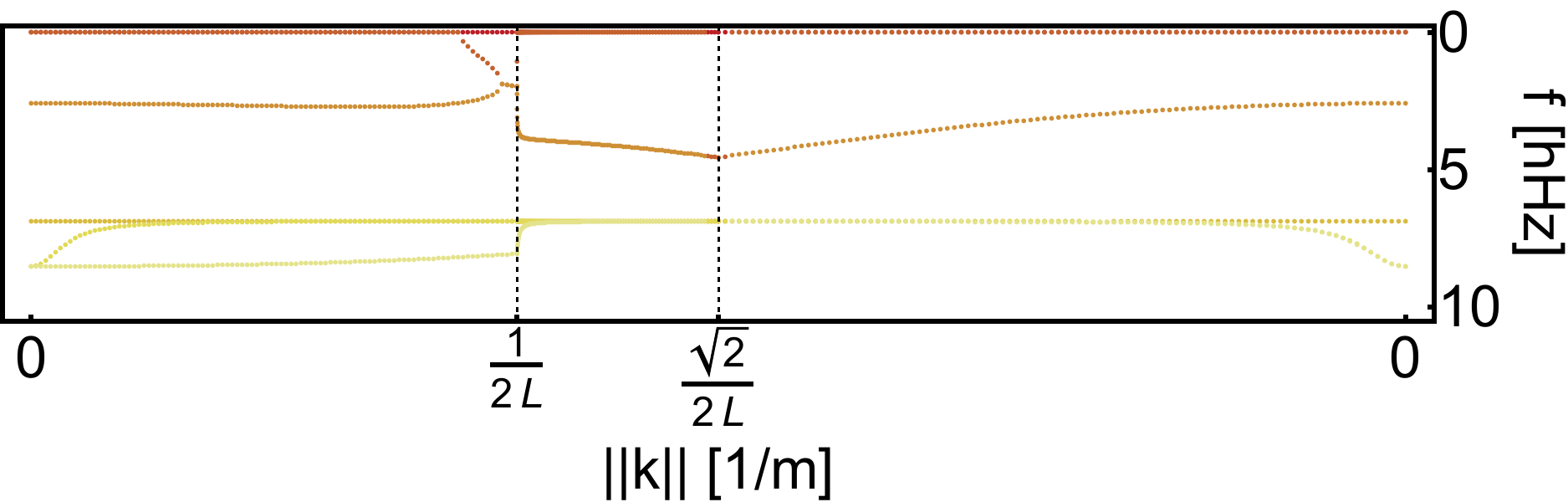}
  \end{minipage}
  \hfill
  \begin{minipage}{0.185\textwidth}
    \centering
    \includegraphics[width=\linewidth]{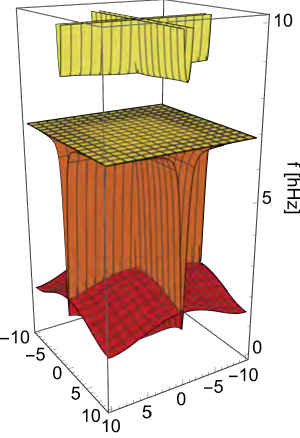}
  \end{minipage}
  \caption{
    (\textit{Top}) Schematic of the four-resonator unit cell and its material and geometrical properties: the size of the unit cell is $L$, the density is $\rho$, the elastic modulus is $E$ and the Poisson's ratio is $\nu$.
    (\textit{Bottom}) Dispersion curves for the four-resonator unit cell for real and imaginary wavenumbers:
    (\textit{left and right}) the first fourteen modes for real wavenumbers spanning the full irreducible Brillouin zone and a detail showing the first six modes;
    (\textit{center}) the first fourteen modes for real (imaginary) wavenumbers spanning the irreducible Brillouin zone contour, and a detail showing the first six modes (note that for the latter the frequency axis has been reversed and the units are hHz).
    The band-gap ranges are highlighted in light green.
  }
  \label{fig:4Reso_unit_cell_disp_curves}
\end{figure}

%%%%%%%%%%%%%%%%%%%%%%%%%%%%%%%%%%%%%%%%%%%%%%%%%%%%%%%%%%%%
The four-resonator unit cell that gives rise to the metamaterial studied in this section was originally designed in \cite{demore2022unfolding}. The unit cell is made of titanium and, at low frequencies, activates a local-resonance mechanism through its four resonators (see Fig.~\ref{fig:4Reso_unit_cell_disp_curves}), leading to the formation of a band-gap \cite{liu2000locally, sugino2016mechanism}.

In addition, broader Bragg-scattering band-gaps \cite{brillouin1953wave} also appear at higher frequency ranges, as shown in Fig.~\ref{fig:4Reso_unit_cell_disp_curves}. The band-gap generation mechanism in each case (Bragg scattering or local resonance) can be identified from the imaginary part of the dispersion plots in Fig.~\ref{fig:4Reso_unit_cell_disp_curves} (for more details on the identification, see \cite{krushynska2017coupling, yilmaz2017dynamics}).
%%%%%%%%%%%%%%%%%%%%%%%%%%%%%%%%%%%%%%%%%%%%%%%%%%%%%%%%%%%%
%%%%%%%%%%%%%%%%%%%%%%%%%%%%%%%%%%%%%%%%%%%%%%%%%%%%%%%%%%%%

%
%
%
%
%%%%%%%%%%%%%%%%%%%%%%%%%%%%%%%%%%%%%%%%%%%%%%%%%%%%%%%%%%%%
%%%%%%%%%%%%%%%%%%%%%%%%%%%%%%%%%%%%%%%%%%%%%%%%%%%%%%%%%%%%
\subsection{RUC choice, Transmissibility and boundary effects}
\label{sec:transm_4reso}
\begin{figure}[!htbp]
  \centering
  \includegraphics[width=0.94\textwidth]{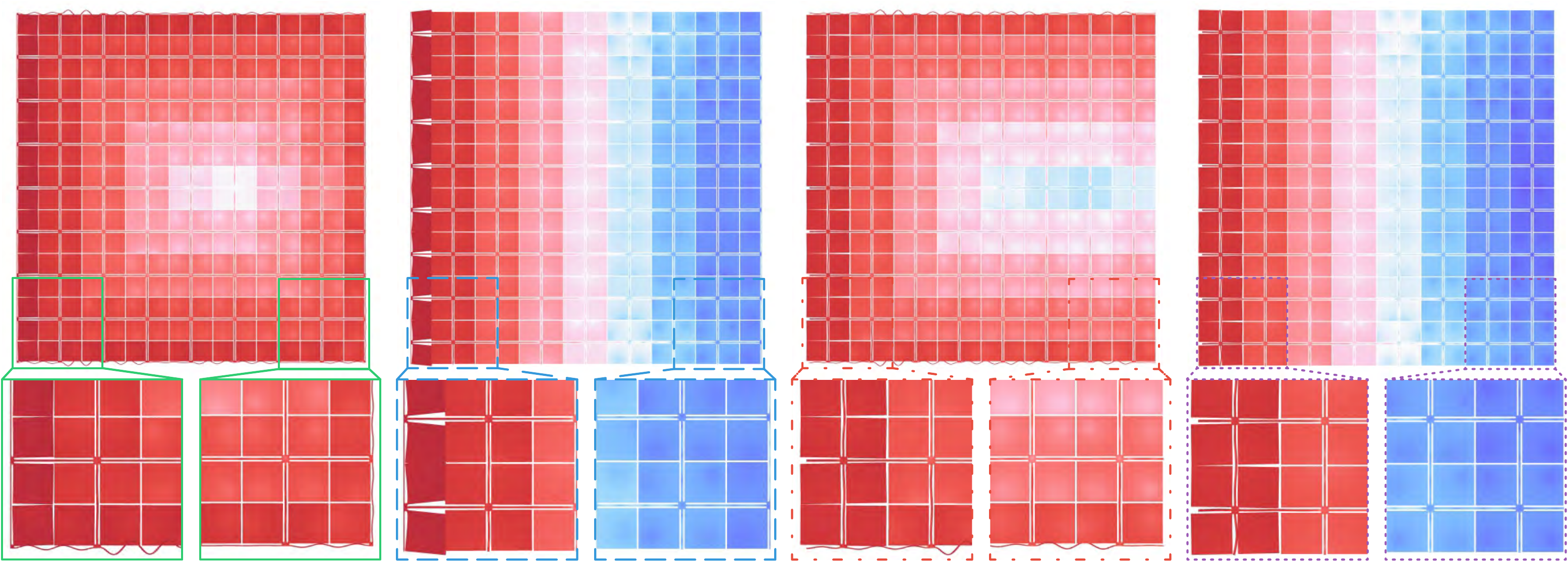}
  \\[10pt]
  \centering
  \includegraphics[width=0.47\textwidth]{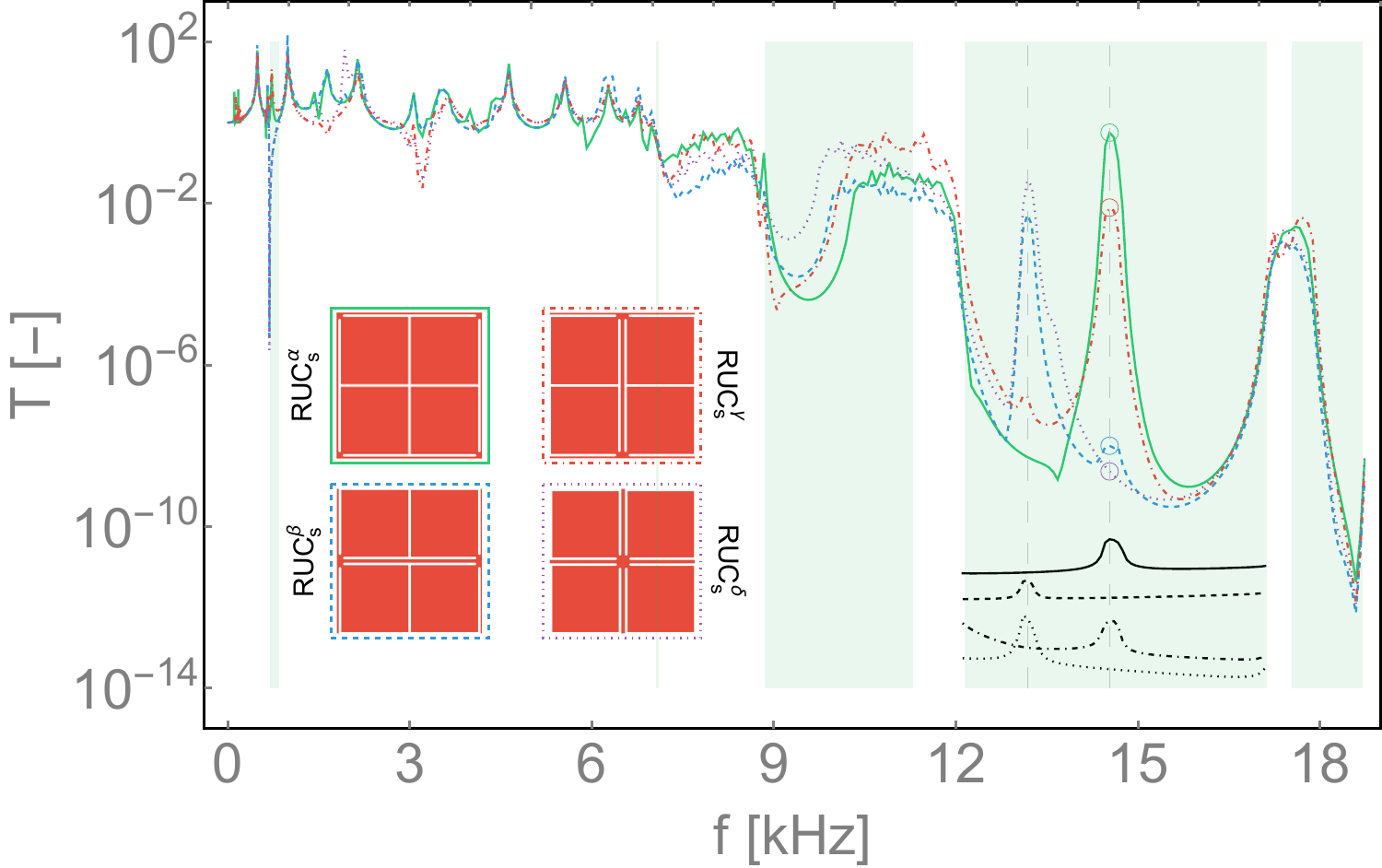}
  \hfill
  \includegraphics[width=0.47\textwidth]{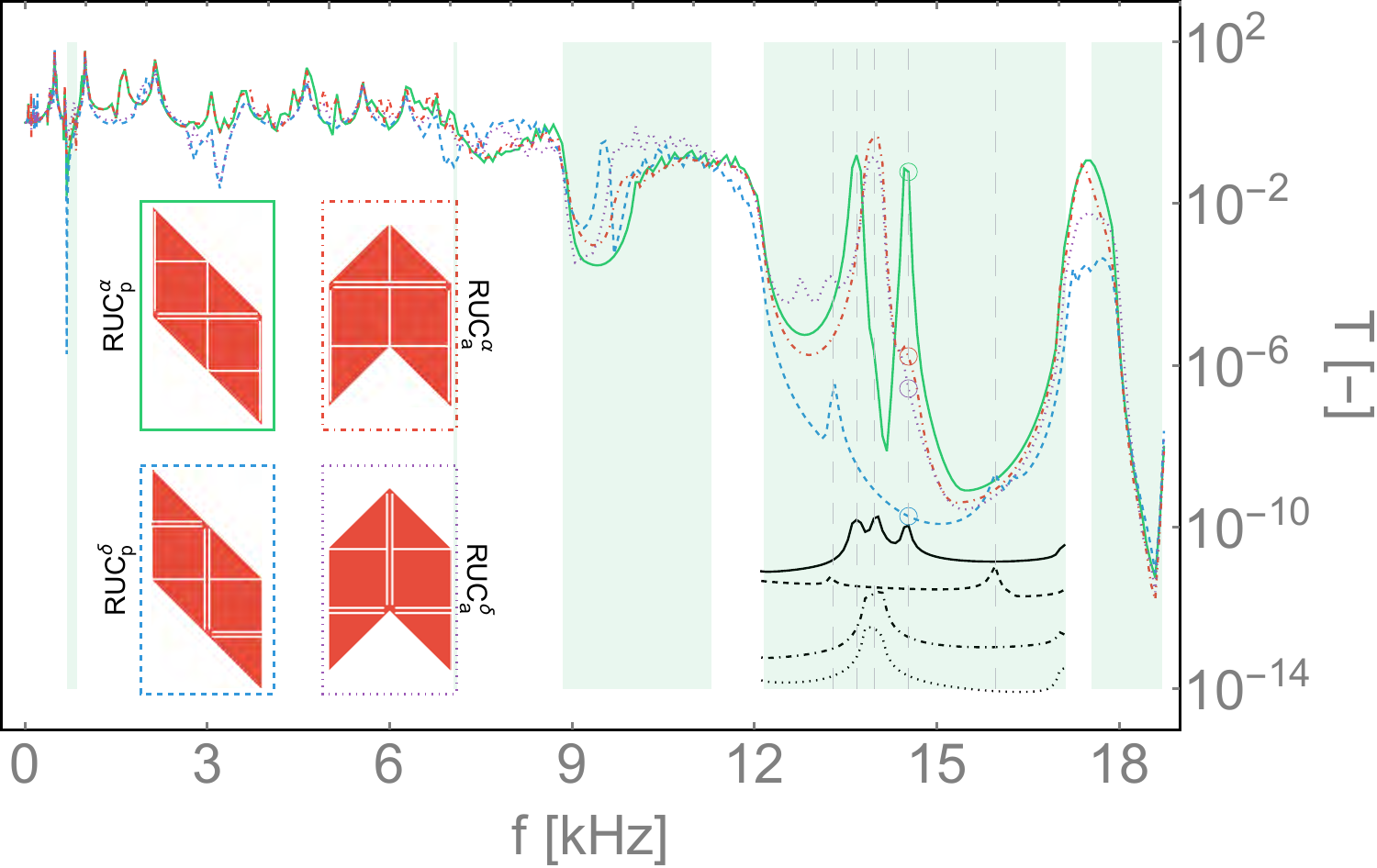}
  \\[10pt]
  \centering
  \includegraphics[width=0.94\textwidth]{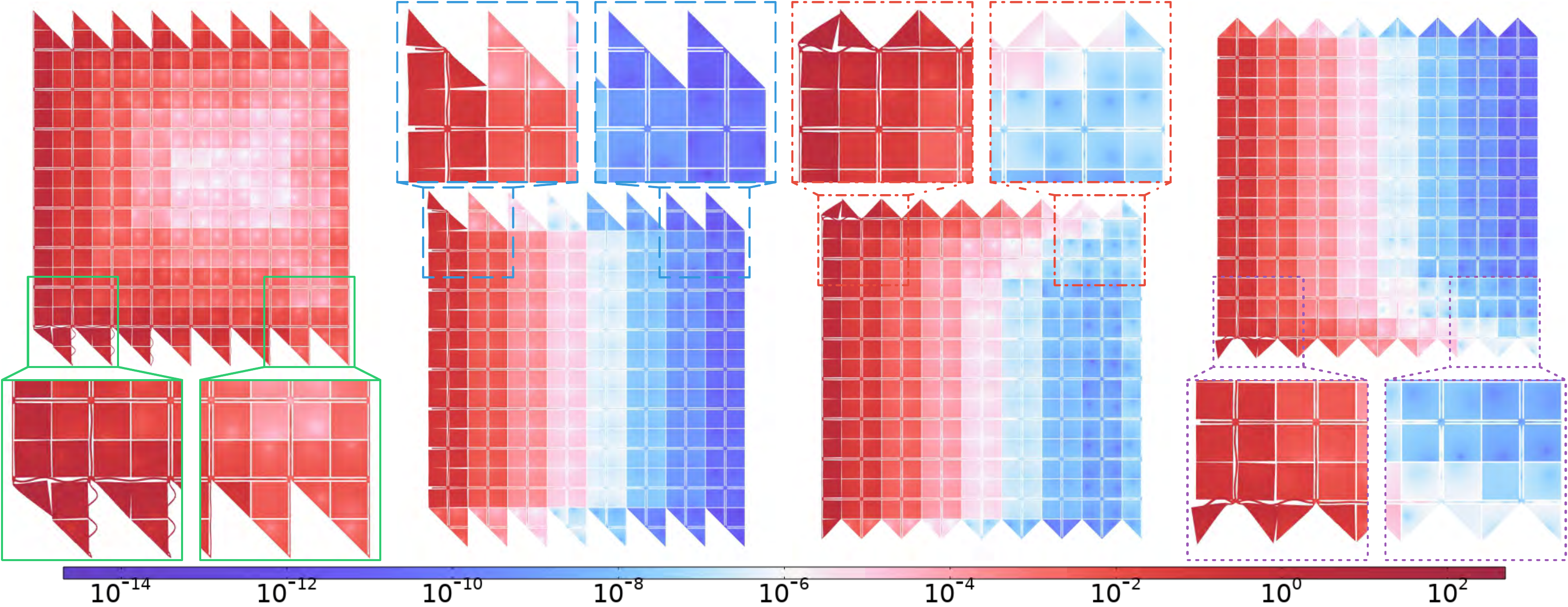}
  \caption{
  (\textit{Top} and \textit{Bottom}) Norm of the dimensionless displacement ($\lVert u \rVert/u_0$) at the frequency $f = 14.533$~[kHz], which lies within the band-gap range, showing boundary details for the considered RUCs.
  The color scale is logarithmic, and the deformation is amplified for visualization purposes.
  (\textit{Center}) Transmissibility curves.
  The band-gap regions are highlighted in light green; circles indicate the frequencies used in the displacement field plots, and black lines denote the average along the top and bottom rows of unit cells of the out-of-plane component of curl $u$.
  }
  \label{fig:4reso_transmission_be}
\end{figure}

%%%%%%%%%%%%%%%%%%%%%%%%%%%%%%%%%%%%%%%%%%%%%%%%%%%%%%%%%%%%
%%%%%%%%%%%%%%%%%%%%%%%%%%%%%%%%%%%%%%%%%%%%%%%%%%%%%%%%%%%%
For this metamaterial, eight RUCs have been identified, and their geometries are shown as insets in Fig.~\ref{fig:4reso_transmission_be}.
Four of them have a square shape, RUC$_{\rm s}^{\star}$, which can be obtained from the unit cell shown in Fig.~\ref{fig:4Reso_unit_cell_disp_curves} by using the following translation vectors in addition to the null vector ($e_{\alpha}=(0,0)^{\rm T}$) to include the original cell itself
\begin{equation}
  e_{\beta} = \frac{1}{2} e_{2} \, ,
  \qquad\qquad
  e_{\gamma} = \frac{1}{2} e_{1} \, ,
  \qquad\qquad
  e_{\delta} = \frac{1}{2} e_{1} + \frac{1}{2} e_{2}\, .
  \label{eq:trans_vect_4reso}
\end{equation}

Two of the other four RUCs, one parallelogram RUC$_{\rm p}^{\star}$ and one arrow-shaped RUC$_{\rm a}^{\star}$, are obtained through the procedure explained in Section \ref{sec:cutNshape} (see the top panels three and four of Fig.~\ref{fig:RUC_proc}), while the remaining two are obtained using the translation vector $e_{\delta}$ reported in Eq.~(\ref{eq:trans_vect_4reso}).

In the top and bottom panels of Fig.~\ref{fig:4reso_transmission_be}, the dimensionless displacement field ($\lVert u \rVert / u_0$) is shown for all eight RUCs at the frequency $f = 14.533$~kHz, which lies within the band-gap range.
For RUC$_{\rm s}^{\alpha}$ and RUC$_{\rm s}^{\gamma}$, the input excitation propagates along the top and bottom boundaries, indicating that the band gap becomes ineffective due to boundary-guided waves.
RUC$_{\rm p}^{\alpha}$ also allows energy to travel along both edges, with slight asymmetry, whereas RUC$_{\rm a}^{\alpha}$ and RUC$_{\rm a}^{\delta}$ exhibit propagation confined to a single boundary.

In contrast, for RUC$_{\rm p}^{\delta}$, wave propagation is strongly suppressed, as expected within the band-gap regime, particularly along the boundaries.
This behavior agrees with the transmissibility curves shown in the central panel of Fig.~\ref{fig:4reso_transmission_be}. Additionally, the transmissibility peaks align with those of the average out-of-plane component of $\mathrm{curl}\,u$, evaluated along the top and bottom rows of unit cells (black curves), following the same pattern as the corresponding transmissibility curves, further corroborating the presence of boundary-guided waves.
Overall, RUC$_{\rm p}^{\delta}$ provides the best performance in terms of transmissibility.

%%%%%%%%%%%%%%%%%%%%%%%%%%%%%%%%%%%%%%%%%%%%%%%%%%%%%%%%%%%%
%%%%%%%%%%%%%%%%%%%%%%%%%%%%%%%%%%%%%%%%%%%%%%%%%%%%%%%%%%%%%

%
%
%
%%%%%%%%%%%%%%%%%%%%%%%%%%%%%%%%%%%%%%%%%%%%%%%%%%%%%%%%%%%%
%%%%%%%%%%%%%%%%%%%%%%%%%%%%%%%%%%%%%%%%%%%%%%%%%%%%%%%%%%%%
\subsection{Discussion}
\label{sec:4reso_discussion}
%%%%%%%%%%%%%%%%%%%%%%%%%%%%%%%%%%%%%%%%%%%%%%%%%%%%%%%%%%%%
%%%%%%%%%%%%%%%%%%%%%%%%%%%%%%%%%%%%%%%%%%%%%%%%%%%%%%%%%%%%
In Fig.~\ref{fig:4reso_transmission_be} of Section~\ref{sec:transm_4reso}, it is shown that, for certain frequency ranges within the band gaps, a finite-size metamaterial built from some RUCs still exhibits nonzero transmissibility.
To better understand this phenomenon, a parametric study of the size of the structure built with RUC$_{\rm s}^{\alpha}$ has been performed, varying the number of unit cells in steps of four in both the vertical and horizontal directions, as reported in Fig.~\ref{fig:4reso_transmission_alpha_size} (see Fig.~\ref{fig:4reso_transmission_beta_gamma_delta_size} in Appendix~\ref{app:4reso_size} for the other RUC$_{\rm s}^{\star}$).

As shown in the top and bottom left panels of Fig.~\ref{fig:4reso_transmission_alpha_size}, the transmissibility spikes can be correlated with the eigenfrequencies of the full structure, calculated using the same boundary conditions reported in Fig.~\ref{fig:BC}.
In addition, an in-depth analysis shows that the ratio $\overline{W}$,
\begin{equation}
  \overline{W} = \frac{W_{\rm bound}}{W_{\rm bulk}},
  \label{eq:ratio_energies}
\end{equation}
defined as the ratio between the total energy localized at the boundary (i.e., the top, bottom, and rightmost unit cells) and the total energy in the bulk (i.e., the remaining unit cells), and evaluated for the modes corresponding to the eigenfrequencies of each structure, as shown in the bottom left panel of Fig.~\ref{fig:4reso_transmission_alpha_size}, exhibits pronounced spikes of several orders of magnitude that correspond to the transmissibility peaks.
%%%%%%%%%%%%%%%%%%%%%%%%%%%%%%%%%%%%%%%%%%%%%%%%%%%%%%%%%%%
\begin{figure}[!htbp]
  \centering
  \includegraphics[width=0.47\textwidth]{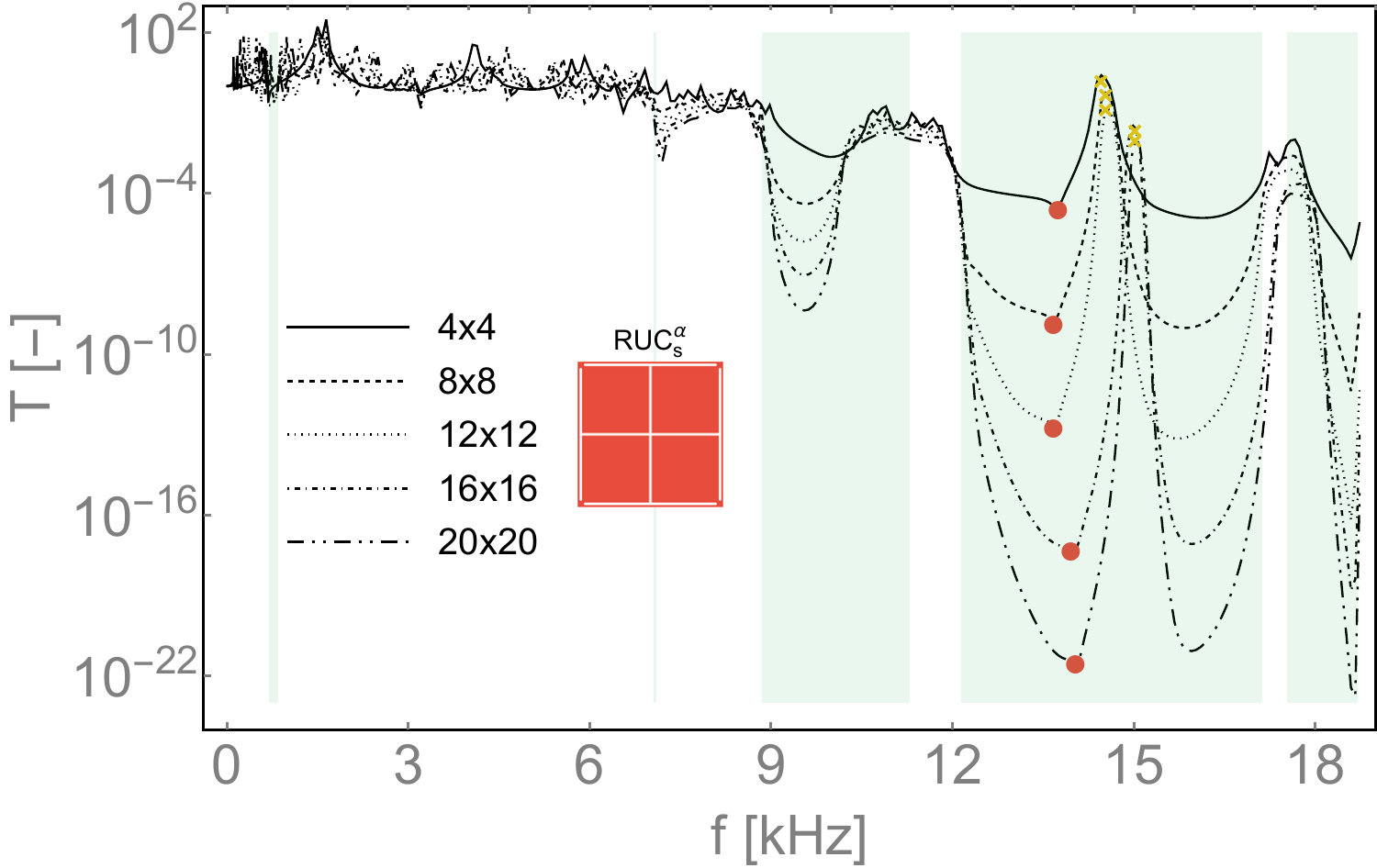}
  \hfill
  \includegraphics[width=0.47\textwidth]{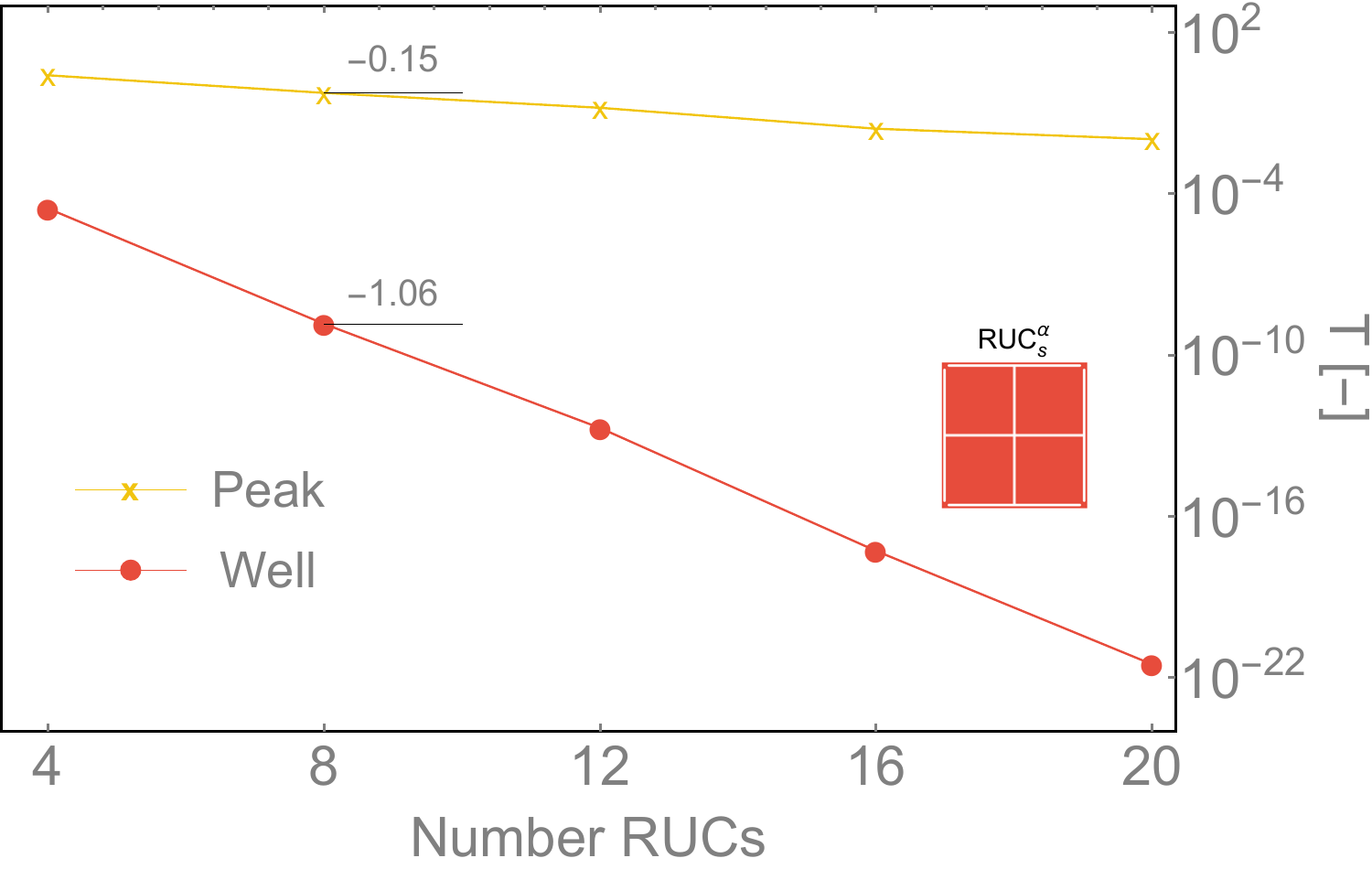}
  \\
  \centering
  \includegraphics[width=0.47\textwidth]{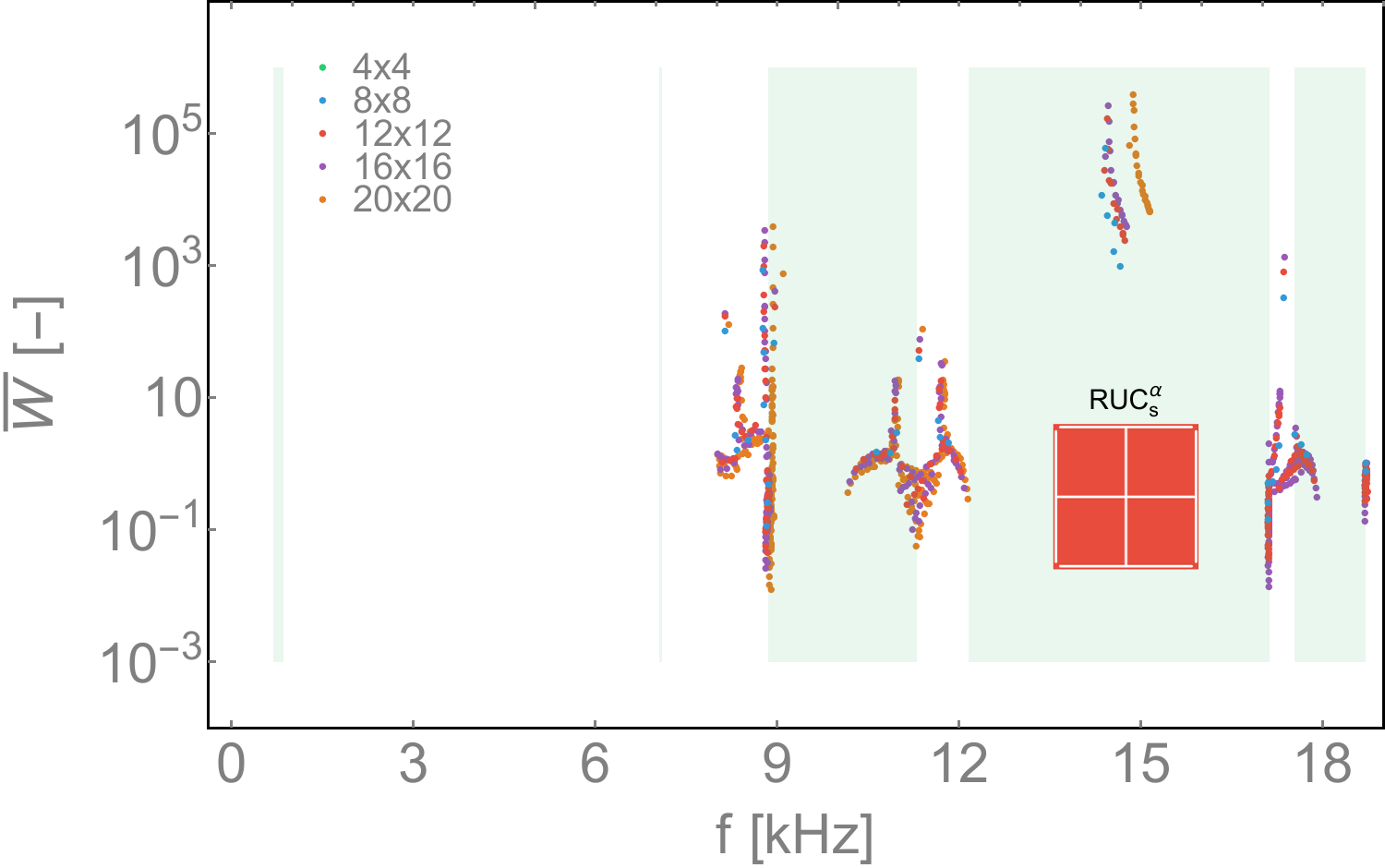}
  \hfill
  \includegraphics[width=0.47\textwidth]{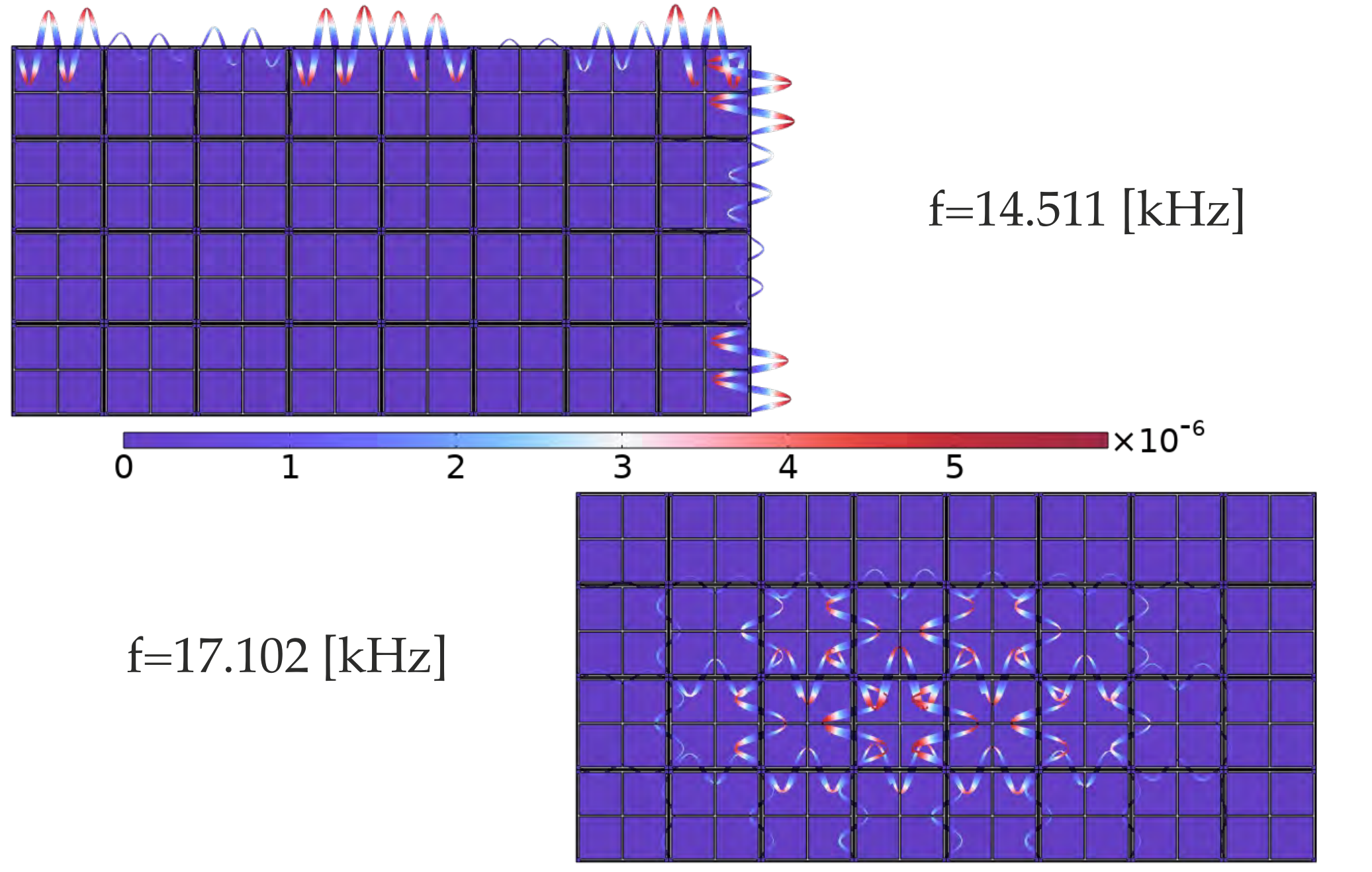}
  \caption{
  (\textit{Top left}) Transmissibility curves for different structure sizes for RUC$_{\rm s}^{\alpha}$, with the transmissibility peaks and wells highlighted by yellow crosses and red dots, respectively;
  (\textit{Top right}) variation of the peaks (yellow curve with crosses) and wells (red curve with dots) as a function of the number of RUCs composing the structure;
  (\textit{Bottom left}) ratio between the energy localized at the top, bottom, and rightmost boundary cells and the energy in the remaining cells in the bulk, denoted by $\overline{W}$, for the eigenfrequencies of each structure;
  (\textit{Bottom right}) norm of the dimensionless displacement field ($\lVert u \rVert/u_0$) for two eigenfrequencies, $f = 14.511$~[kHz] and $f = 17.102$~[kHz] (due to the symmetry of this test, only half of the structure is reported).
  }
  \label{fig:4reso_transmission_alpha_size}
\end{figure}
%%%%%%%%%%%%%%%%%%%%%%%%%%%%%%%%%%%%%%%%%%%%%%%%%%%%%%%%%%%%%
This corroborates the hypothesis that the nonzero transmissibility within the band-gap range is due to boundary-guided waves associated with the natural vibration modes of the structure.
Boundary effects can also be investigated through a Bloch-Floquet analysis performed on the full structure, with periodic boundary conditions applied only in the horizontal directions.
However, this \emph{ad hoc} approach is less accurate, is only meaningful when compared with Fig.~\ref{fig:BC} (see Appendix~\ref{app:4reso_size}), and fails to distinguish between RUCs related by a horizontal shift.

%%%%%%%%%%%%%%%%%%%%%%%%%%%%%%%%%%%%%%%%%%%%%%%%%%%%%%%%%%%%
\begin{figure}[!htbp]
  \centering
  \includegraphics[width=0.47\textwidth]{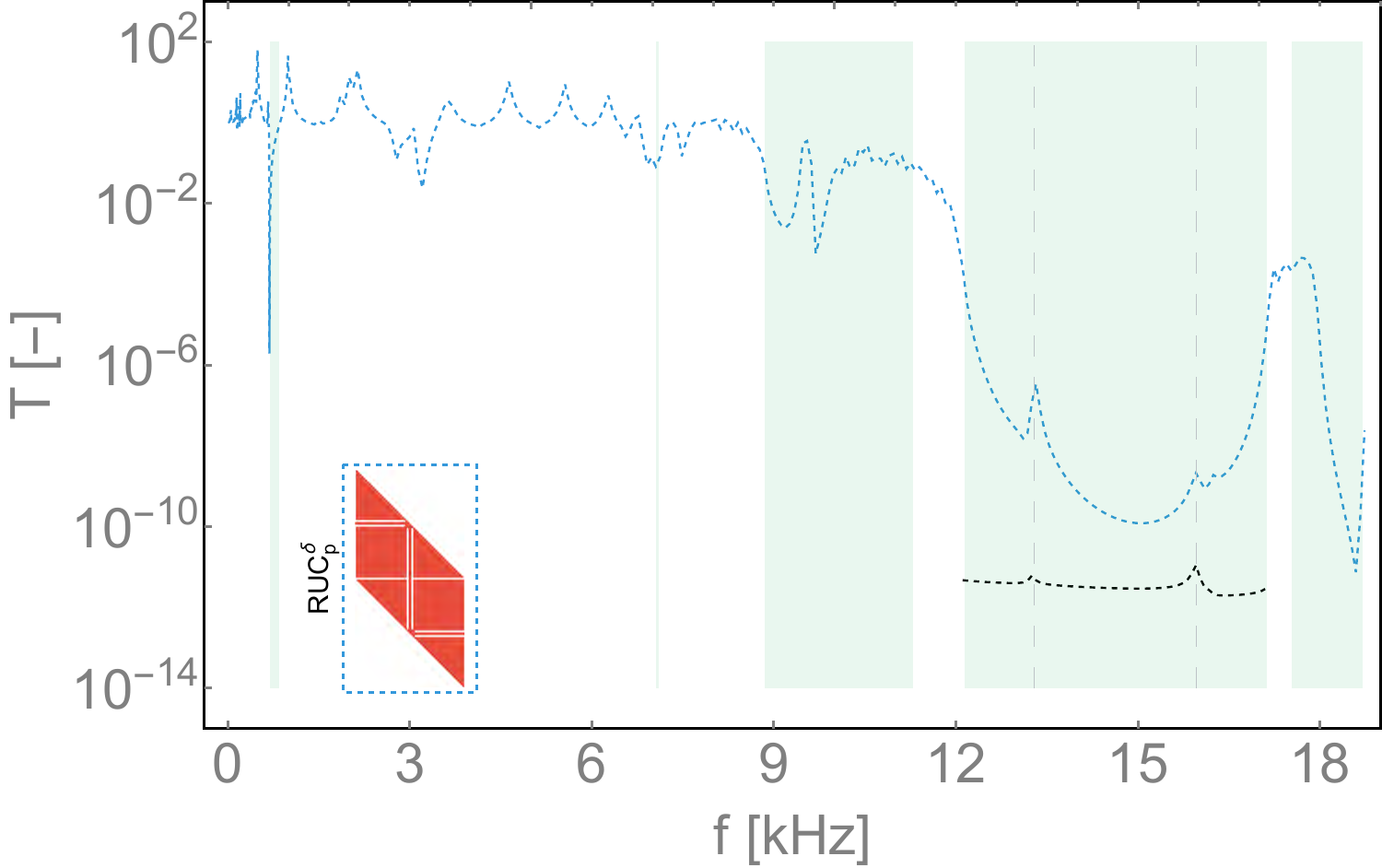}
  \hfill
  \includegraphics[width=0.47\textwidth]{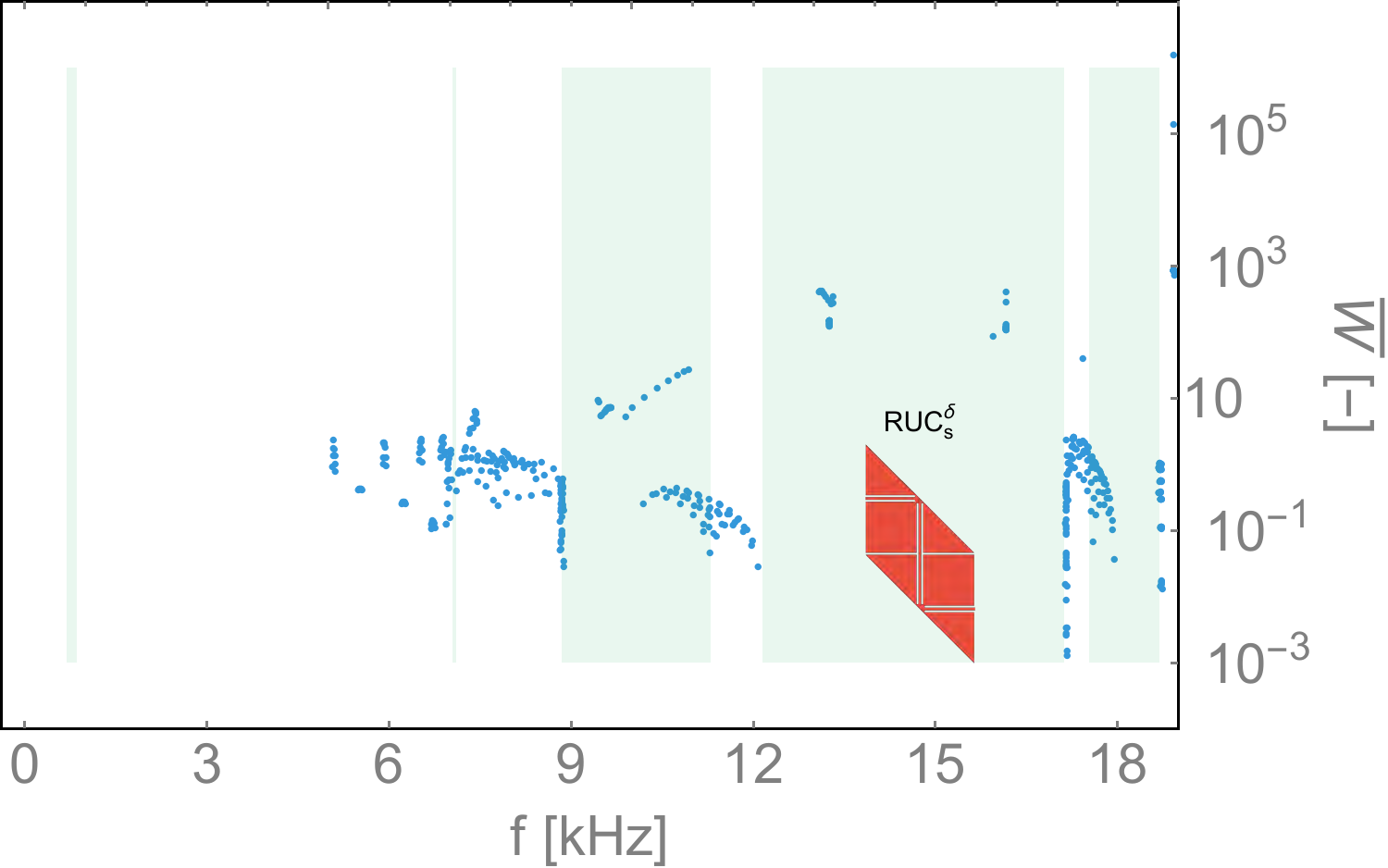}
  \caption{
    (\textit{Left}) Transmissibility curve for the RUC$_{\rm p}^{\delta}$ and
    (\textit{right}) ratio between the energy localized at the top, bottom, and rightmost boundary cells and the energy in the remaining cells in the bulk, denoted by $\overline{W}$, for the eigenfrequencies of an $8\times 8$ structure;
  }
  \label{fig:4reso_transmission_para_delta}
\end{figure}
%%%%%%%%%%%%%%%%%%%%%%%%%%%%%%%%%%%%%%%%%%%%%%%%%%%%%%%%%%%%%

In particular, to qualitatively characterize this phenomenon occurring within the 12-17~kHz band-gap range, the surface average of the out-of-plane component of $\mathrm{curl}\,u$ (the only nonzero component) has been evaluated over the top, bottom, and rightmost unit cells and is shown in black, following the same pattern as the respective RUC transmissibility curves in Fig.~\ref{fig:4reso_transmission_be}.
In this case, the peaks coincide with the transmissibility peaks, further corroborating the interpretation that the nonzero transmissibility in the band-gap range is due to boundary-guided waves.

It can also be observed in the bottom left panel of Fig.~\ref{fig:4reso_transmission_alpha_size} that there is a concentration of eigenfrequencies at the end of the second band gap (around $f = 12$~kHz).
This explains why the transmissibility curves increase sharply to a maximum despite being well within the band-gap range.

The top right panel of Fig.~\ref{fig:4reso_transmission_alpha_size} shows how the peaks (yellow curve with crosses) and wells (red curve with dots) vary as a function of the number of RUCs composing the structure. The wells quickly approach zero transmissibility, as expected for a band-gap frequency, whereas the peaks decrease in intensity by less than one order of magnitude compared to the wells.
This suggests that simply increasing the sample size to mitigate boundary effects is not a reliable strategy in every case.

In the bottom right panel, the dimensionless displacement field ($\lVert u \rVert / u_0$) is shown for two eigenfrequencies: $f = 14.511$~kHz (inside a band gap and near a peak) and $f = 17.102$~kHz (outside a band gap). The first clearly shows displacement (energy) focusing on the free top and right boundaries (and the bottom, due to symmetry), which is responsible for the reduced effectiveness within the band gap, while the second shows energy stored primarily in the bulk (only half of the structure is shown due to symmetry).

From Fig.~\ref{fig:4reso_transmission_para_delta}, it can be seen for RUC$_{\rm p}^{\delta}$ that the eigenfrequencies correspond to the transmissibility spikes.
However, the ratio of energy localized at the boundary to that in the bulk is modest, and very few eigenfrequencies appear.
Furthermore, the results indicate that a continuous structural path along the top or bottom edges of the chosen RUC enables eigenfrequencies to emerge inside the band gaps, causing boundary effects.
In contrast, when this path is absent, transmission is suppressed.
These findings are, however, limited to the specific metamaterial samples, RUC designs, and tests conducted in this study.

%
%
%
%%%%%%%%%%%%%%%%%%%%%%%%%%%%%%%%%%%%%%%%%%%%%%%%%%%%%%%%%%%%
%%%%%%%%%%%%%%%%%%%%%%%%%%%%%%%%%%%%%%%%%%%%%%%%%%%%%%%%%%%%
\section{Square-circular-hole unit cell metamaterial}
\label{sec:circ}
%%%%%%%%%%%%%%%%%%%%%%%%%%%%%%%%%%%%%%%%%%%%%%%%%%%%%%%%%%%%
%%%%%%%%%%%%%%%%%%%%%%%%%%%%%%%%%%%%%%%%%%%%%%%%%%%%%%%%%%%%

\begin{figure}[!htbp]
  \centering
  \includegraphics[width=0.8\textwidth]{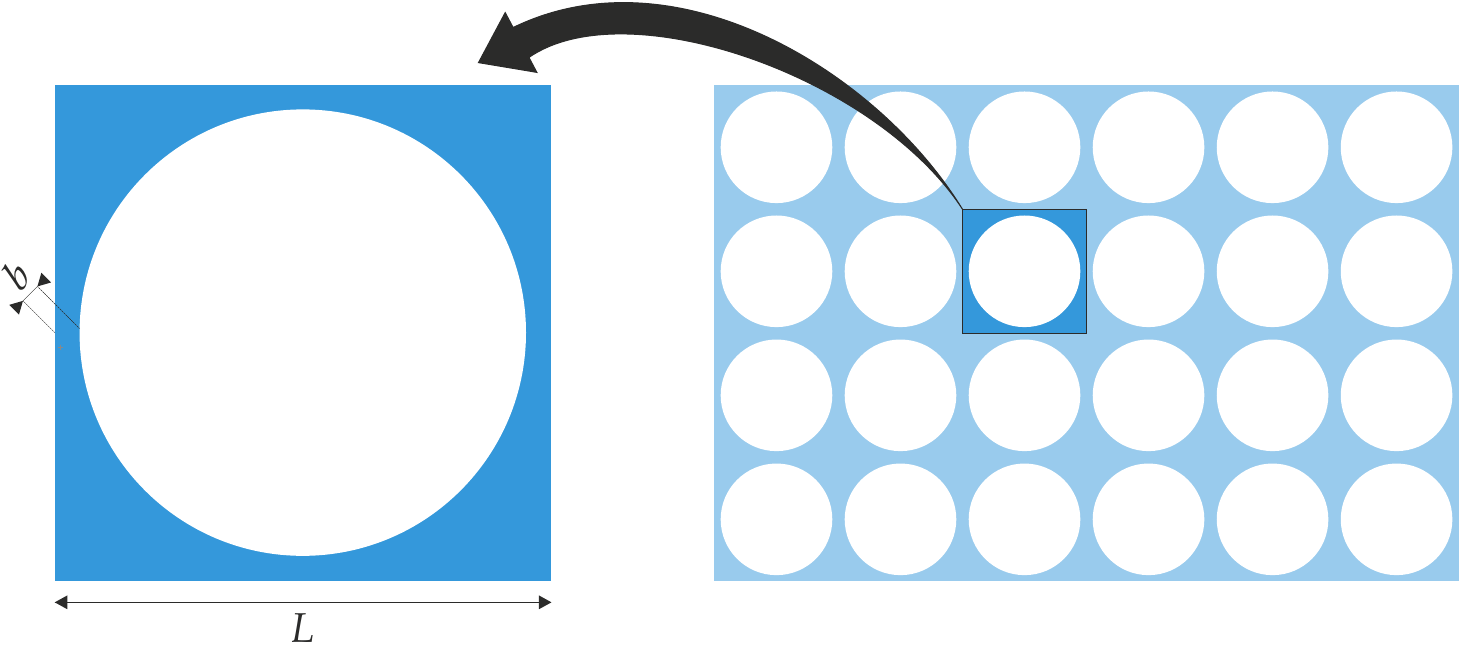}\\*
  \vspace*{12pt}
  \renewcommand{\arraystretch}{1.3}
  \centering
  \begin{tabular}{ccccccc}
    \hline
    $L$ [mm] & $b$ [mm] & $\rho$  [kg/m$^3$] & $E$ [GPa] & $\nu$ [-] & $\eta$ [-]
    \\
    \hline
    50       & 2.5      & 1210               & 1.49      & 0.42      & 0.1
    \\
    \hline
  \end{tabular}
  \\[10pt]
  \begin{minipage}{0.185\textwidth}
    \centering
    \includegraphics[width=\linewidth]{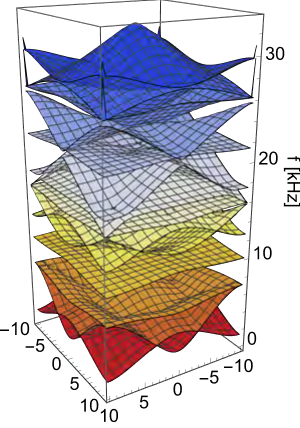}
  \end{minipage}
  \hfill
  \begin{minipage}{0.305\textwidth}
    \centering
    \includegraphics[width=\linewidth]{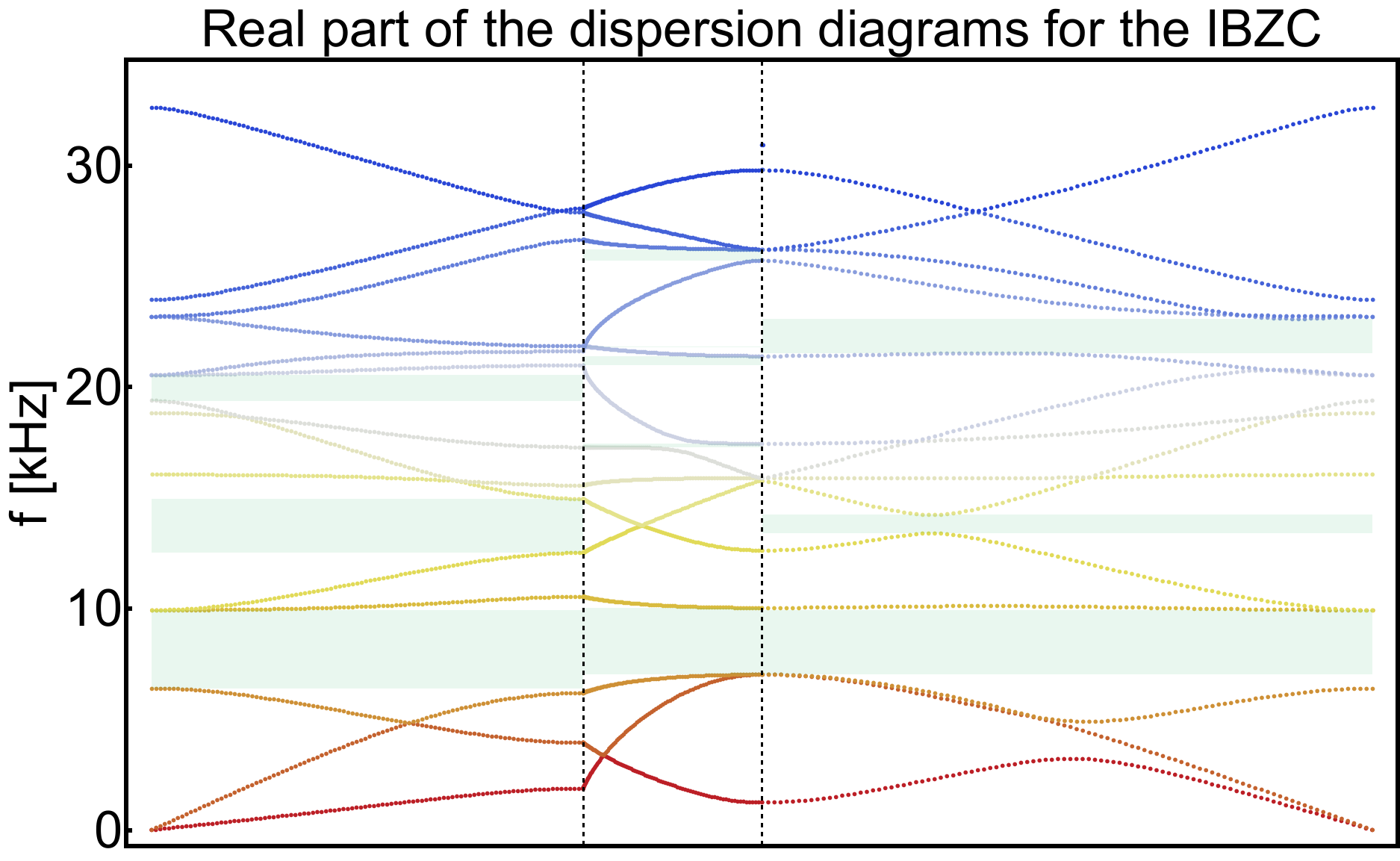}\\*
    \includegraphics[width=\linewidth]{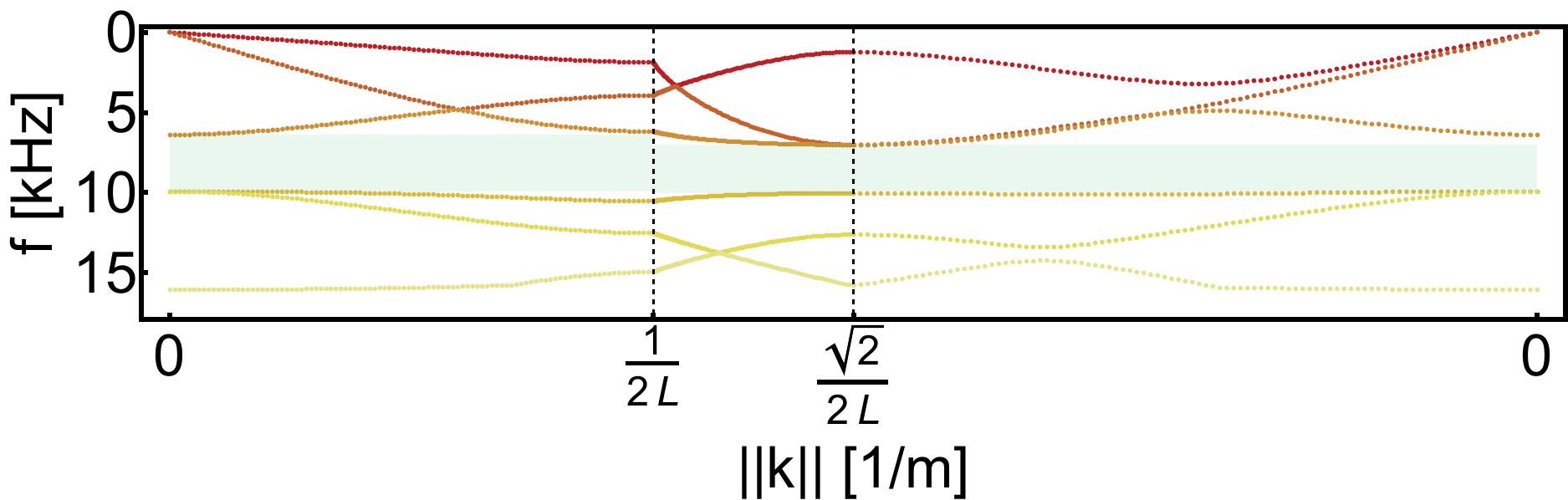}
  \end{minipage}
  \hfill
  \begin{minipage}{0.305\textwidth}
    \centering
    \includegraphics[width=\linewidth]{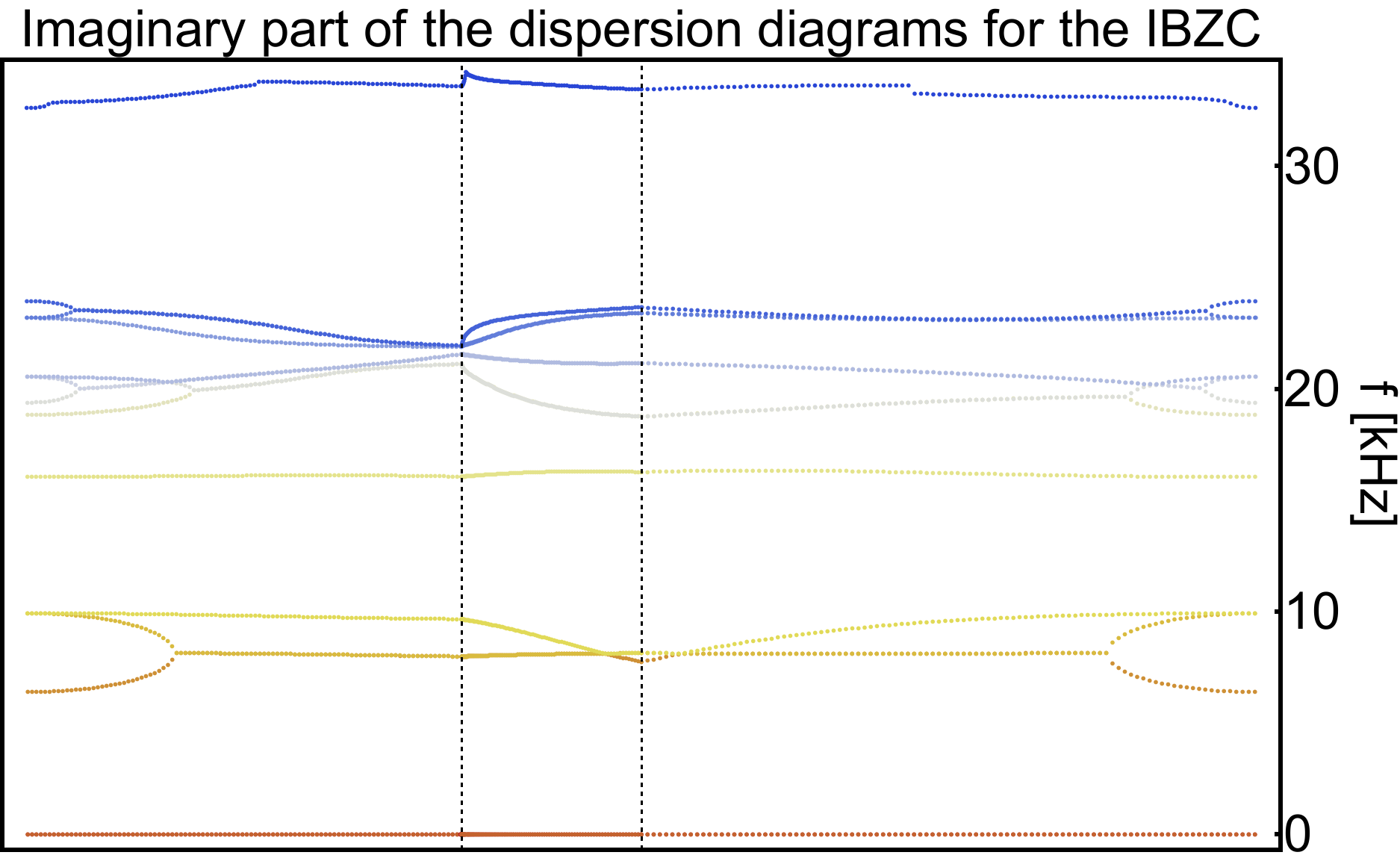}\\*
    \includegraphics[width=\linewidth]{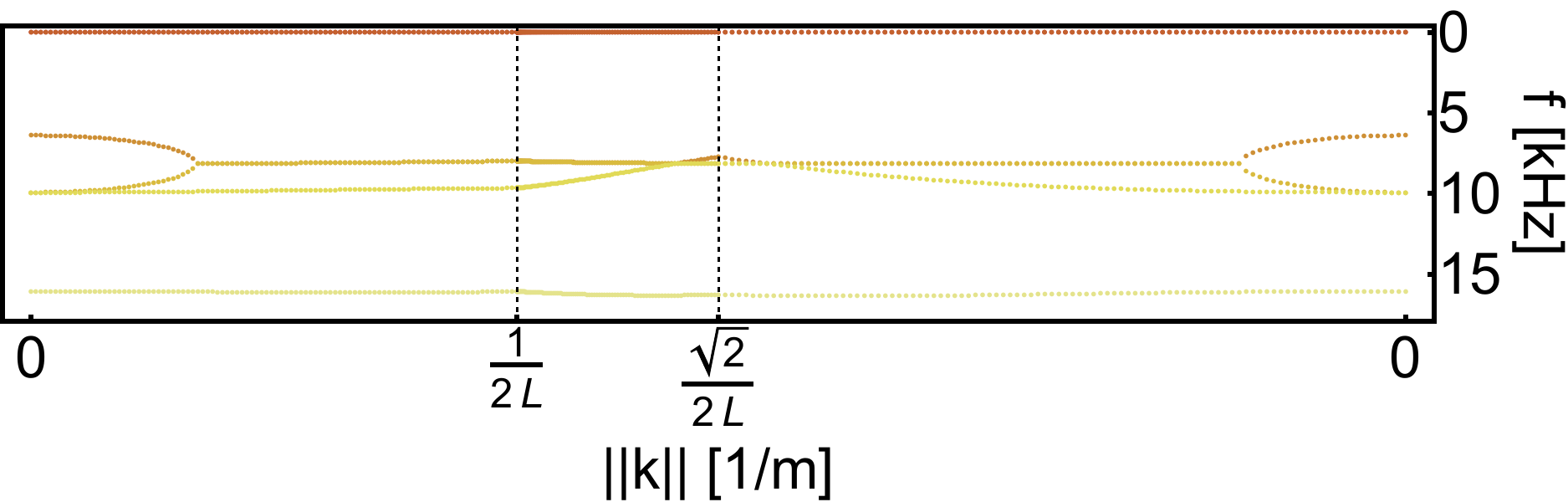}
  \end{minipage}
  \hfill
  \begin{minipage}{0.185\textwidth}
    \centering
    \includegraphics[width=\linewidth]{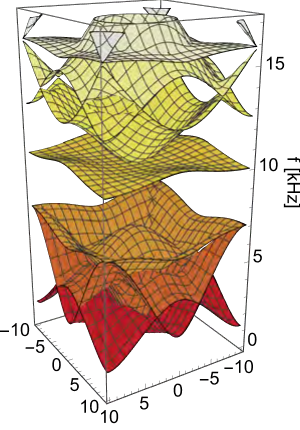}
  \end{minipage}
  \caption{
    (\textit{Top}) Schematic of the square-circular-hole unit cell and its material and geometrical properties: the size of the unit cell is $L$, the density is $\rho$, the elastic modulus is $E$ and the Poisson's ratio is $\nu$.
    (\textit{Bottom}) Dispersion curves for the square-circular-hole unit cell for real and imaginary wavenumbers:
    (\textit{left and right}) the first fourteen modes for real wavenumbers spanning the full irreducible Brillouin zone and a detail showing the first six modes;
    (\textit{center}) the first fourteen modes for real (imaginary) wavenumbers spanning the irreducible Brillouin zone contour, and a detail showing the first six modes (note that for the latter the frequency axis has been reversed and the units are hHz).
    The band-gap ranges are highlighted in light green.
  }
  \label{fig:circ_unit_cell_disp_curves}
\end{figure}

The square-circular-hole unit cell is a classical photopolymer design that enables the formation of band-gaps primarily through a Bragg-scattering mechanism (see Fig.~\ref{fig:circ_unit_cell_disp_curves}).
%%%%%%%%%%%%%%%%%%%%%%%%%%%%%%%%%%%%%%%%%%%%%%%%%%%%%%%%%%%%
%%%%%%%%%%%%%%%%%%%%%%%%%%%%%%%%%%%%%%%%%%%%%%%%%%%%%%%%%%%%

%
%
%
%
%%%%%%%%%%%%%%%%%%%%%%%%%%%%%%%%%%%%%%%%%%%%%%%%%%%%%%%%%%%%
%%%%%%%%%%%%%%%%%%%%%%%%%%%%%%%%%%%%%%%%%%%%%%%%%%%%%%%%%%%%
\subsection{RUC choice, Transmissibility and boundary effects}
\label{sec:transm_circ}
\begin{figure}[!htbp]
  \centering
  \includegraphics[width=0.94\textwidth]{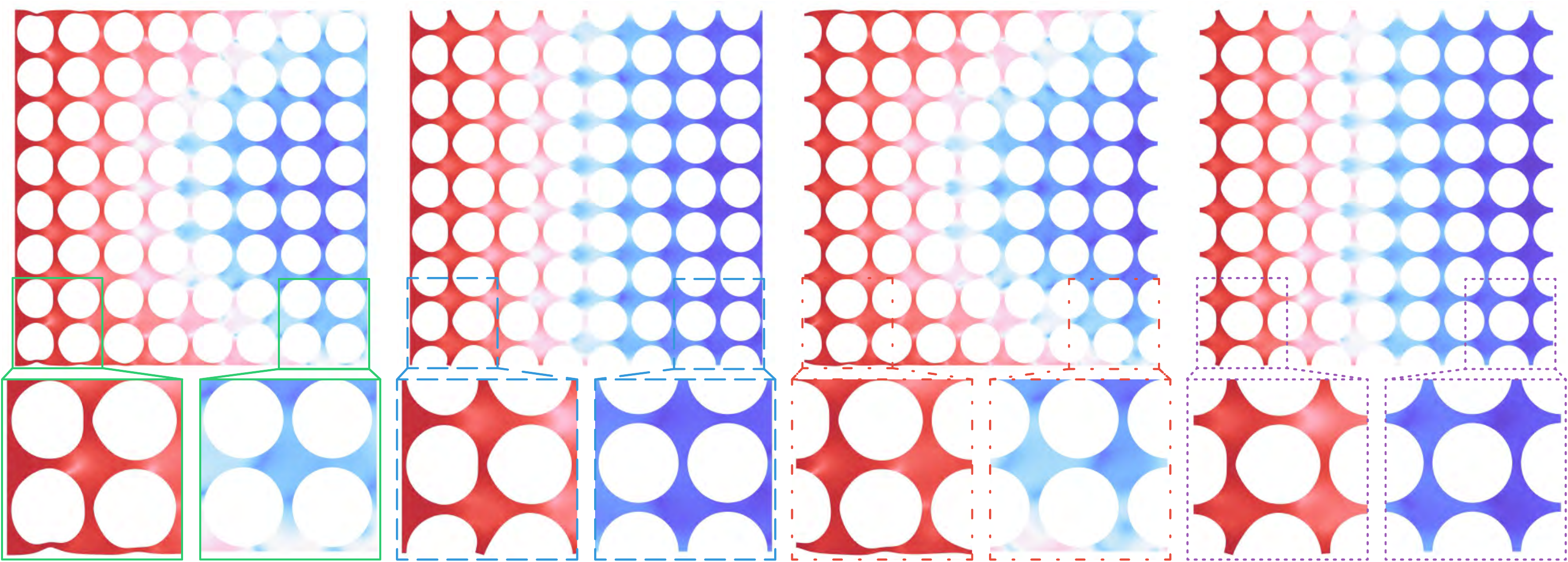}
  \\[10pt]
  \centering
  \includegraphics[width=0.47\textwidth]{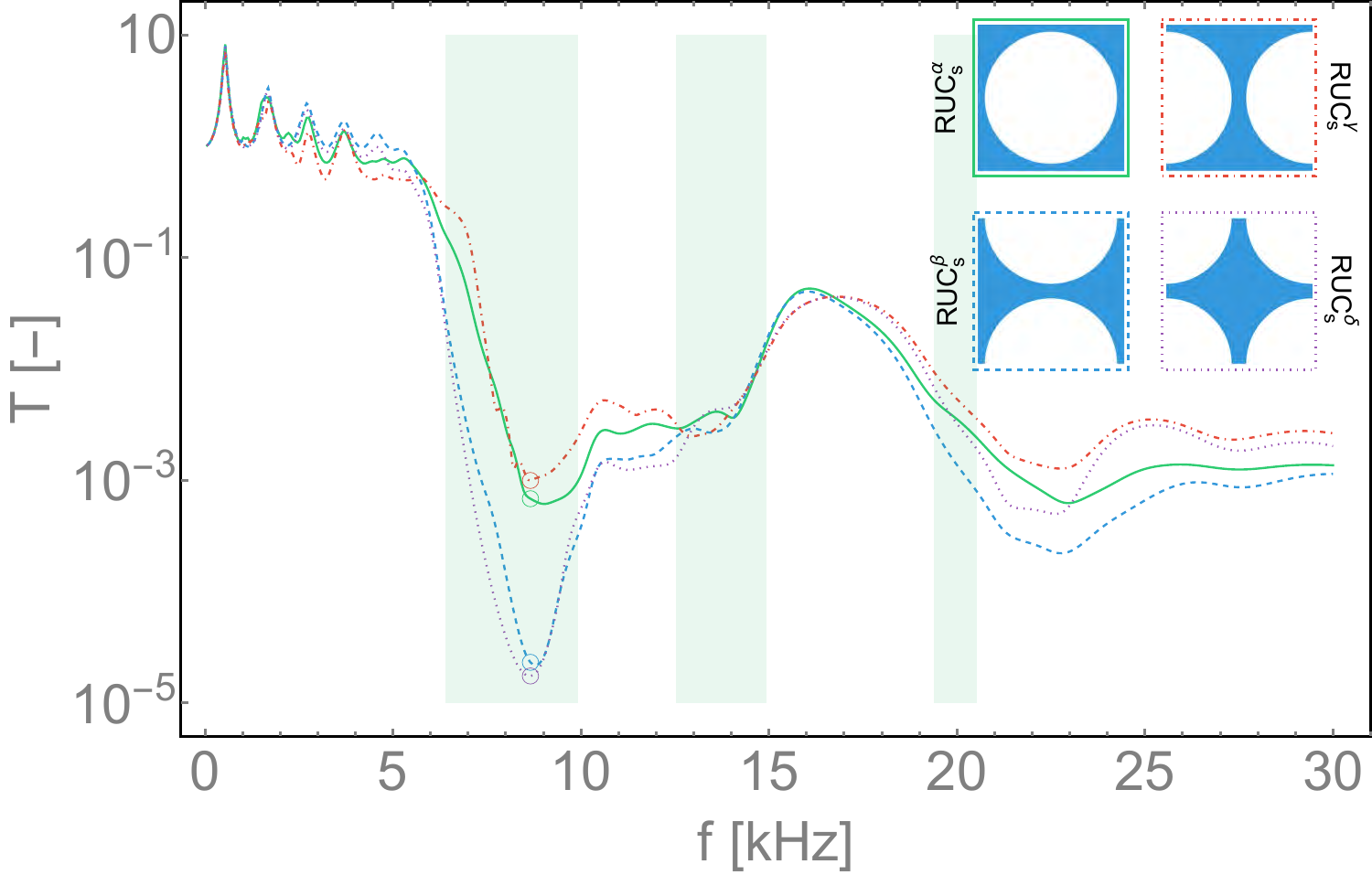}
  \hfill
  \includegraphics[width=0.47\textwidth]{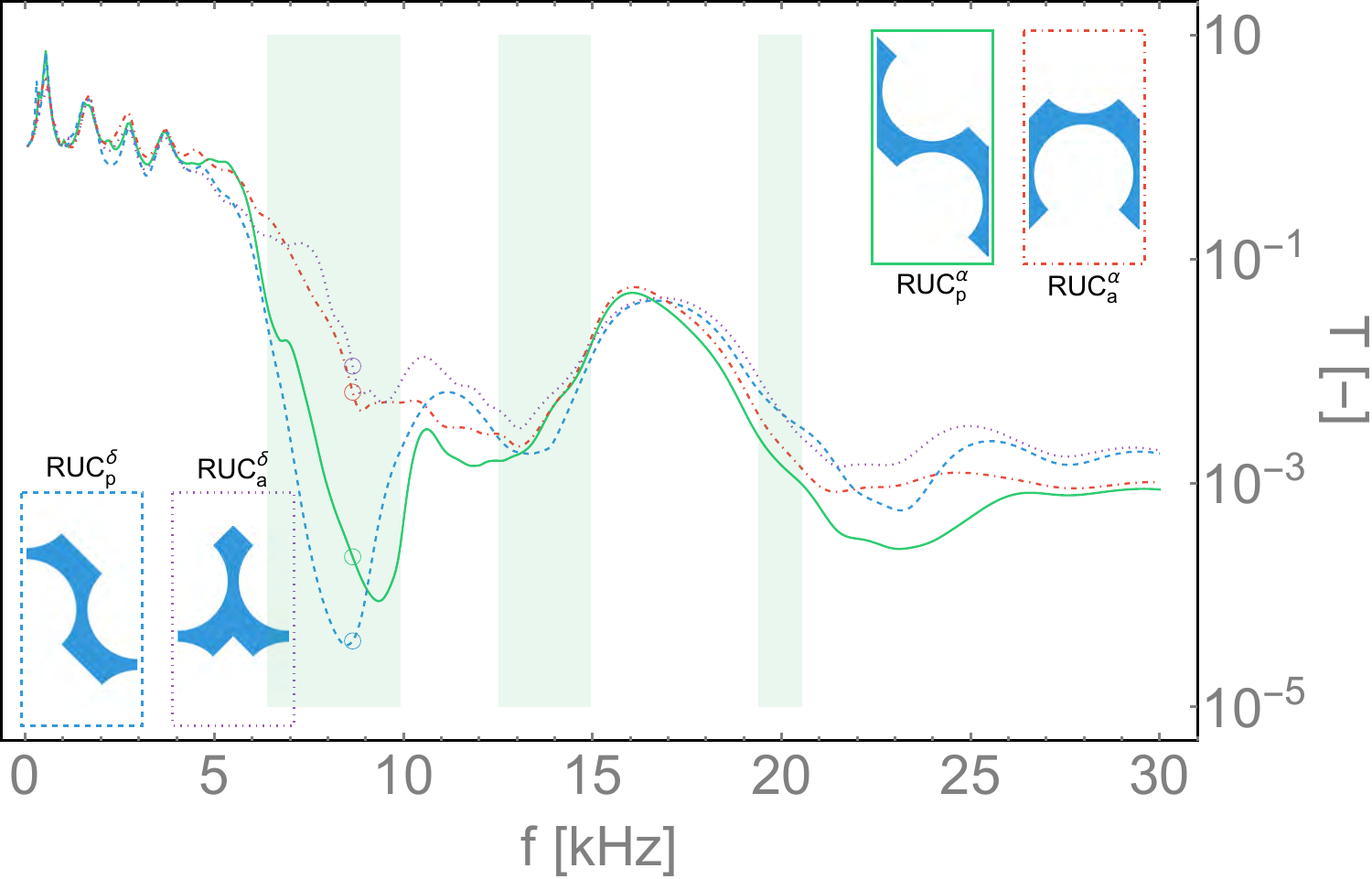}
  \\[10pt]
  \centering
  \includegraphics[width=0.94\textwidth]{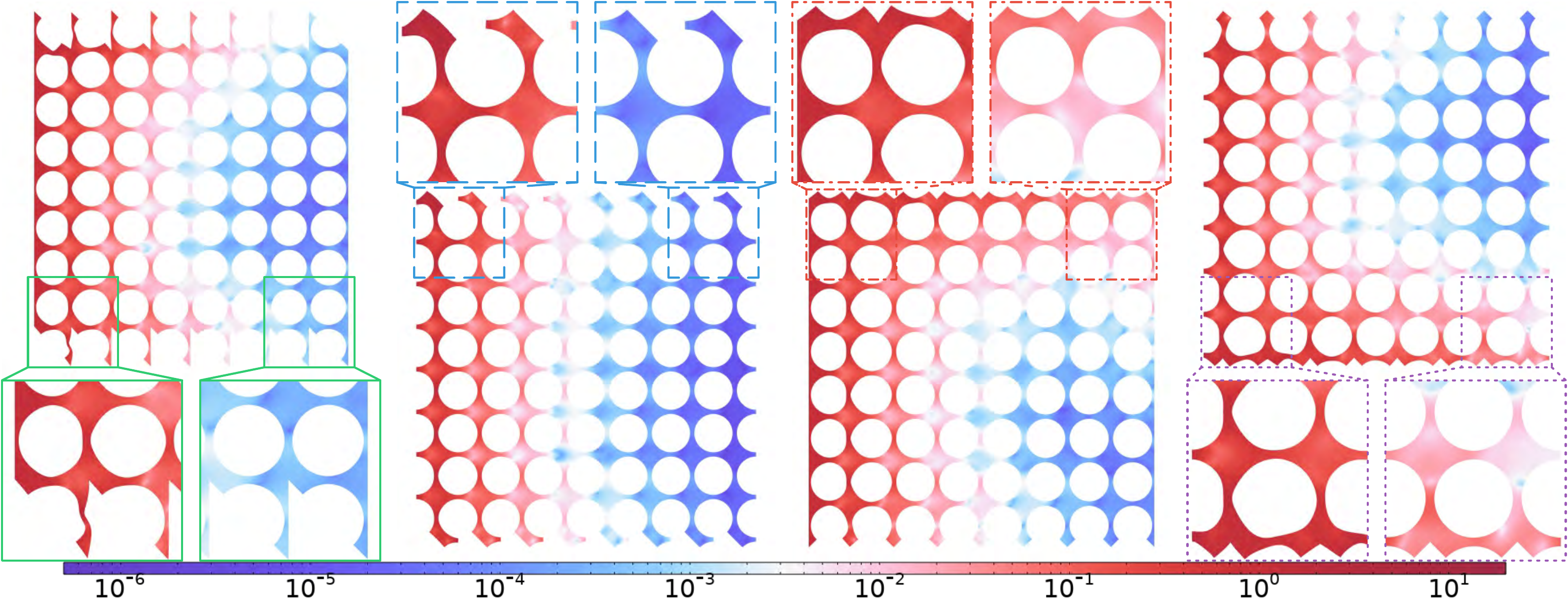}
  \caption{
  (\textit{Top} and \textit{Bottom}) Norm of the dimensionless displacement ($\lVert u \rVert/u_0$) at the frequency $f = 8.657$~[kHz], which lies within the band-gap range, showing boundary details for the considered RUCs.
  (\textit{Center}) Transmissibility curves.
  The band-gap regions are highlighted in light green; circles indicate the frequencies used in the displacement field plots.
  The color scale is logarithmic, and the deformation is amplified for visualization purposes.
  }
  \label{fig:circ_transmission_be}
\end{figure}

%%%%%%%%%%%%%%%%%%%%%%%%%%%%%%%%%%%%%%%%%%%%%%%%%%%%%%%%%%%%
%%%%%%%%%%%%%%%%%%%%%%%%%%%%%%%%%%%%%%%%%%%%%%%%%%%%%%%%%%%%
As in Section \ref{sec:transm_4reso}, eight RUCs have been identified, and their geometries are shown as insets in Fig.~\ref{fig:circ_transmission_be}. Four of them have a square shape, RUC$_{\rm s}^{\star}$; two are parallelogram, RUC$_{\rm p}^{\star}$; and two are arrow-shaped, RUC$_{\rm a}^{\star}$.

%%%%%%%%%%%%%%%%%%%%%%%%%%%%%%%%%%%%%%%%%%%%%%%%%%%%%%%%%%%%
%%%%%%%%%%%%%%%%%%%%%%%%%%%%%%%%%%%%%%%%%%%%%%%%%%%%%%%%%%%%%

%
%
%
%%%%%%%%%%%%%%%%%%%%%%%%%%%%%%%%%%%%%%%%%%%%%%%%%%%%%%%%%%%%
%%%%%%%%%%%%%%%%%%%%%%%%%%%%%%%%%%%%%%%%%%%%%%%%%%%%%%%%%%%%
\subsection{Discussion}
\label{sec:circ_discussion}
%%%%%%%%%%%%%%%%%%%%%%%%%%%%%%%%%%%%%%%%%%%%%%%%%%%%%%%%%%%%
%%%%%%%%%%%%%%%%%%%%%%%%%%%%%%%%%%%%%%%%%%%%%%%%%%%%%%%%%%%%

In Fig.~\ref{fig:circ_transmission_be} of Section~\ref{sec:transm_circ}, the transmissibility for certain RUCs within specific band-gap frequency ranges is higher than expected.
To better understand this phenomenon, a parametric study of the size of the structure for RUC$_{\rm s}^{\alpha}$, in which the number of unit cells is varied in steps of four in both the vertical and horizontal directions, is shown in Fig.~\ref{fig:circ_transmission_size} (see Fig.~\ref{fig:circ_transmission_beta_gamma_delta_size} in Appendix~\ref{app:circ_size} for the other RUC$_{\rm s}^{\star}$).

As shown in the top and bottom left panels of Fig.~\ref{fig:circ_transmission_size}, the transmissibility wells can be correlated with the eigenfrequencies of the full structure, calculated using the same boundary conditions reported in Fig.~\ref{fig:BC}.
In addition, an in-depth analysis shows that the ratio $\overline{W}$, defined in Eq.~(\ref{eq:ratio_energies}) as the total energy in the top, bottom, and rightmost boundary unit cells divided by the total energy in the remaining bulk unit cells, and shown in the bottom left panel of Fig.~\ref{fig:circ_transmission_size}, clearly aligns with the increase in transmissibility.

This further corroborates the interpretation that transmissibility in the band-gap range is strongly influenced by boundary-guided waves.
%%%%%%%%%%%%%%%%%%%%%%%%%%%%%%%%%%%%%%%%%%%%%%%%%%%%%%%%%%%%
\begin{figure}[!htbp]
  \centering
  \includegraphics[width=0.47\textwidth]{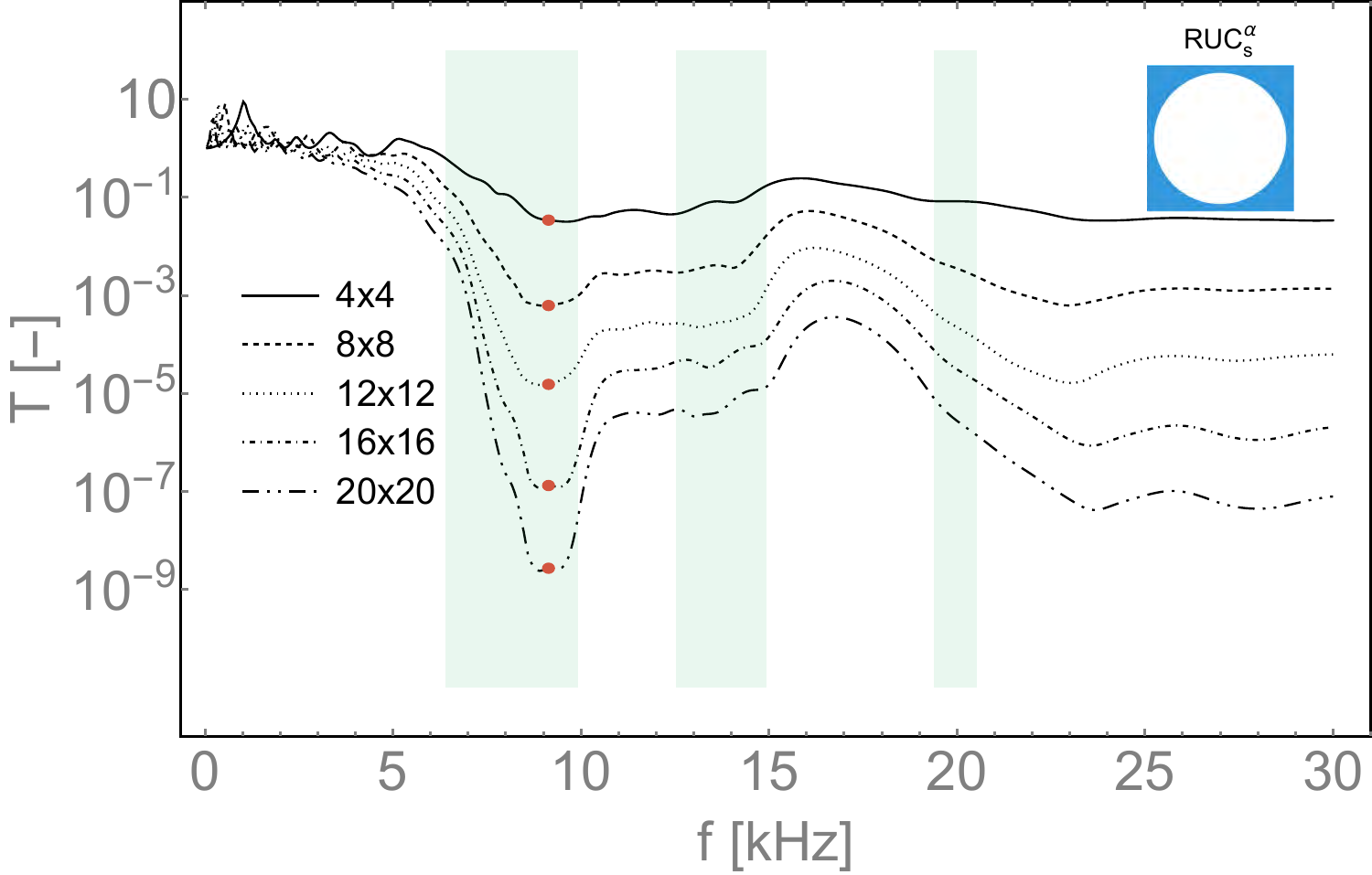}
  \hfill
  \includegraphics[width=0.47\textwidth]{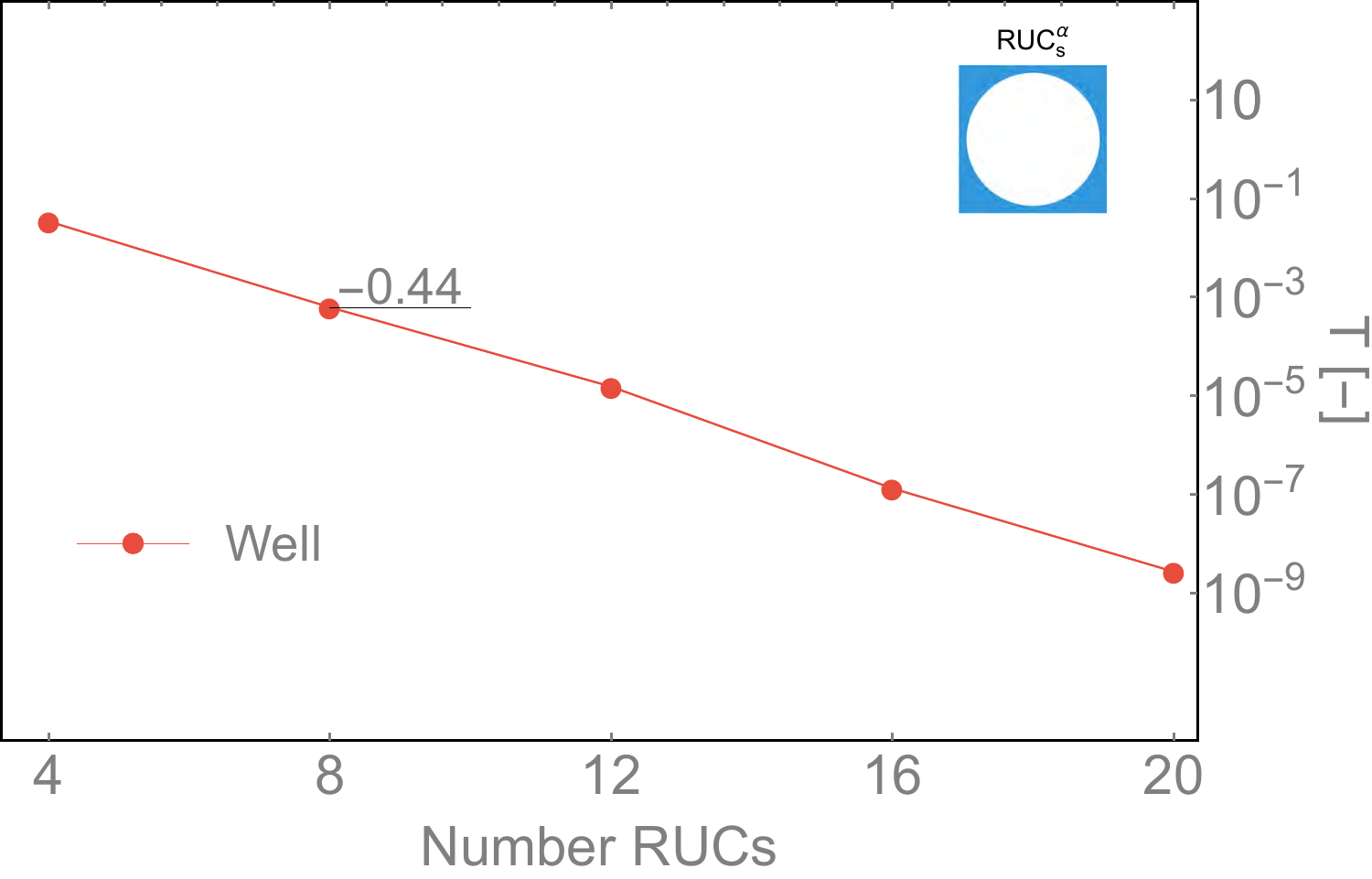}
  \\
  \includegraphics[width=0.47\textwidth]{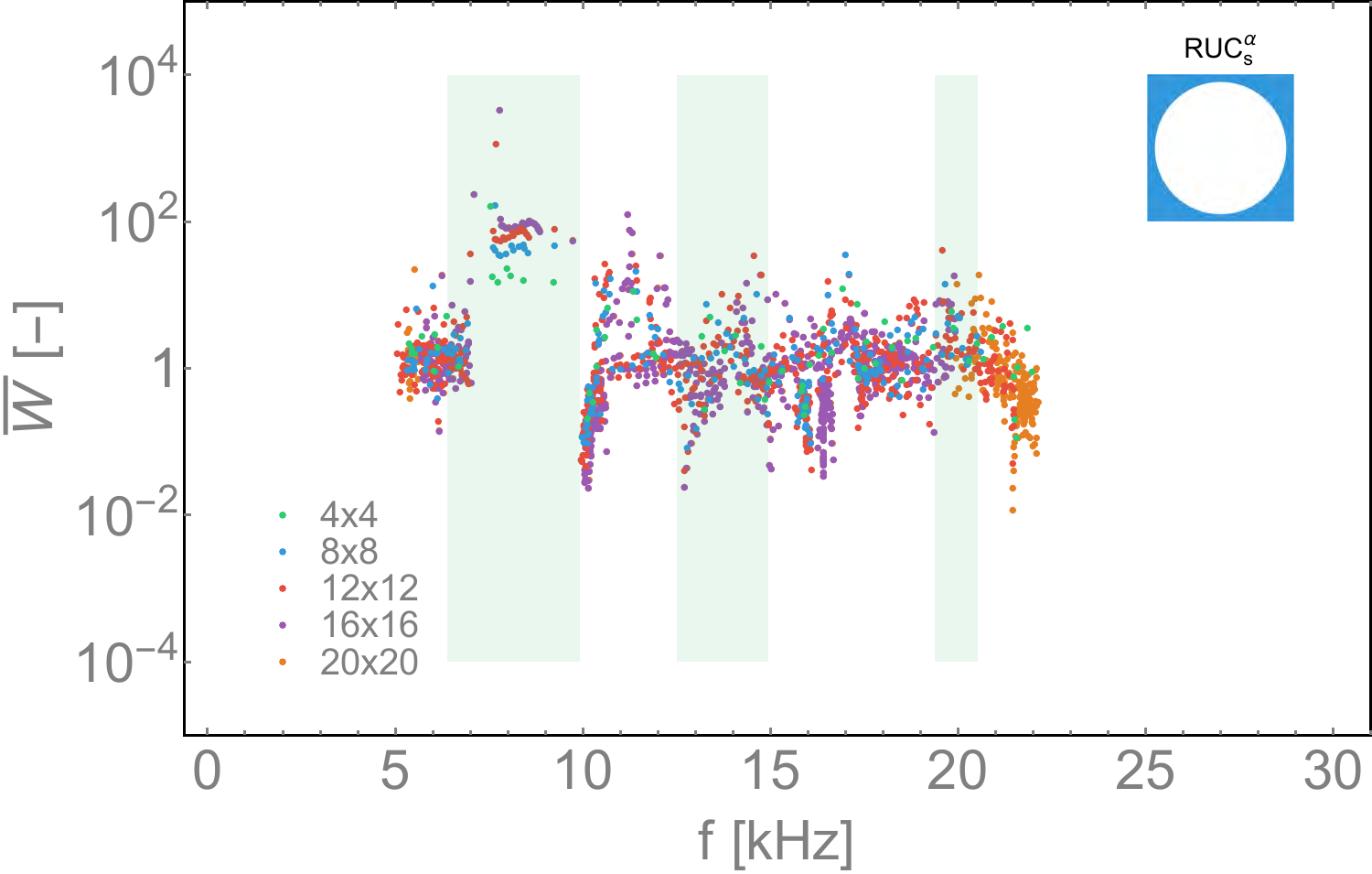}
  \hfill
  \includegraphics[width=0.47\textwidth]{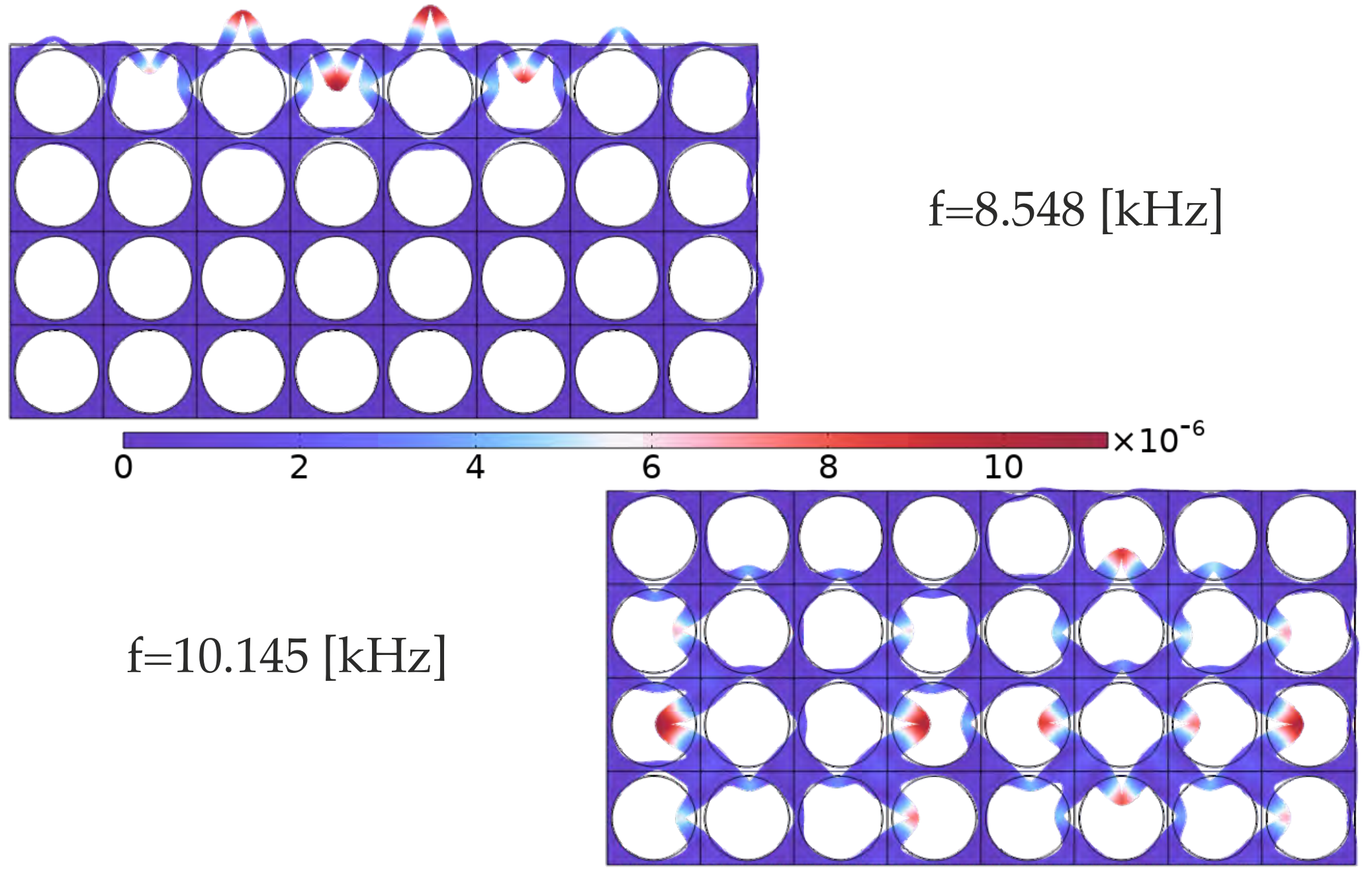}
  \caption{
  (\textit{Top left}) Transmissibility curves for different structure sizes for RUC$_{\rm s}^{\alpha}$, with the transmissibility wells highlighted by red dots;
  (\textit{Top right}) variation of wells (red curve with dots) as a function of the number of RUCs composing the structure;
  (\textit{Bottom left}) ratio between the energy localized at the top, bottom, and rightmost boundary cells and the energy in the remaining cells in the bulk, denoted by $\overline{W}$, for the eigenfrequencies of each structure;
  (\textit{Bottom right}) norm of the dimensionless displacement field ($\lVert u \rVert/u_0$) for two eigenfrequencies, $f = 8.511$~[kHz] and $f = 10.145$~[kHz]  (due to the symmetry of this test, only half of the structure is reported).
  }
  \label{fig:circ_transmission_size}
\end{figure}
%%%%%%%%%%%%%%%%%%%%%%%%%%%%%%%%%%%%%%%%%%%%%%%%%%%%%%%%%%%%%
It can also be seen in the bottom left panel of Fig.~\ref{fig:circ_transmission_size} that there is a concentration of eigenfrequencies at the beginning of the first band gap, around $f = 6$~kHz, which explains why the transmissibility curves decrease only gradually to a minimum, even when already well inside the band-gap range.

As in Section~\ref{sec:4reso_discussion}, an \emph{ad hoc} Bloch-Floquet analysis can be performed to provide information about localized effects (see Appendix~\ref{app:circ_size}), although it is less accurate and fails to distinguish between RUCs related by a horizontal shift.

The top right panel of Fig.~\ref{fig:circ_transmission_size} shows how the wells (red curve with dots) quickly approach zero transmissibility as the number of RUCs composing the structure increases. This suggests that, for this metamaterial, increasing the sample size to mitigate boundary effects is effective, since the energy focused at the boundary is not significantly higher than that in the bulk within the band-gap range.

In the bottom right panel, the dimensionless displacement field ($\lVert u \rVert / u_0$) is shown for two eigenfrequencies: $f = 8.548$~kHz (inside a band gap) and $f = 10.145$~kHz (outside a band gap). The first clearly shows displacement (energy) focusing on the free top and right boundaries, which is responsible for the reduced effectiveness of the band gap, while the second shows energy stored primarily in the bulk (only half of the structure is shown due to symmetry).

For this metamaterial, it is also observed that a continuous structural path along the top or bottom boundaries of the RUC allows eigenfrequencies to emerge within the band-gap regions, inducing boundary effects. In contrast, the absence of such a path inhibits transmission. However, this observation is specific to the metamaterial samples, RUC configurations, and tests considered in this study.

It should also be noted that, since the damping $\eta$ characterizing the square-circular-hole unit cell is several orders of magnitude larger than that of the four-resonator unit cell, no peaks appear for the square-circular-hole metamaterial.
The high damping dissipates the boundary-guided waves before they reach the output boundary.

%
%
%
%
%%%%%%%%%%%%%%%%%%%%%%%%%%%%%%%%%%%%%%%%%%%%%%%%%%%%%%%%%%%%
%%%%%%%%%%%%%%%%%%%%%%%%%%%%%%%%%%%%%%%%%%%%%%%%%%%%%%%%%%%%
\section{A non integer square-circular-hole unit cells structure: a four-leaf clover sample}
\label{sec:clover}
\begin{figure}[!htbp]
  \centering
  \centering
  \includegraphics[width=0.85\textwidth]{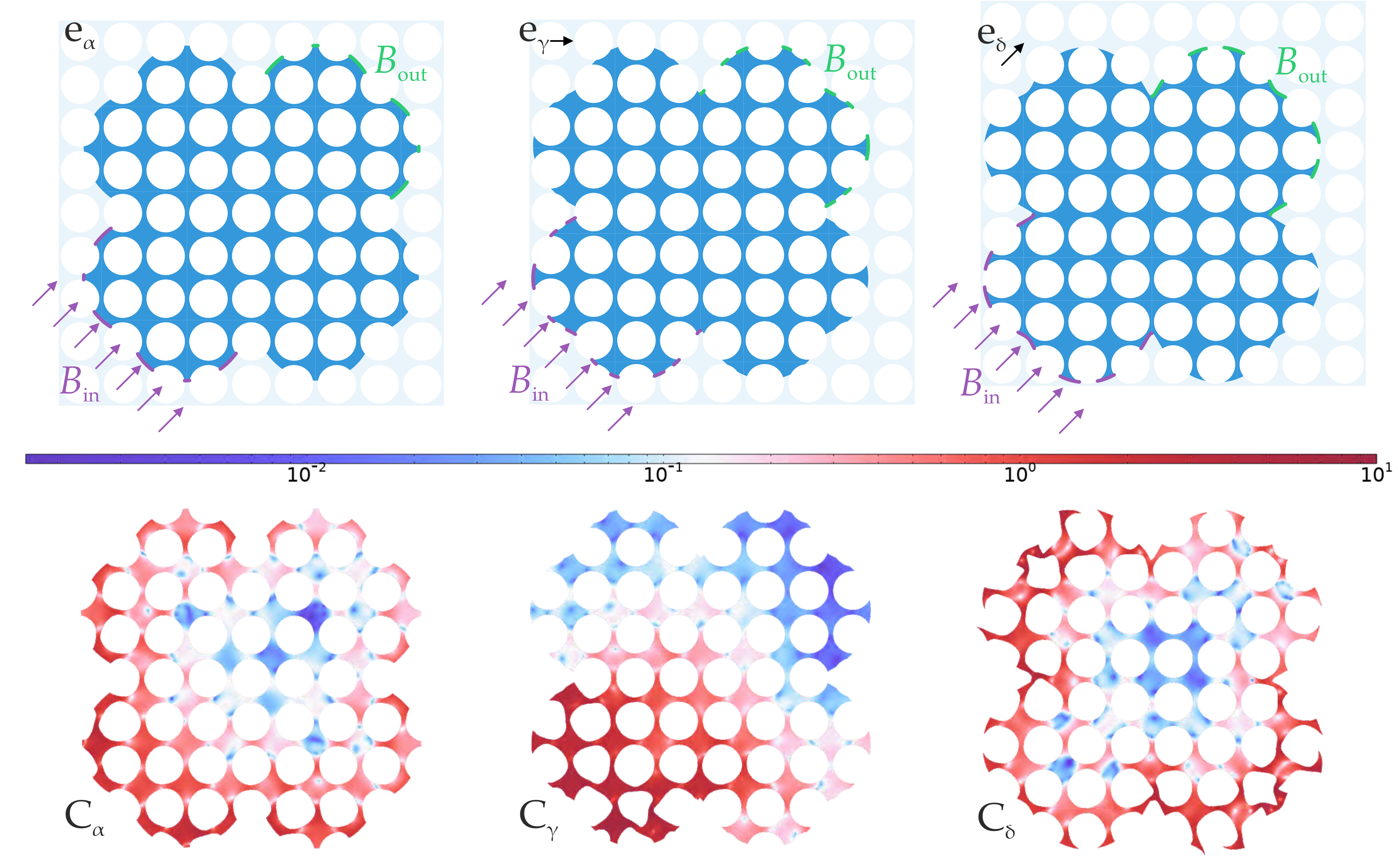}
  \caption{
    (\textit{Top}) Three four-leaf clover shape finite-size structure $C_{\alpha}$, $C_{\gamma}$, and $C_{\delta}$ that have been cut out from the square-circular-hole metamaterial after shifting it by the vectors $e_{\alpha}$, $e_{\gamma}$, and $e_{\delta}$, respectively.
    The displacement $\overline{u}_{\rm in} = \overline{u}_{0} (e_1 + e_2)/L$ is applied on the boundary highlighted in purple, while the output displacement $\overline{u}_{\rm out}$ is measured on the boundary highlighted in green;
    (\textit{Bottom}) norm of the dimensionless displacement field for each cut at the band-gap frequency $f = 8.130$~kHz.
  }
  \label{fig:circ_shape_defo_clover}
\end{figure}

%%%%%%%%%%%%%%%%%%%%%%%%%%%%%%%%%%%%%%%%%%%%%%%%%%%%%%%%%%%%
%%%%%%%%%%%%%%%%%%%%%%%%%%%%%%%%%%%%%%%%%%%%%%%%%%%%%%%%%%%%
As a final test, a structure that is not an integer multiple of unit cells is considered. The chosen shape for the structure is a four-leaf clover obtained from three different cuts, namely $C_{\alpha}$, $C_{\gamma}$, and $C_{\delta}$, generated by shifting the square--circular-hole unit cell along the vectors $e_{\alpha}$, $e_{\gamma}$, and $e_{\delta}$, respectively\footnote{The structure obtained from $e_{\beta}$ is the mirror reflection of the one obtained from $e_{\gamma}$ and is therefore equivalent} (see Fig.~\ref{fig:circ_shape_defo_clover}).

This test aims to highlight the complexities arising from a non-regular, non-periodic boundary and to investigate propagation properties along the $45^\circ$ direction. In Fig.~\ref{fig:circ_shape_defo_clover}, the boundary where the displacement $\overline{u}_{\rm in} = \overline{u}_{0}(e_1 + e_2)/L$ is applied is highlighted in purple, while the boundary where the output displacement $\overline{u}_{\rm out}$ is measured is highlighted in green.
%%%%%%%%%%%%%%%%%%%%%%%%%%%%%%%%%%%%%%%%%%%%%%%%%%%%%%%%%%%%
%%%%%%%%%%%%%%%%%%%%%%%%%%%%%%%%%%%%%%%%%%%%%%%%%%%%%%%%%%%%%

\begin{figure}[!htbp]
  \centering
  \includegraphics[width=0.47\textwidth]{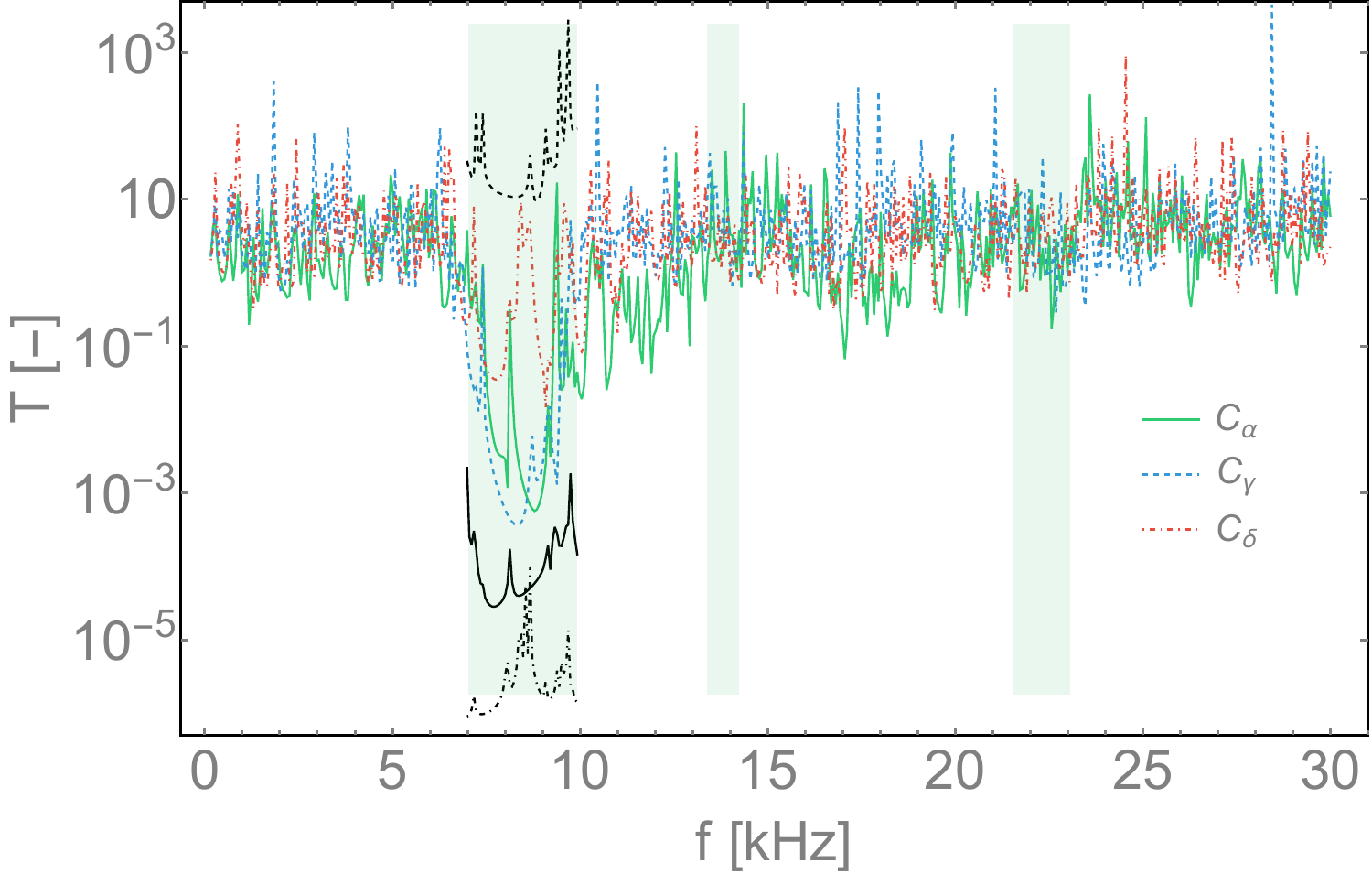}
  \hfill
  \includegraphics[width=0.47\textwidth]{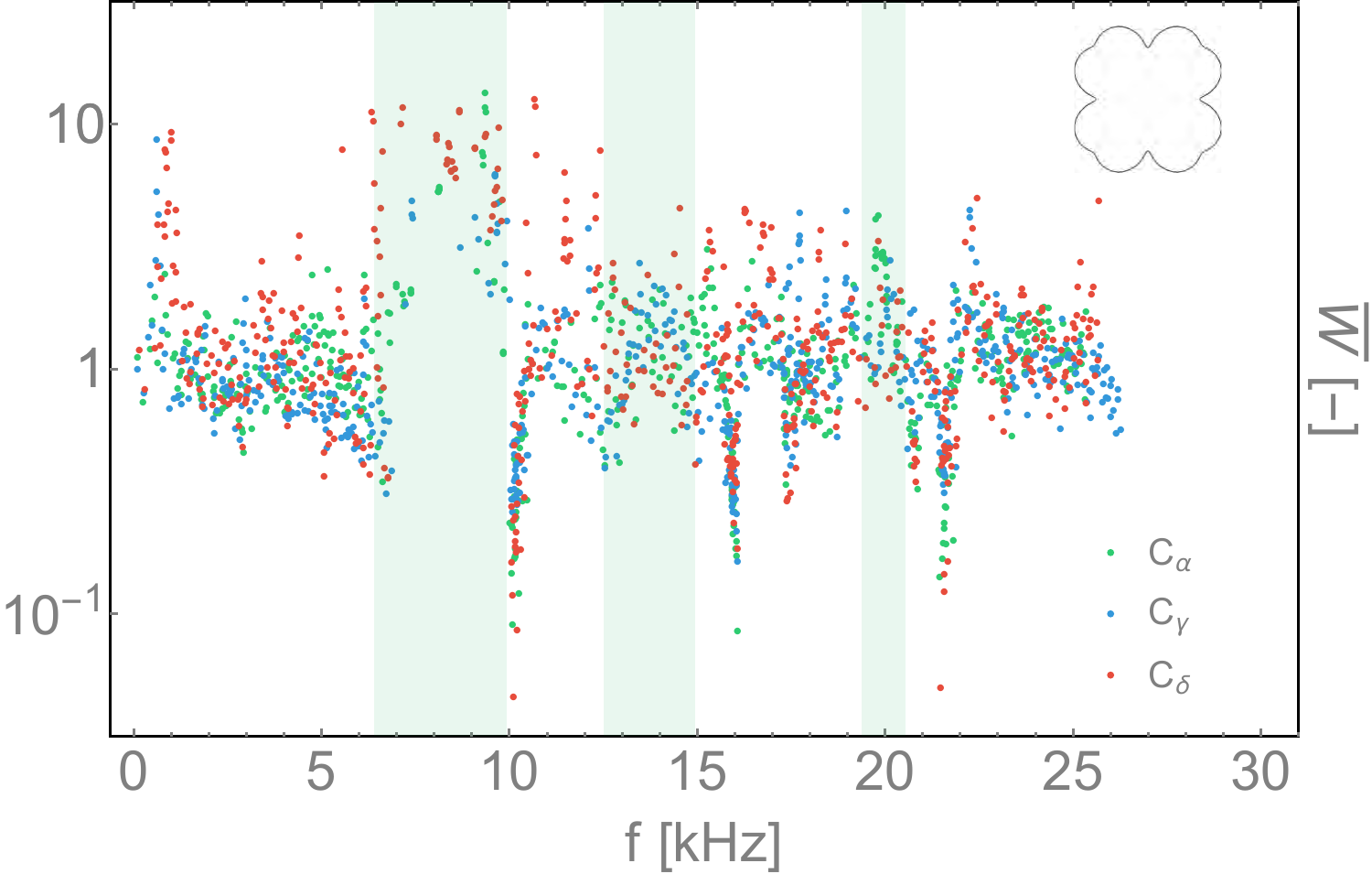}
  \\[10pt]
  \includegraphics[width=0.85\textwidth]{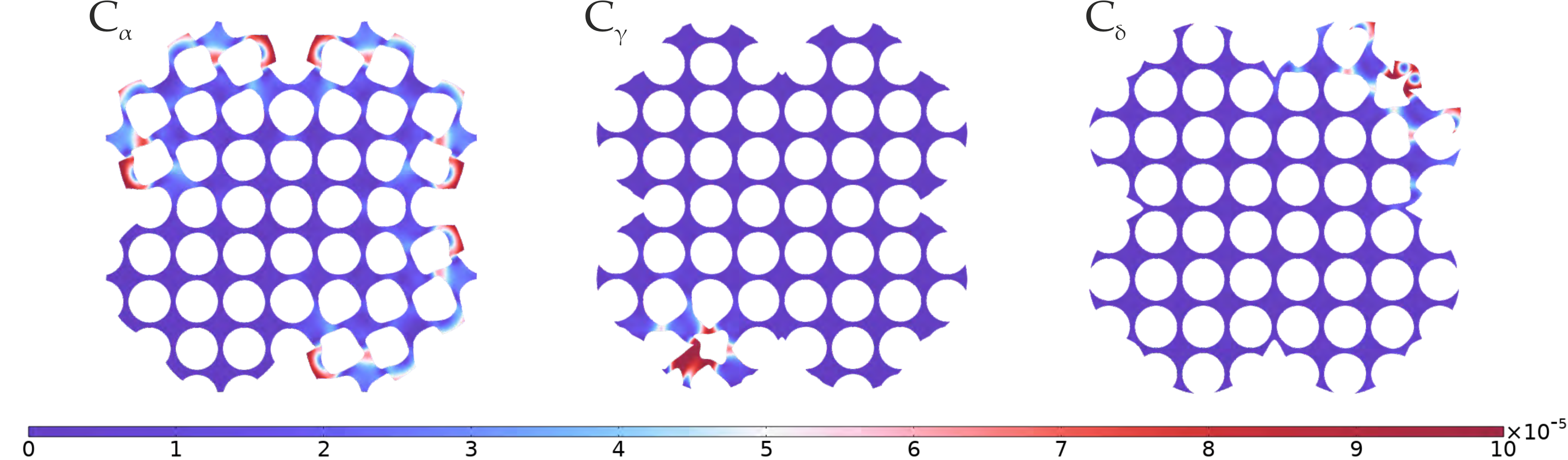}
  \caption{
  (\textit{Top}) On the left, the transmissibility curves for the three structures.
  The black lines denote the average out-of-plane component of curl $u$ on the boundary unit cells that do not share a boundary with $B_{\rm i}$ and $B_{\rm o}$, while on the right, the ratio between the energy localized at the boundary and the energy in the bulk, denoted by $\overline{W}$ and defined in Eq.(\ref{eq:ratio_energies}).
  (\textit{Bottom}) norm of the dimensionless displacement field ($\lVert u \rVert/u_0$) for the cuts $C_{\alpha}$, $C_{\gamma}$, and $C_{\delta}$, for the eigenfrequencies $f=\{8.139,8.683,8.063\}$~[kHz], respectively.
  }
  \label{fig:circ_transmission_clover}
\end{figure}

From Fig.~\ref{fig:circ_transmission_clover}, the transmissibility curves for $C_{\alpha}$, $C_{\gamma}$, and $C_{\delta}$ can be observed.
It is clear that the two structures symmetric with respect to the loading direction ($C_{\alpha}$ and $C_{\delta}$) exhibit significant transmission spikes in the first band gap (around $f = 7$~kHz), while the transmission for $C_{\gamma}$ in the same band-gap range remains low, making it the best choice in this context.

It is nevertheless important to stress that breaking symmetry is not a sufficient condition for reducing transmissibility, as can be seen in Fig.~\ref{fig:4reso_transmission_be} and Fig.~\ref{fig:circ_transmission_be}.
%%%%%%%%%%%%%%%%%%%%%%%%%%%%%%%%%%%%%%%%%%%%%%%%%%%%%%%%%%%%
%%%%%%%%%%%%%%%%%%%%%%%%%%%%%%%%%%%%%%%%%%%%%%%%%%%%%%%%%%%%%
It can be observed in Fig.~\ref{fig:circ_transmission_clover} that both the boundary-averaged out-of-plane component of $\mathrm{curl}\,u$ and the distribution of the most energetic eigenfrequencies align with the transmission peaks within the first band-gap region. This further supports the conclusion that the reduced effectiveness of the band gap in finite-size samples is correlated with eigenfrequencies in that range that trigger boundary-guided waves.

%
%
%
%
%%%%%%%%%%%%%%%%%%%%%%%%%%%%%%%%%%%%%%%%%%%%%%%%%%%%%%%%%%%%
%%%%%%%%%%%%%%%%%%%%%%%%%%%%%%%%%%%%%%%%%%%%%%%%%%%%%%%%%%%%
\section{Conclusions}
\label{sec:conclusions}
%%%%%%%%%%%%%%%%%%%%%%%%%%%%%%%%%%%%%%%%%%%%%%%%%%%%%%%%%%%%
%%%%%%%%%%%%%%%%%%%%%%%%%%%%%%%%%%%%%%%%%%%%%%%%%%%%%%%%%%%%
The objective of this study is to provide a comprehensive analysis of the mechanisms governing transmissibility within the band-gap range and to clarify the role played by the selection of different Representative Unit Cells (RUCs) as the fundamental building blocks of a finite-sized structure.

The results demonstrate that spikes in transmissibility within the band-gap range are directly associated with the presence of eigenfrequencies of the full structure under identical boundary conditions (see Fig.~\ref{fig:4reso_transmission_alpha_size}). These transmissibility peaks also systematically coincide with peaks in the average out-of-plane component of $\mathrm{curl}\,u$ evaluated along the boundary of the finite-size structure, indicating that the observed loss of attenuation is due to boundary-guided wave phenomena (see Fig.~\ref{fig:4reso_transmission_be}). In these configurations, transmissibility is effectively compromised by localized boundary effects rather than by bulk propagation mechanisms.

Importantly, these resonant features cannot be readily suppressed by simply increasing the specimen size, as shown in Fig.~\ref{fig:4reso_transmission_alpha_size}. Although increasing the structural dimensions improves the minimum transmissibility within the band gap, the decay of the resonance-induced spikes occurs at a significantly slower rate, making size scaling alone an inefficient mitigation strategy. Additionally, it is observed that the presence of a clear and continuous structural path along either the top or bottom boundaries of a selected RUC allows eigenfrequencies to appear within the band-gap regions, thereby inducing boundary effects. Conversely, the absence of such a path results in suppressed transmission. However, this observation remains specific to the particular metamaterial samples, RUC configurations, and tests conducted in this study, and a broader generalization of these findings is not suggested at this time.

To address these limitations, alternative structural RUCs can be selected to eliminate or substantially reduce eigenfrequencies within band-gap ranges. This approach provides a more targeted means of restoring robust attenuation performance.

Based on the results presented here, the proposed procedure to identify an effective RUC involves: (i) evaluating the eigenfrequencies (and performing a directional Bloch--Floquet analysis, if needed), (ii) computing $\overline{W}$ to assess energy localization, (iii) selecting the best-performing RUC, and (iv) validating its performance through a limited frequency-domain analysis.

Future work should therefore focus on implementing optimization schemes to systematically identify the RUC configuration that maximizes band-gap efficiency. Such optimization must account for all propagation directions to ensure omnidirectional performance and guarantee consistent attenuation characteristics across all directions.
%%%%%%%%%%%%%%%%%%%%%%%%%%%%%%%%%%%%%%%%%%%%%%%%%%%%%%%%%%%%%%%%%%%%%%%%%%%%%%%%%%%%%%%%%%%%%%%%

%%%%%%%%%%%%%%%%%%%%%%%%%%%%%%%%%%%%%%%%%%%%%%%%%%%%%%%%%%%%%%%%%%%%%%%%%%%%%%%%
%%%%%%%%%%%%%%%%%%%%%%%%%%%%%%%%%%%%%%%%%%%%%%%%%%%%%%%%%%%%%%%%%%%%%%%%%%%%%%%%
\vspace{10pt}
\begingroup
\setstretch{1}
\scriptsize
\noindent
\textbf{Acknowledgements:}
Gianluca Rizzi acknowledges support from the European Commission through the funding of the ERC Consolidator Grant META-LEGO, N$^\circ$ 101001759 (Prof. Angela Madeo).
\\
\noindent
\textbf{Conflict of interest:} The authors declare that they have no conflicts of interest in this work.
\\
\noindent
\textbf{Declaration of generative AI and AI-assisted technologies in the writing process:}
During the preparation of this work the authors used AI-assisted technologies in order to improve grammar and English language. After using this tool/service, the authors reviewed and edited the content as needed and take full responsibility for the content of the publication.
\endgroup
%%%%%%%%%%%%%%%%%%%%%%%%%%%%%%%%%%%%%%%%%%%%%%%%%%%%%%%%%%%%%%%%%%%%%%%%%%%%%%%%
%%%%%%%%%%%%%%%%%%%%%%%%%%%%%%%%%%%%%%%%%%%%%%%%%%%%%%%%%%%%%%%%%%%%%%%%%%%%%%%%
\vspace{-5pt}
\begingroup
\setstretch{1}
\setlength\bibitemsep{2pt}
\printbibliography
\endgroup
%%%%%%%%%%%%%%%%%%%%%%%%%%%%%%%%%%%%%%%%%%%%%%%%%%%%%%%%%%%%%%%%%%%%%%%%%%%%%%%%
%%%%%%%%%%%%%%%%%%%%%%%%%%%%%%%%%%%%%%%%%%%%%%%%%%%%%%%%%%%%%%%%%%%%%%%%%%%%%%%%

\clearpage
%%%%%%%%%%%%%%%%%%%%%%%%%%%%%%%%%%%%%%%%%%%%%%%%%%%%%%%%%%%%
%%%%%%%%%%%%%%%%%%%%%%%%%%%%%%%%%%%%%%%%%%%%%%%%%%%%%%%%%%%%
\appendix
%%%%%%%%%%%%%%%%%%%%%%%%%%%%%%%%%%%%%%%%%%%%%%%%%%%%%%%%%%%%
%%%%%%%%%%%%%%%%%%%%%%%%%%%%%%%%%%%%%%%%%%%%%%%%%%%%%%%%%%%%
\section{Four-resonator RUCs: size-dependent parametric study}
\label{app:4reso_size}
%%%%%%%%%%%%%%%%%%%%%%%%%%%%%%%%%%%%%%%%%%%%%%%%%%%%%%%%%%%%
%%%%%%%%%%%%%%%%%%%%%%%%%%%%%%%%%%%%%%%%%%%%%%%%%%%%%%%%%%%%
%%%%%%%%%%%%%%%%%%%%%%%%%%%%%%%%%%%%%%%%%%%%%%%%%%%%%%%%%%%
\begin{figure}[!ht]
  \centering
  \begin{minipage}{\textwidth}
    \centering
    \includegraphics[width=0.28\textwidth]{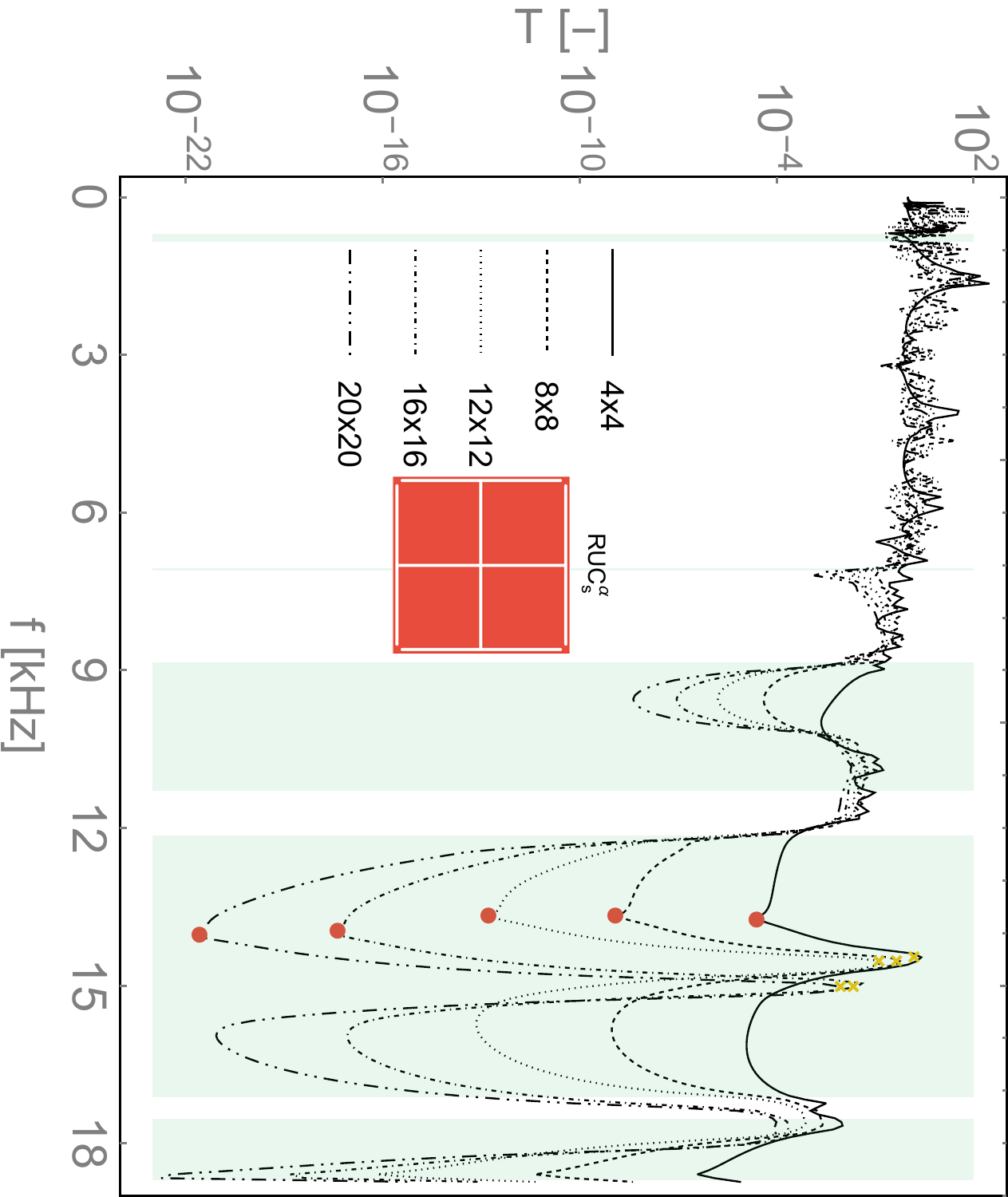}
    \hfill
    \includegraphics[width=0.28\textwidth]{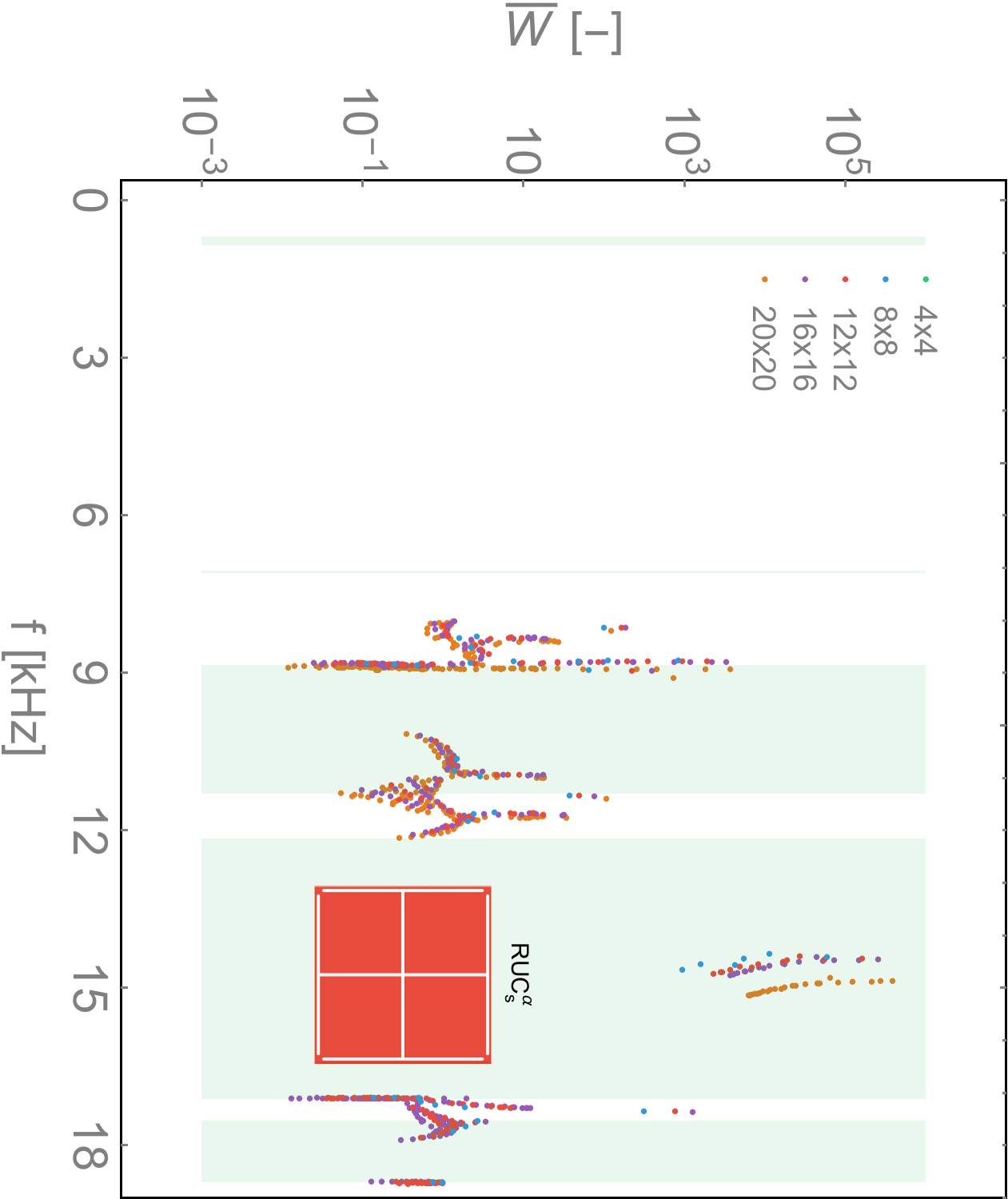}
    \hfill
    \includegraphics[width=0.28\textwidth]{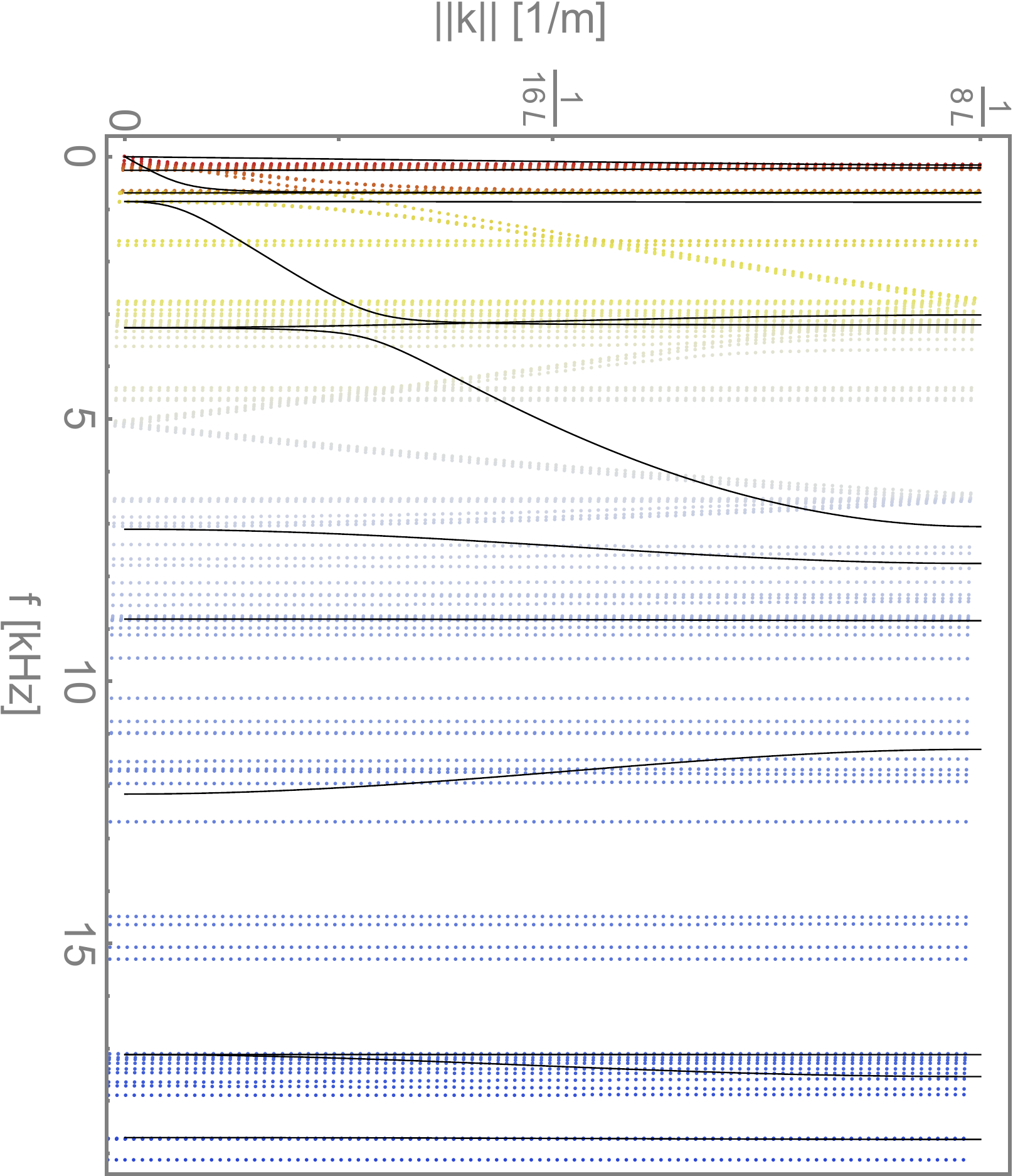}
  \end{minipage}
  \\
  \begin{minipage}{\textwidth}
    \centering
    \includegraphics[width=0.28\textwidth]{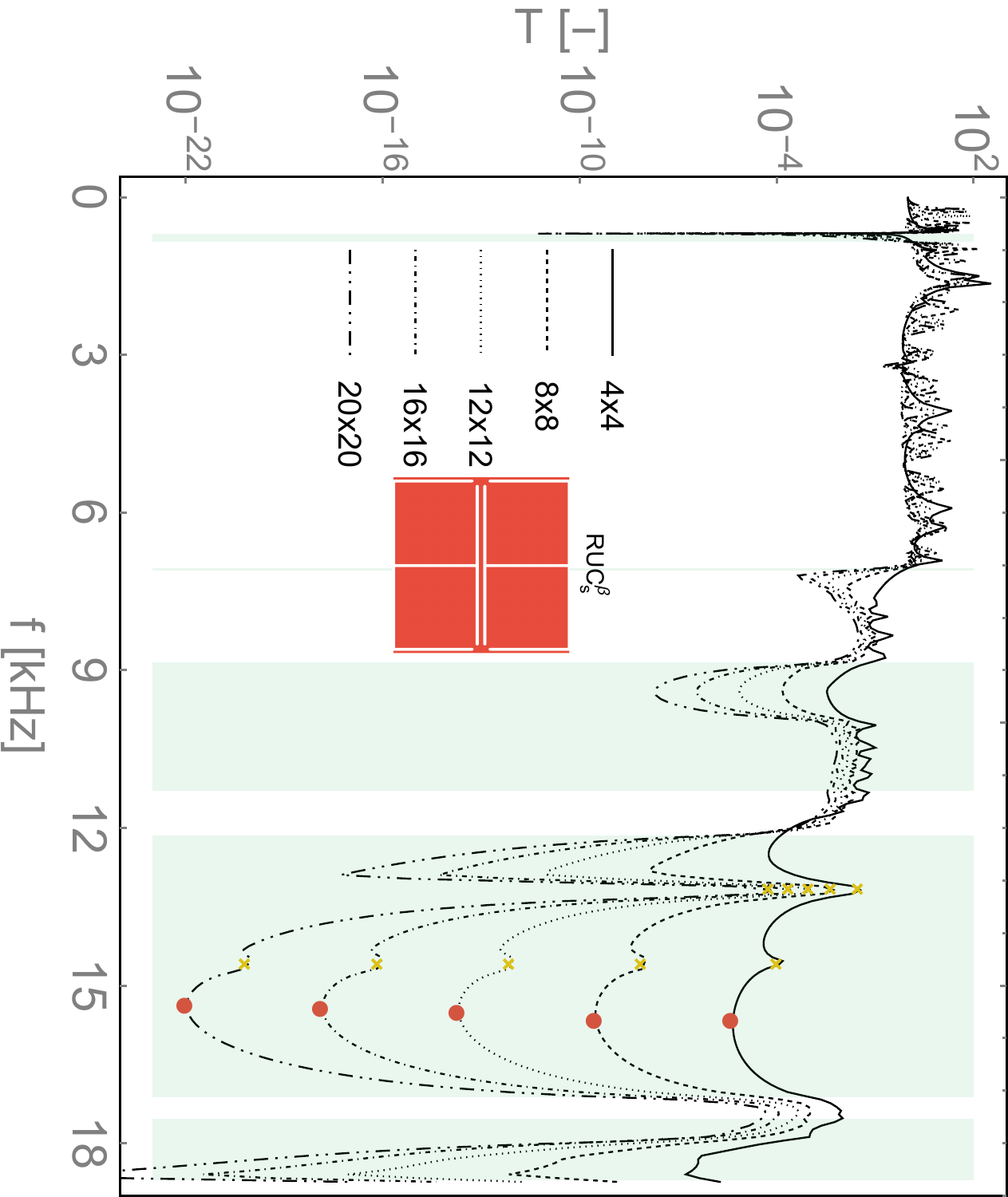}
    \hfill
    \includegraphics[width=0.28\textwidth]{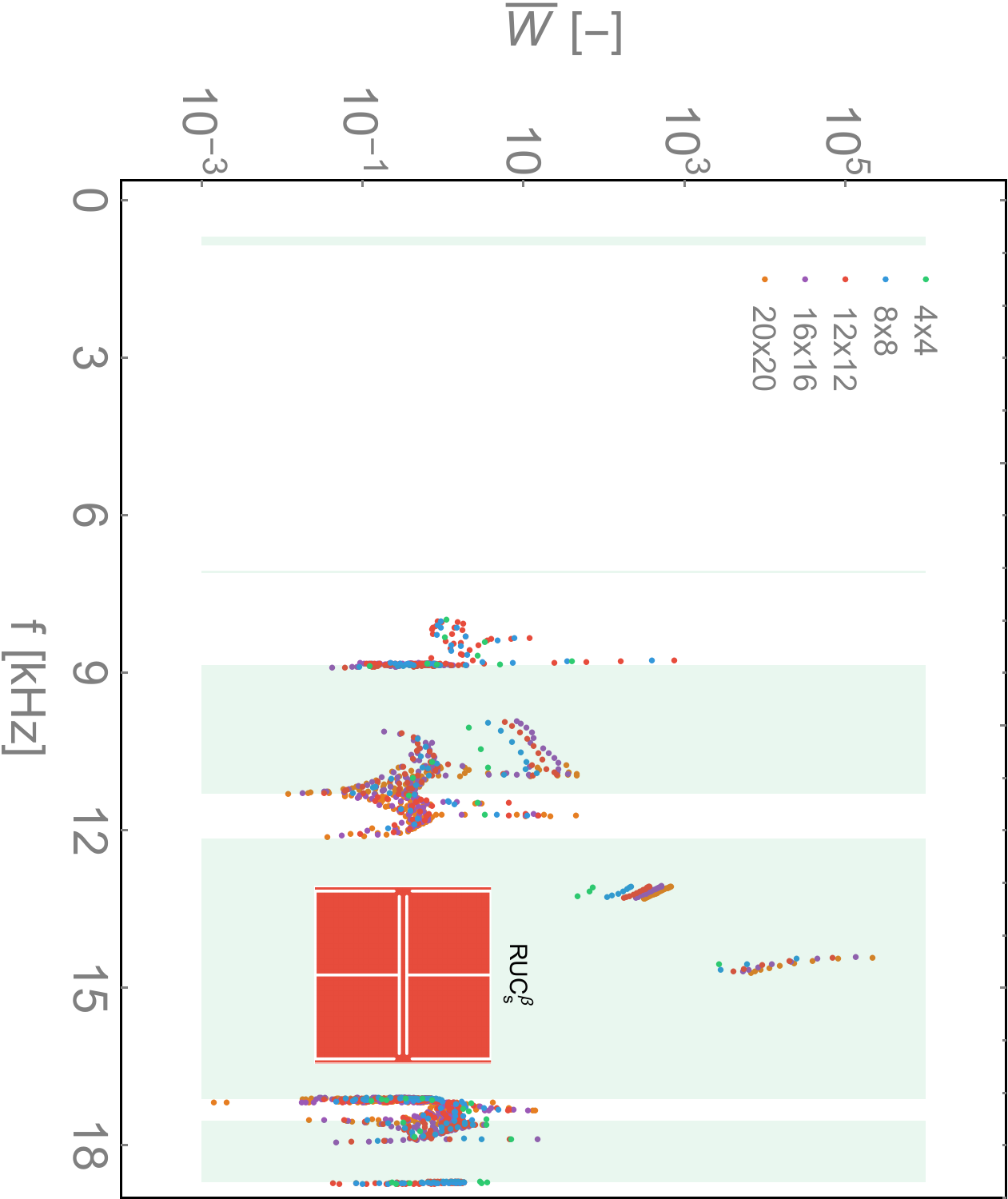}
    \hfill
    \includegraphics[width=0.28\textwidth]{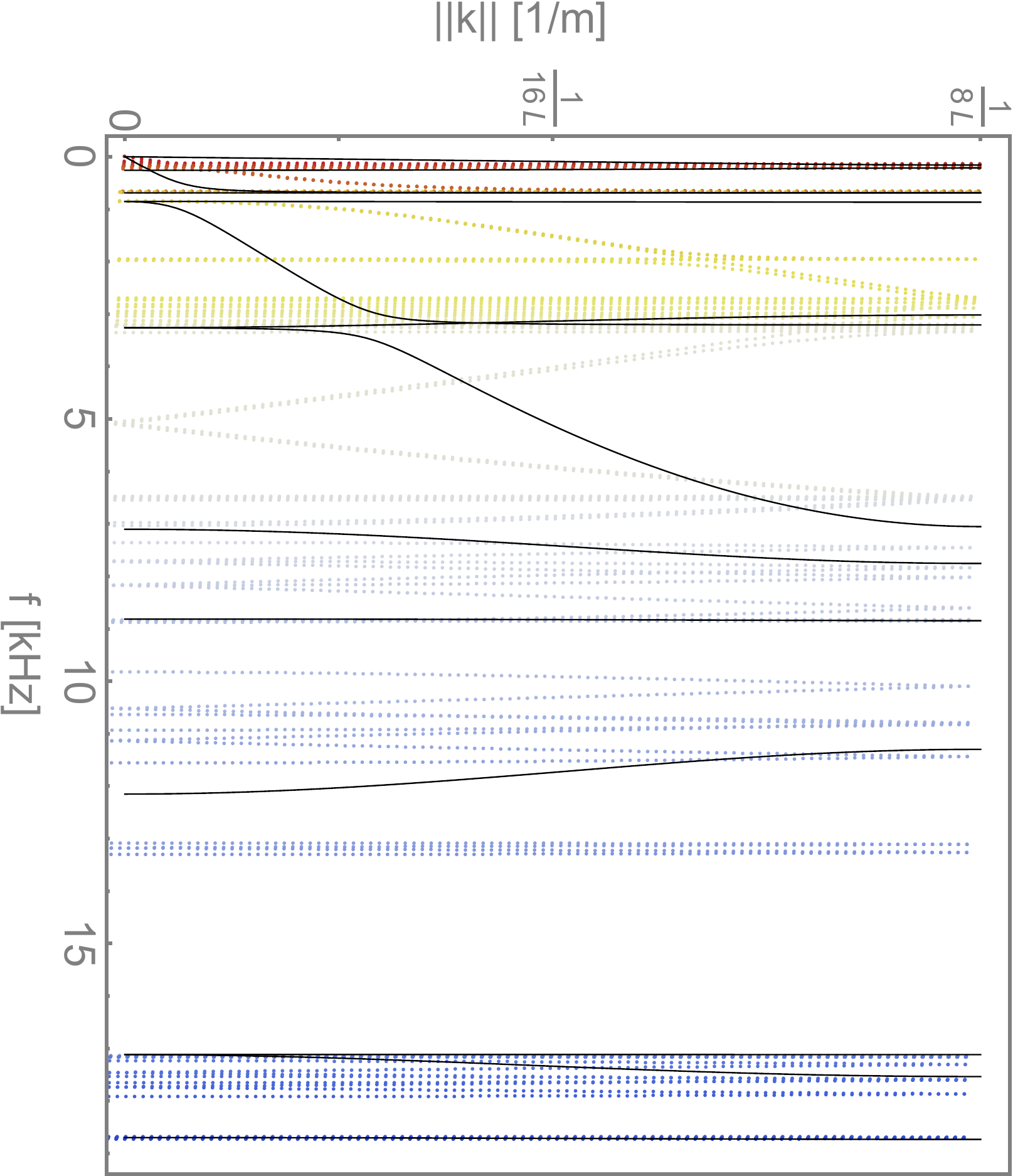}
  \end{minipage}
  \\
  \begin{minipage}{\textwidth}
    \centering
    \includegraphics[width=0.28\textwidth]{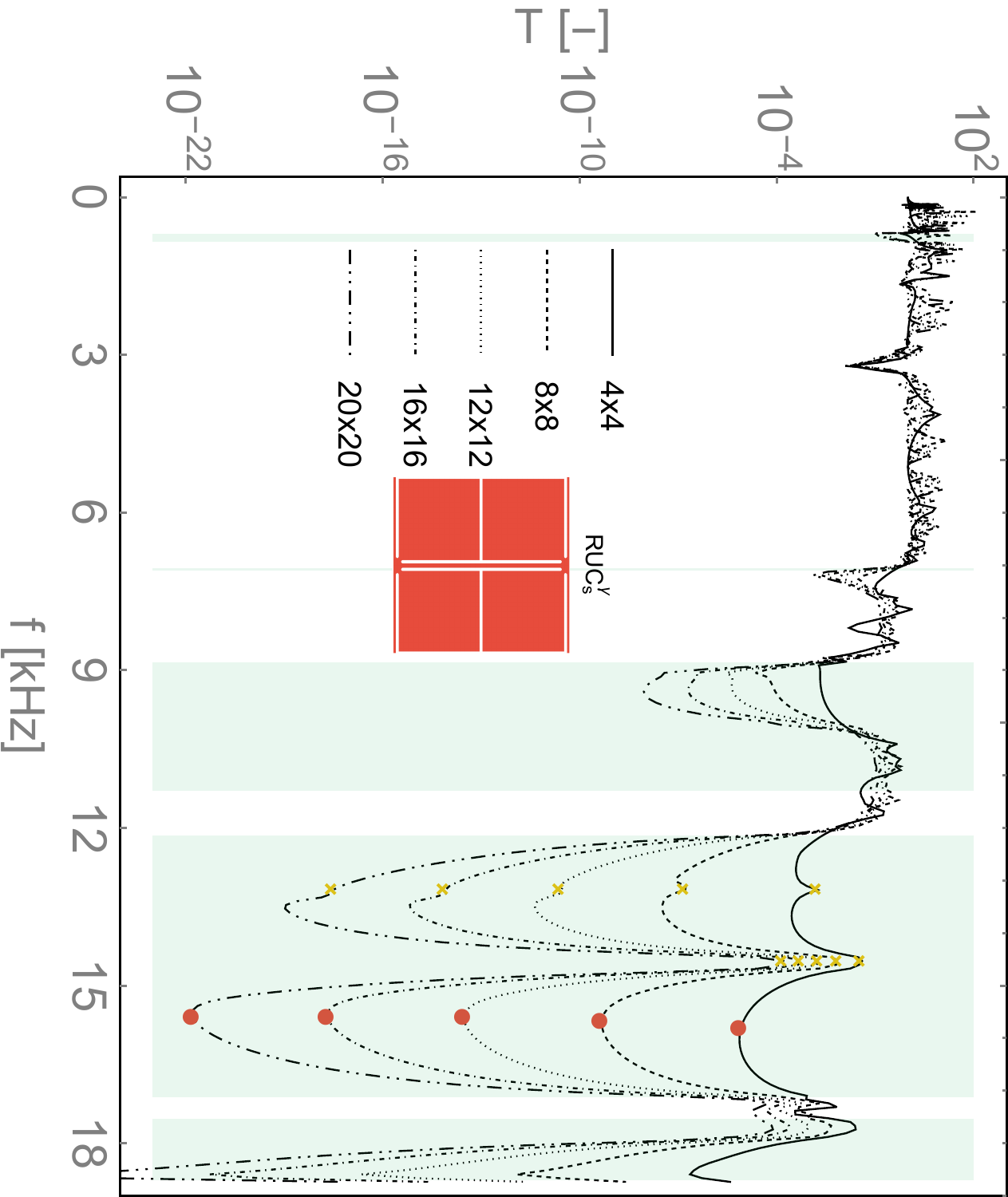}
    \hfill
    \includegraphics[width=0.28\textwidth]{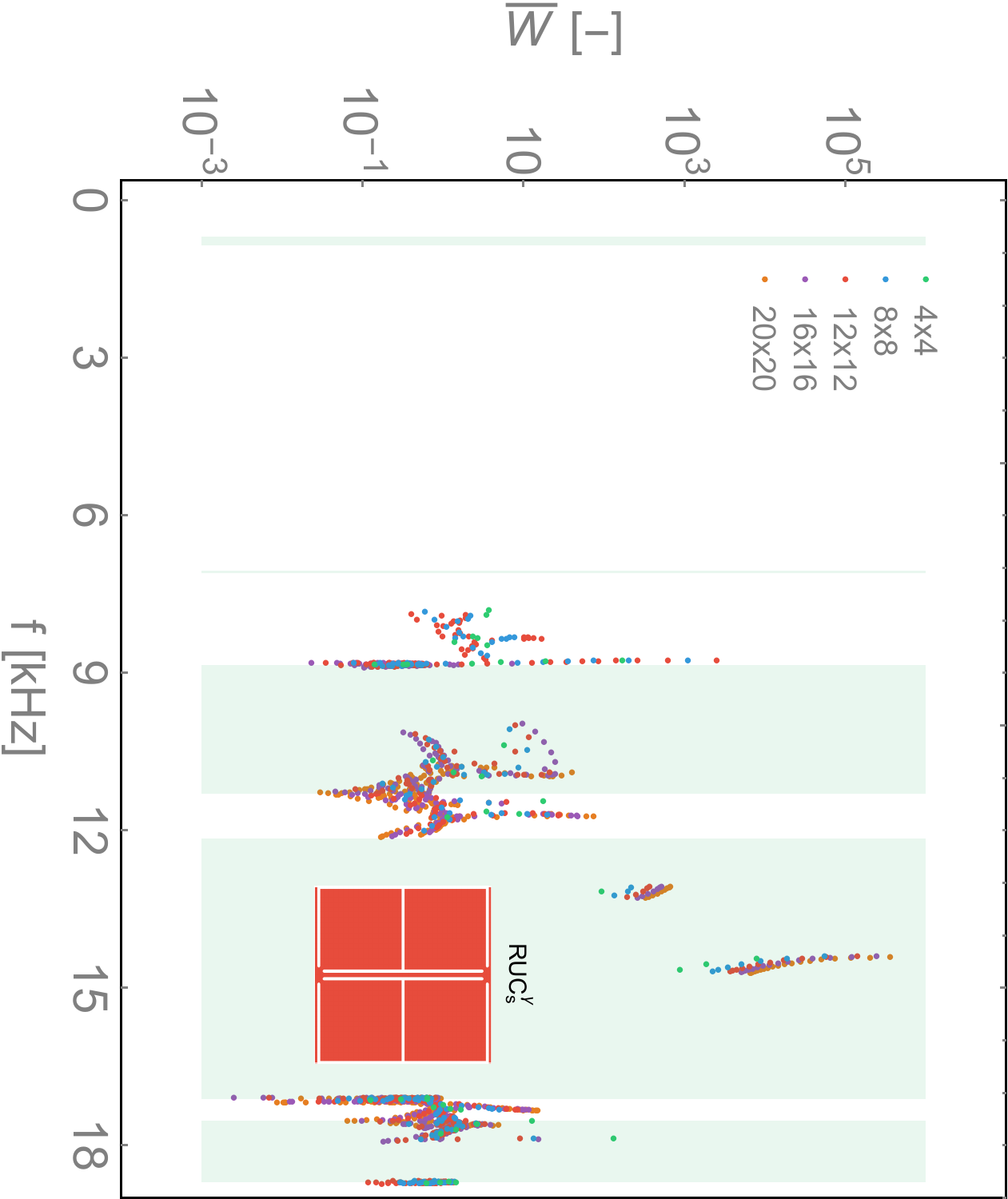}
    \hfill
    \includegraphics[width=0.28\textwidth]{Figures/4Reso_disp_curves_RE_14_alpha_gamma_only_LR.pdf}
  \end{minipage}
  \\
  \begin{minipage}{\textwidth}
    \centering
    \includegraphics[width=0.28\textwidth]{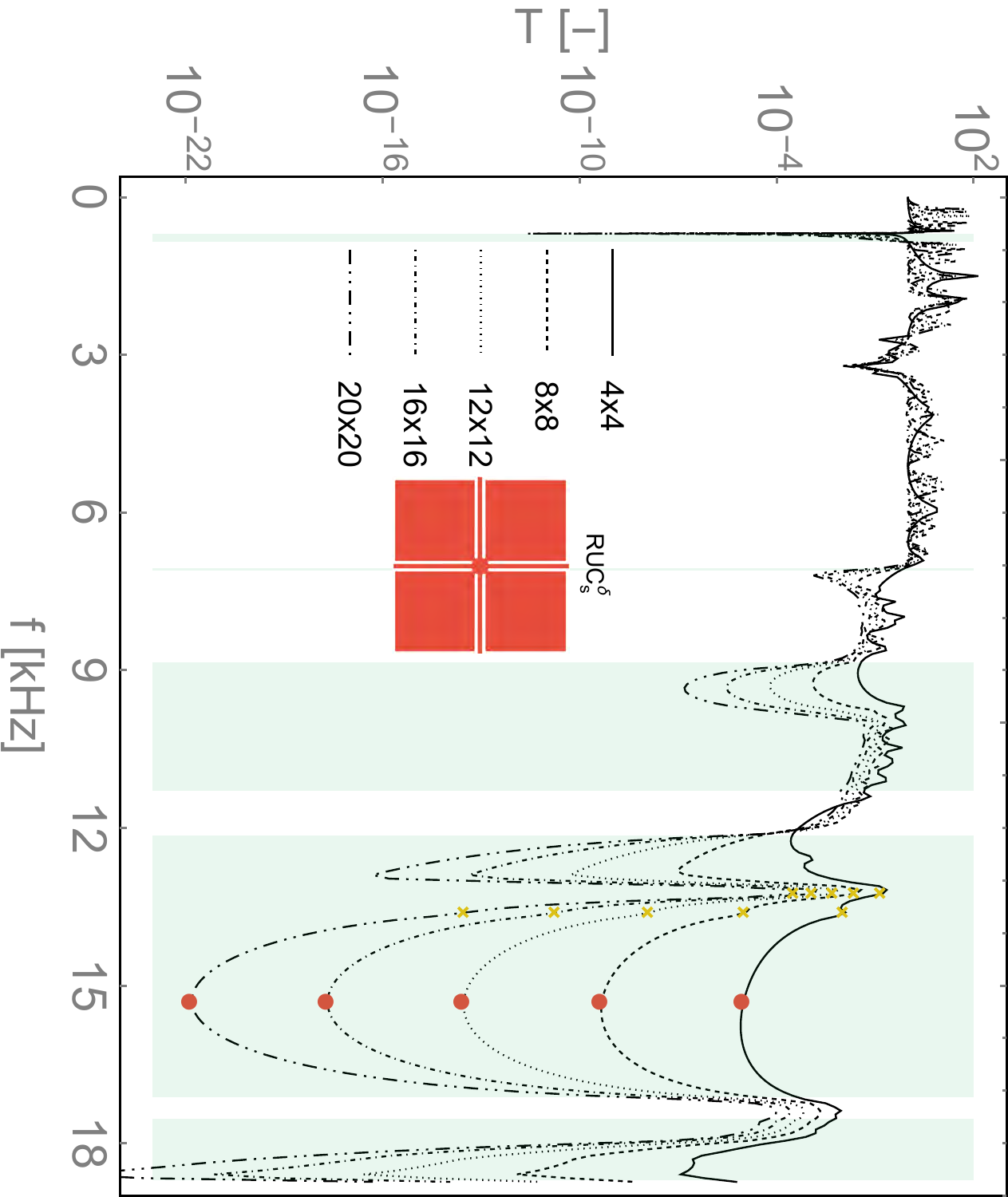}
    \hfill
    \includegraphics[width=0.28\textwidth]{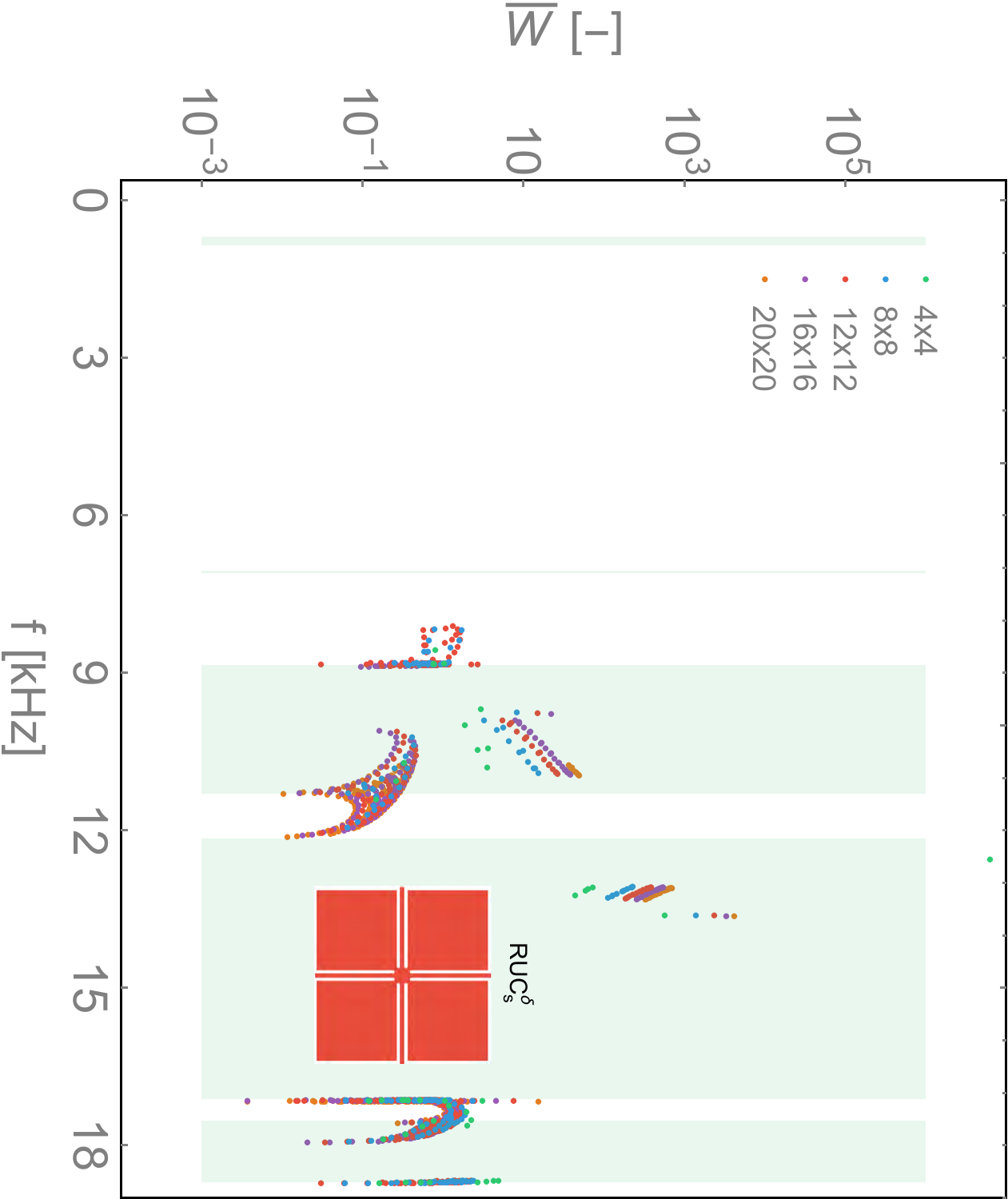}
    \hfill
    \includegraphics[width=0.28\textwidth]{Figures/4Reso_disp_curves_RE_14_beta_delta_only_LR.pdf}
  \end{minipage}
  \caption{
    (\textit{First column}) Transmissibility and
    (\textit{second column}) eigenfrequency for a parametric study of the RUC$_{\rm s}^{\star}$ of the four-resonator metamaterial, with the structure size increasing from $4 \times 4$ to $20 \times 20$ in increments of four;
    (\textit{third column}) dispersion curves for a $4 \times 4$ structure with Floquet conditions applied exclusively in the horizontal direction (the black lines represent the classical BF analysis for a single unit cell).
  }
  \label{fig:4reso_transmission_beta_gamma_delta_size}
\end{figure}
%%%%%%%%%%%%%%%%%%%%%%%%%%%%%%%%%%%%%%%%%%%%%%%%%%%%%%%%%%%
%%%%%%%%%%%%%%%%%%%%%%%%%%%%%%%%%%%%%%%%%%%%%%%%%%%%%%%%%%%%
%%%%%%%%%%%%%%%%%%%%%%%%%%%%%%%%%%%%%%%%%%%%%%%%%%%%%%%%%%%%
\section{Square-circular-hole RUCs: size-dependent parametric study}
\label{app:circ_size}
%%%%%%%%%%%%%%%%%%%%%%%%%%%%%%%%%%%%%%%%%%%%%%%%%%%%%%%%%%%%
%%%%%%%%%%%%%%%%%%%%%%%%%%%%%%%%%%%%%%%%%%%%%%%%%%%%%%%%%%%%
%%%%%%%%%%%%%%%%%%%%%%%%%%%%%%%%%%%%%%%%%%%%%%%%%%%%%%%%%%%
\begin{figure}[!ht]
  \centering
  \begin{minipage}{\textwidth}
    \centering
    \includegraphics[width=0.28\textwidth]{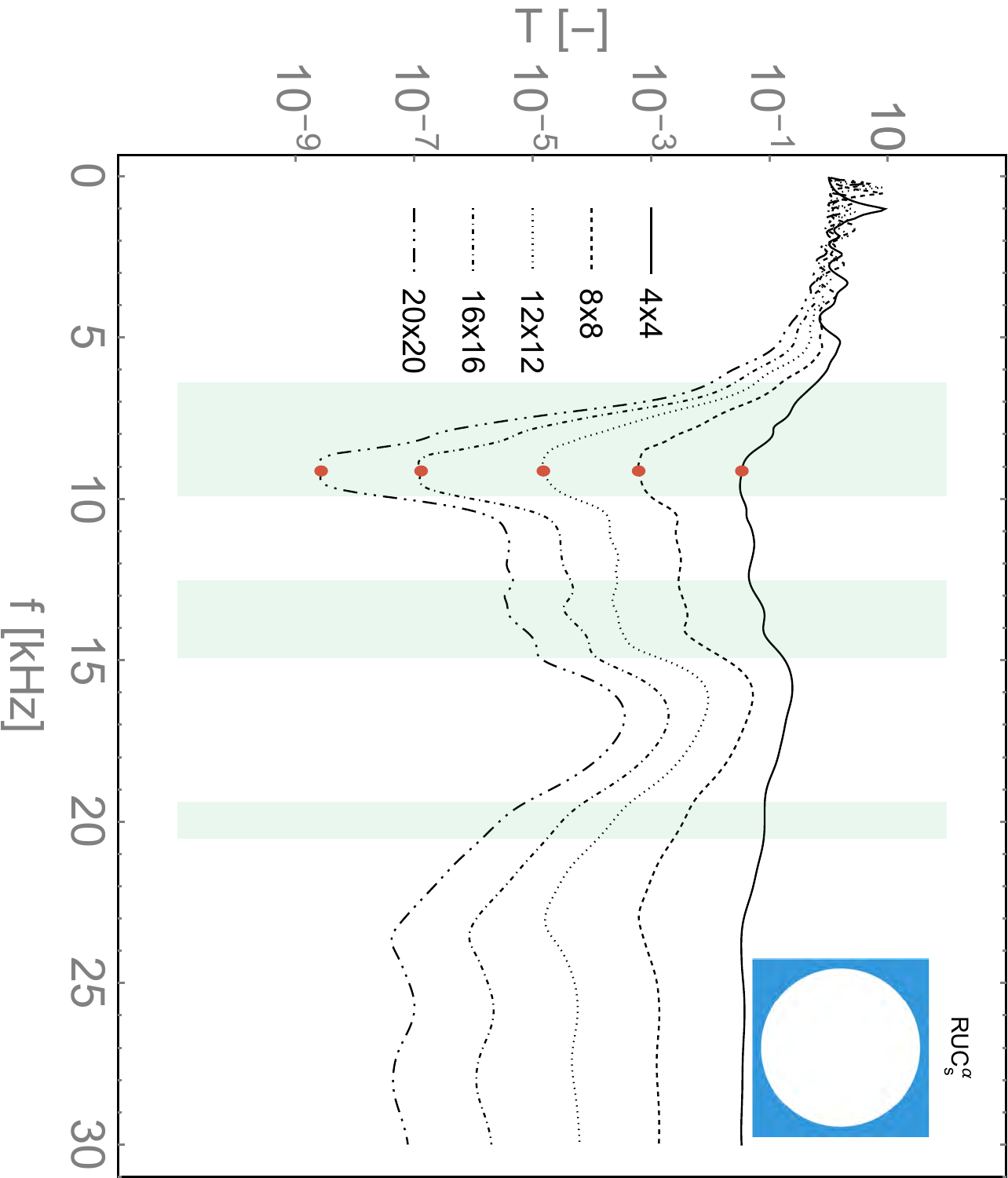}
    \hfill
    \includegraphics[width=0.28\textwidth]{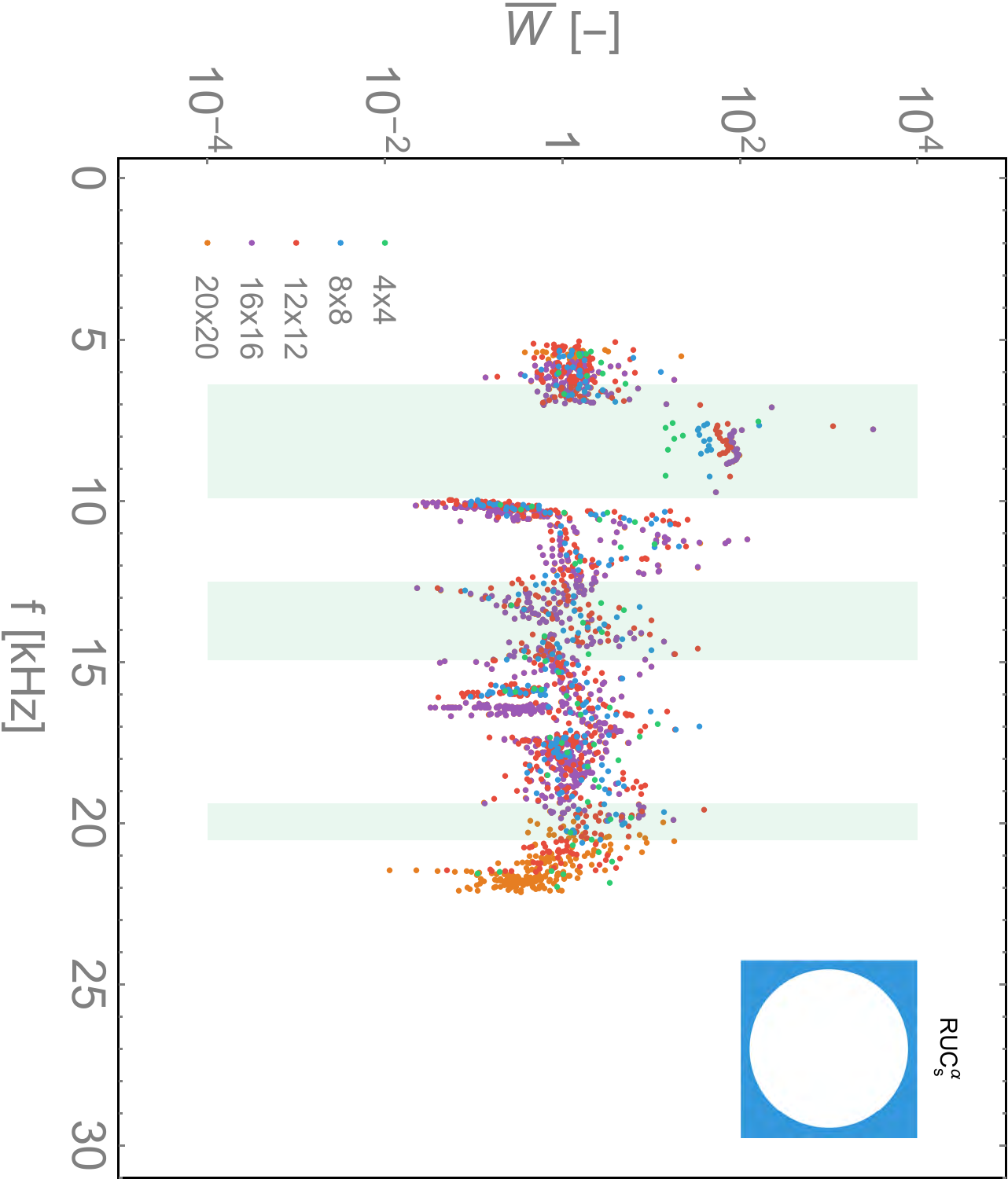}
    \hfill
    \includegraphics[width=0.28\textwidth]{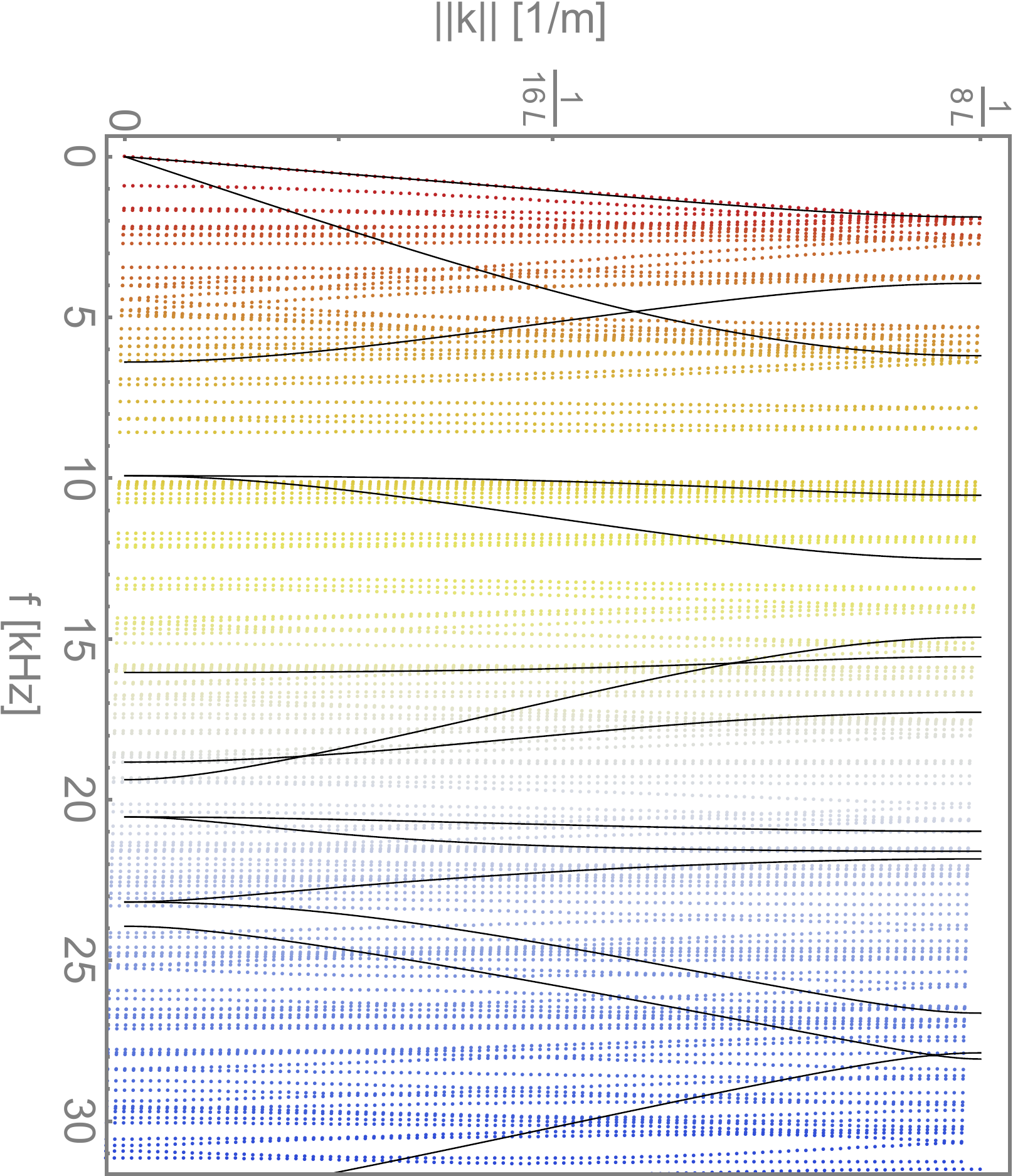}
  \end{minipage}
  \\
  \begin{minipage}{\textwidth}
    \centering
    \includegraphics[width=0.28\textwidth]{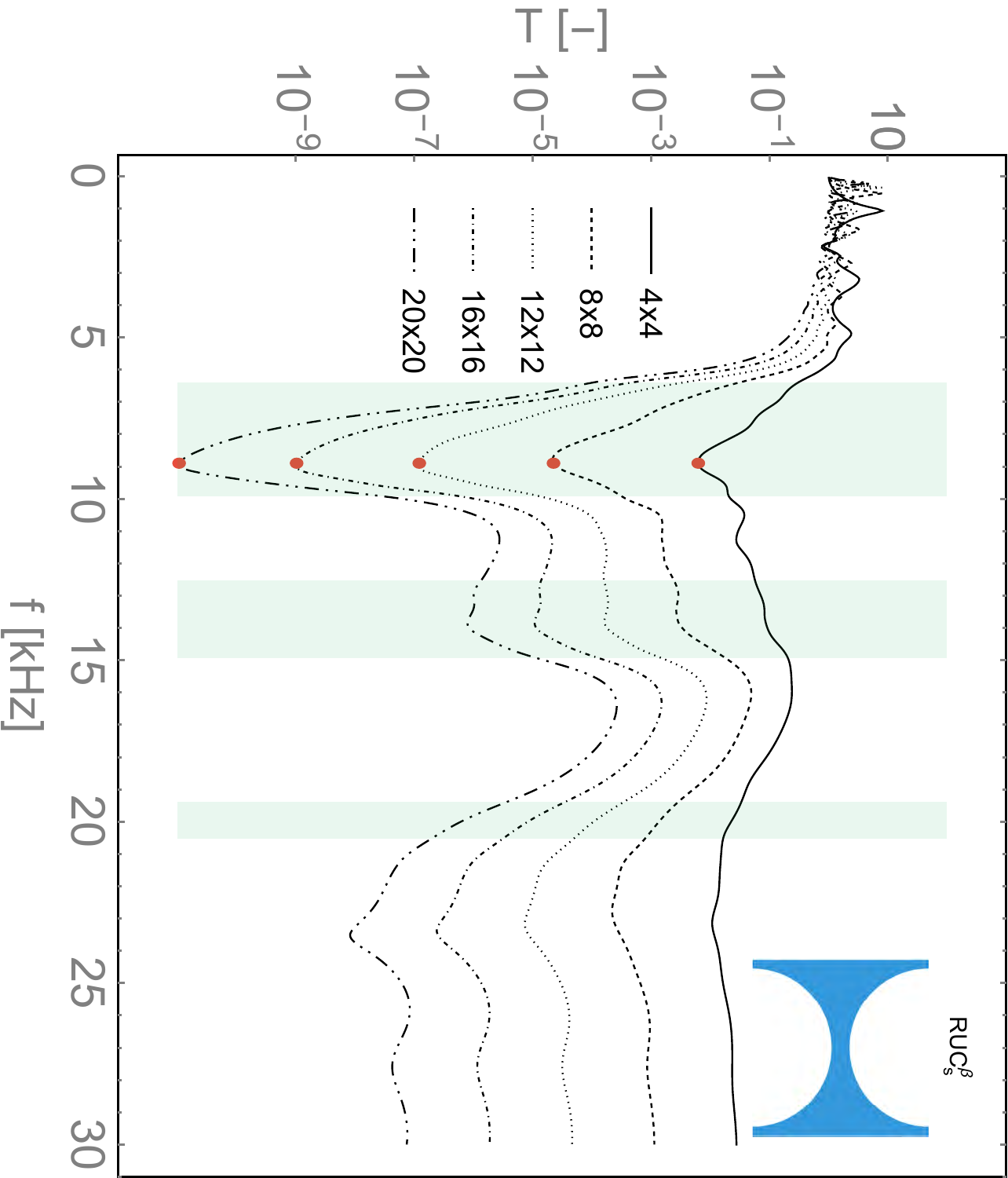}
    \hfill
    \includegraphics[width=0.28\textwidth]{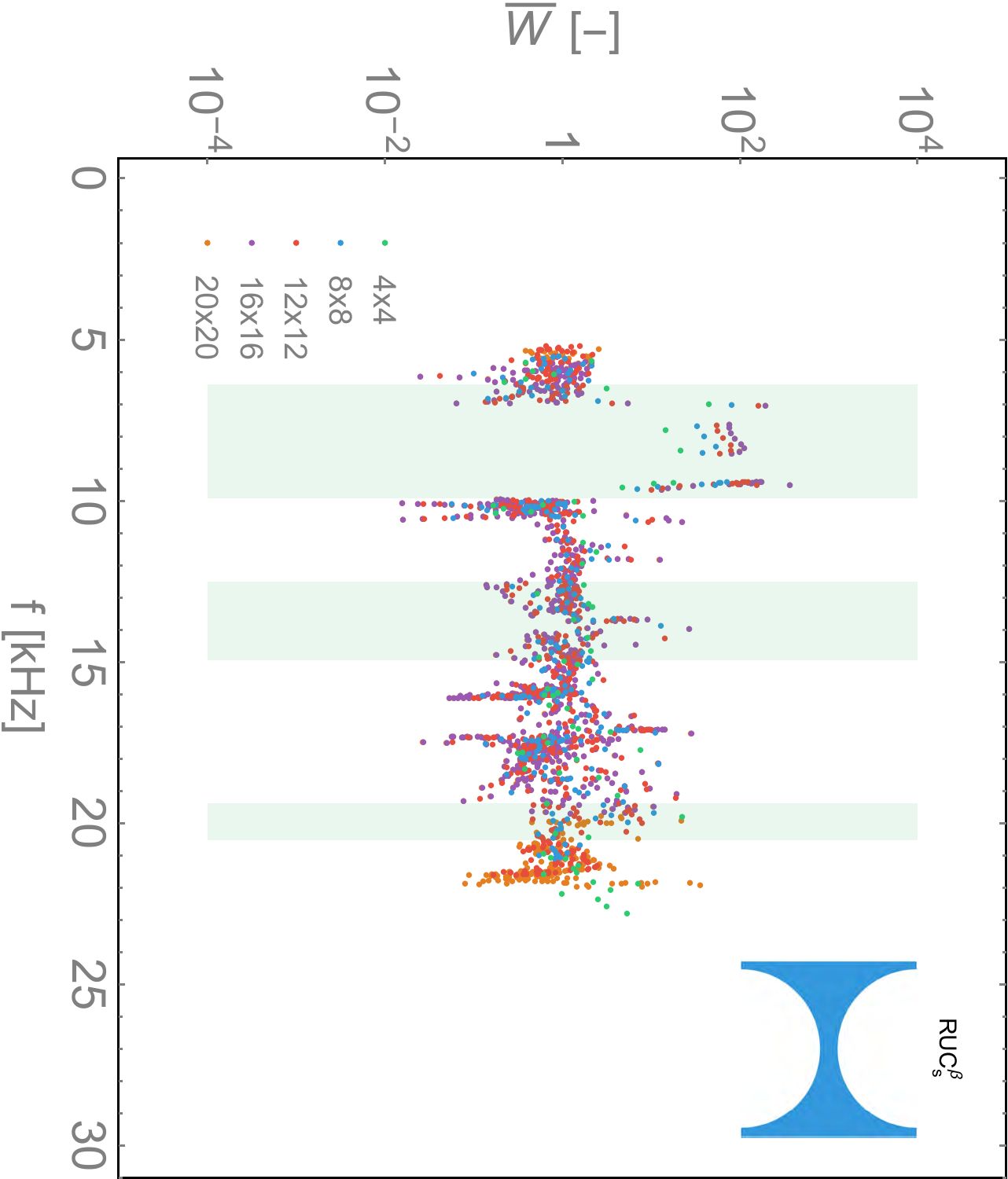}
    \hfill
    \includegraphics[width=0.28\textwidth]{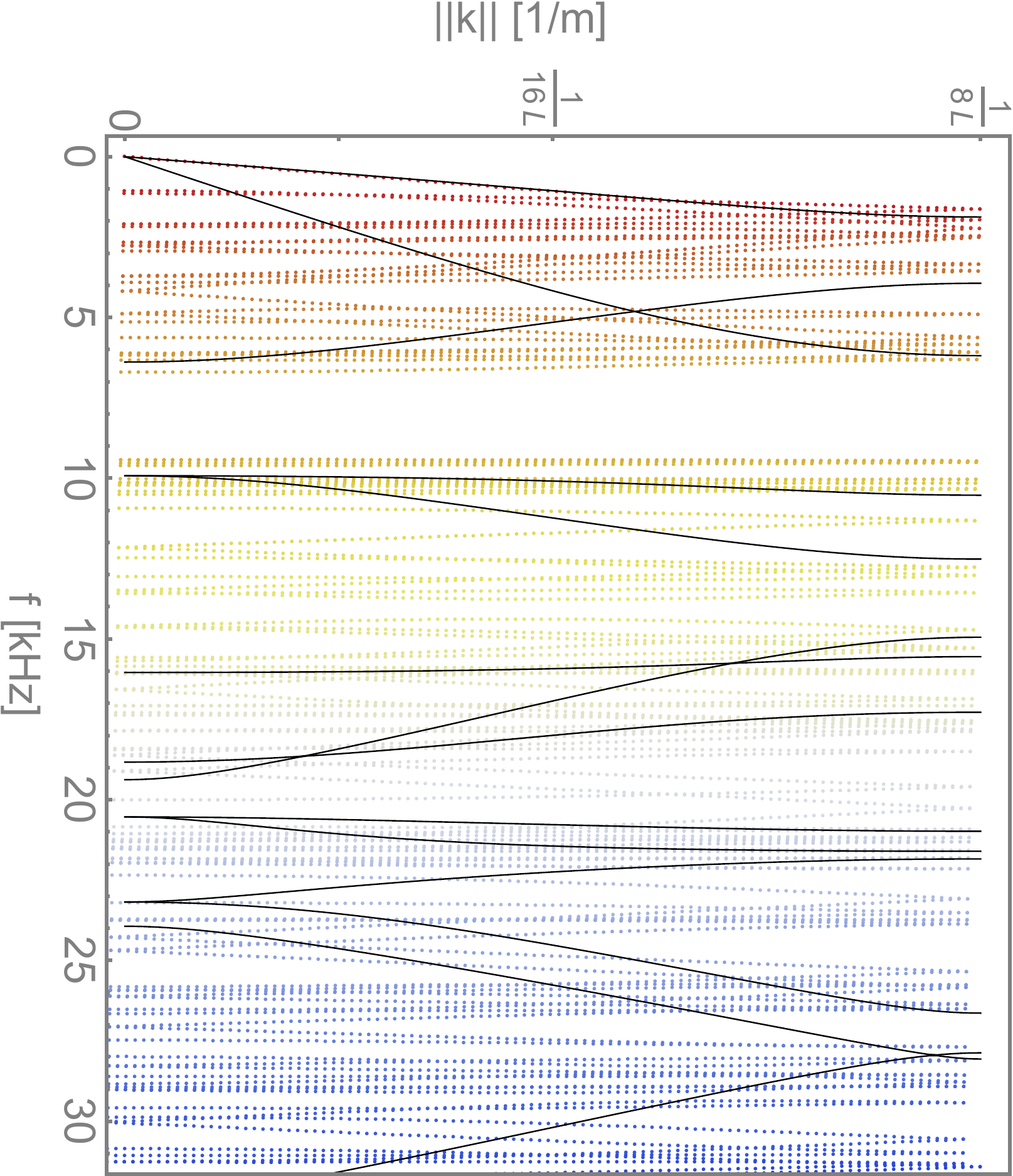}
  \end{minipage}
  \\
  \begin{minipage}{\textwidth}
    \centering
    \includegraphics[width=0.28\textwidth]{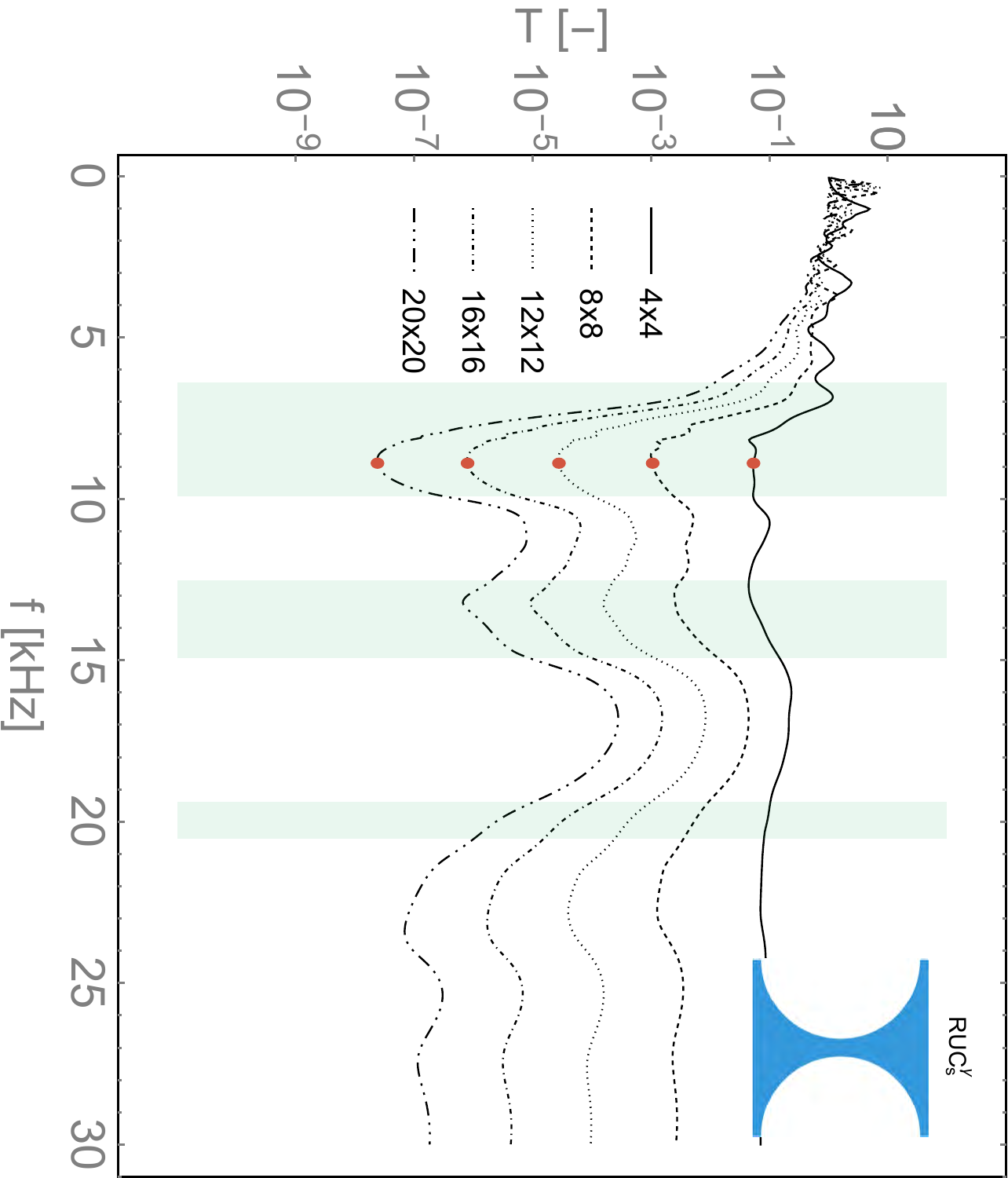}
    \hfill
    \includegraphics[width=0.28\textwidth]{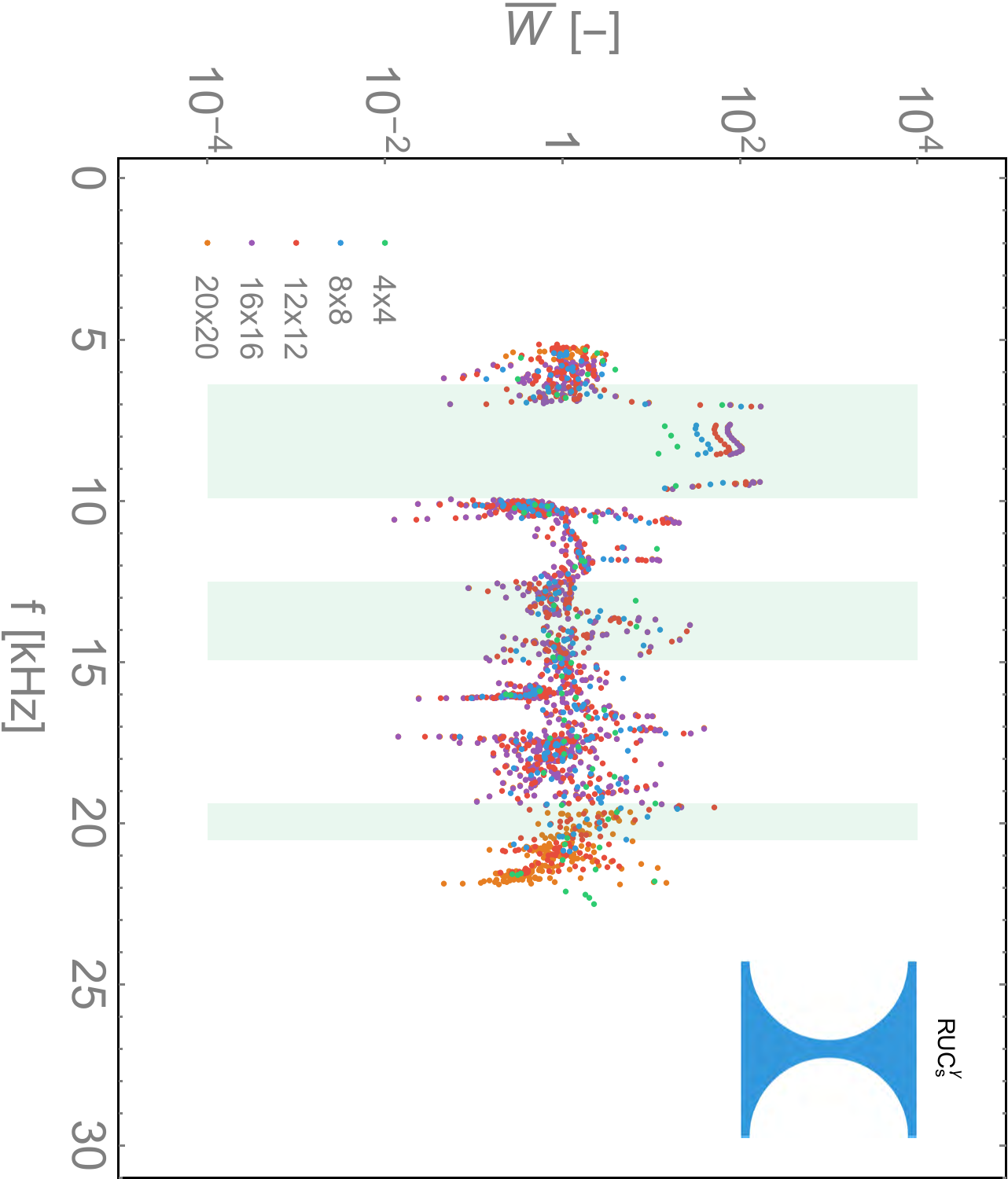}
    \hfill
    \includegraphics[width=0.28\textwidth]{Figures/Circ_disp_curves_RE_14_alpha_gamma_only_LR.pdf}
  \end{minipage}
  \\
  \begin{minipage}{\textwidth}
    \centering
    \includegraphics[width=0.28\textwidth]{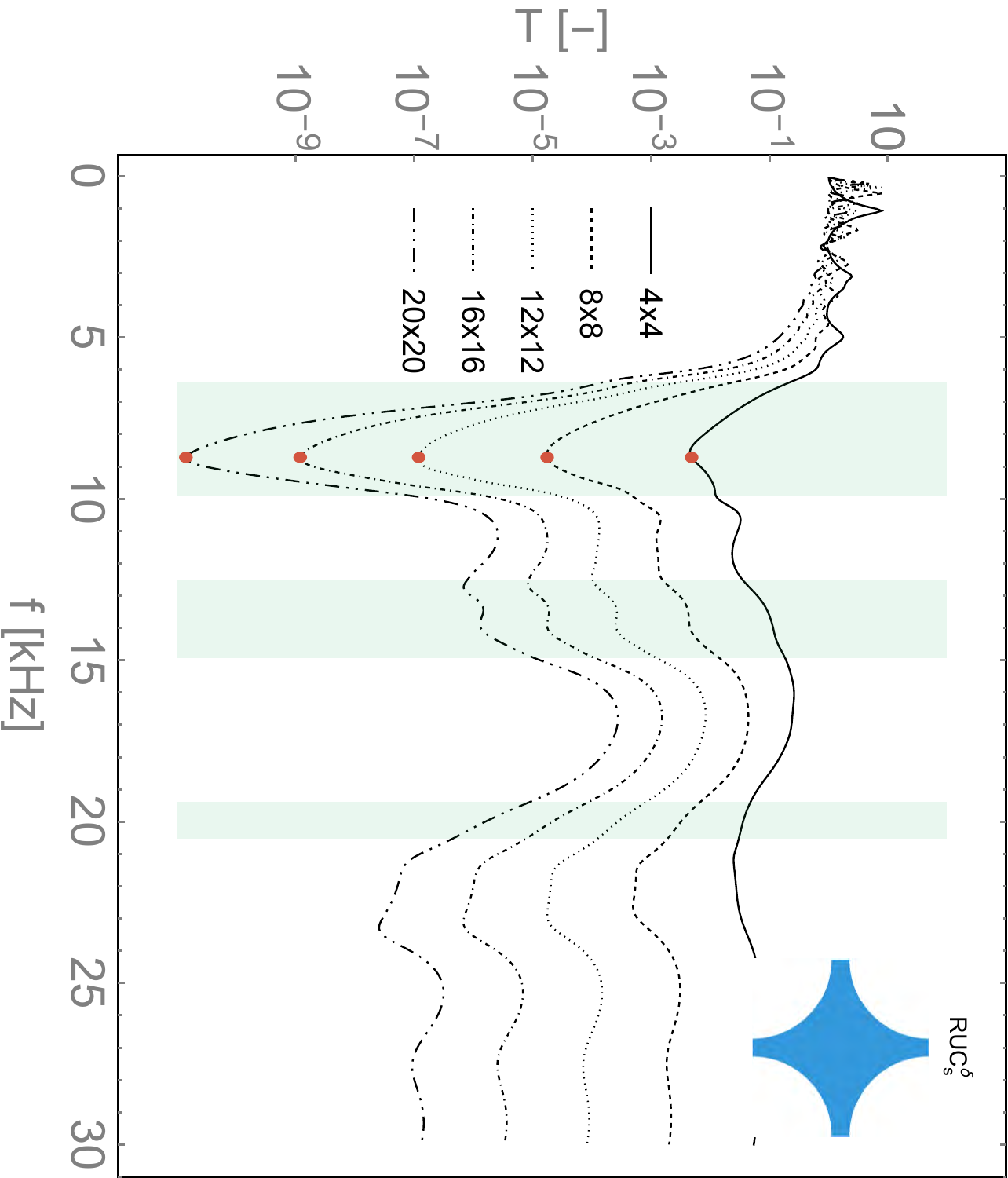}
    \hfill
    \includegraphics[width=0.28\textwidth]{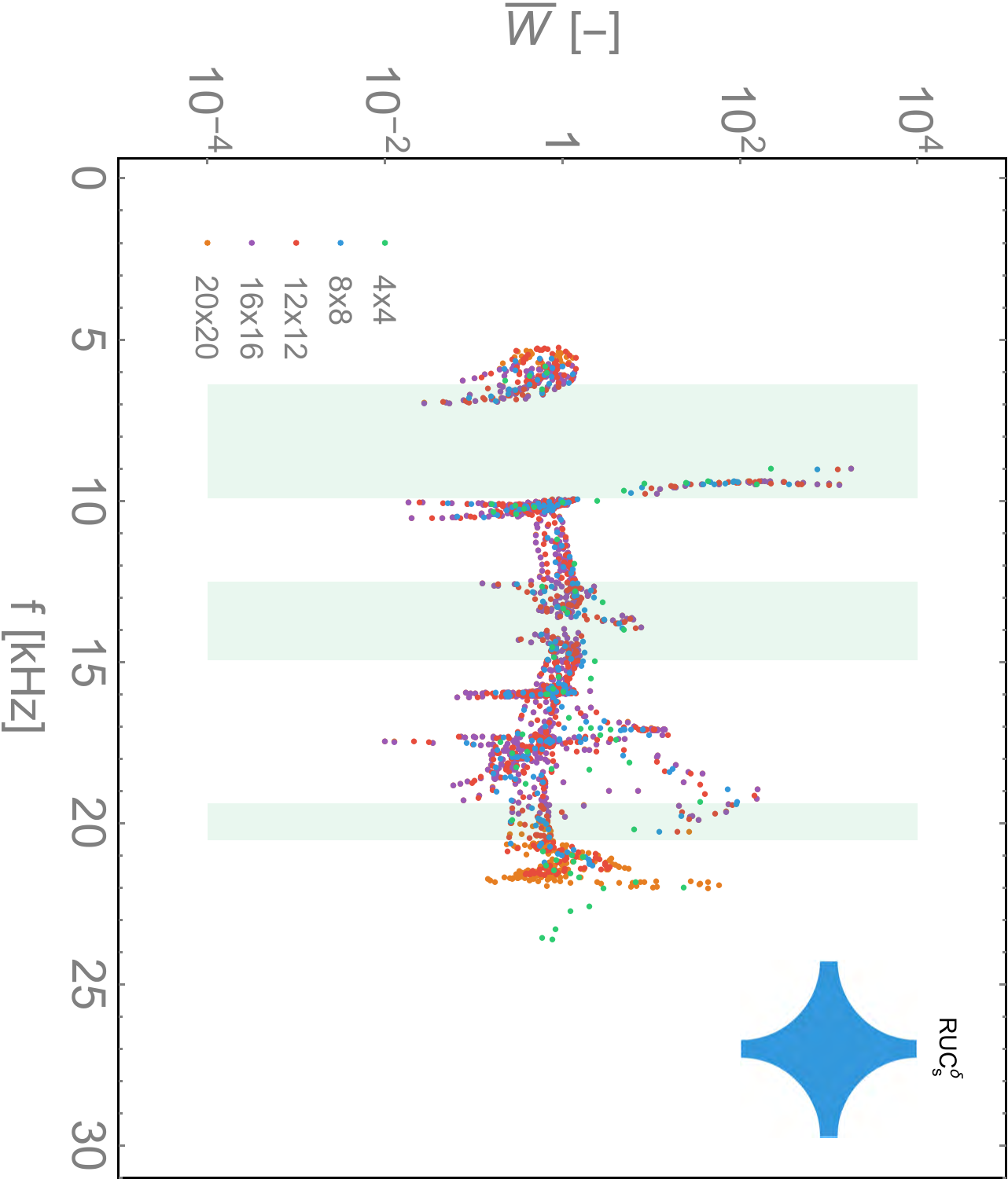}
    \hfill
    \includegraphics[width=0.28\textwidth]{Figures/Circ_disp_curves_RE_14_beta_delta_only_LR.pdf}
  \end{minipage}
  \caption{
    (\textit{First column}) Transmissibility and
    (\textit{second column}) eigenfrequency for a parametric study of the RUC$_{\rm s}^{\star}$ of the square-circular-hole metamaterial, with the structure size increasing from $4 \times 4$ to $20 \times 20$ in increments of four;
    (\textit{third column}) dispersion curves for a $4 \times 4$ structure with Floquet conditions applied exclusively in the horizontal direction (the black lines represent the classical BF analysis for a single unit cell).
  }
  \label{fig:circ_transmission_beta_gamma_delta_size}
\end{figure}
%%%%%%%%%%%%%%%%%%%%%%%%%%%%%%%%%%%%%%%%%%%%%%%%%%%%%%%%%%%
%%%%%%%%%%%%%%%%%%%%%%%%%%%%%%%%%%%%%%%%%%%%%%%%%%%%%%%%%%%%%%%%%%%%%%%%%%%%%%%%
\end{document}